\documentclass[twocolumn,english,letterpaper,superscriptaddress,prb,aps,showpacs]{revtex4}
\usepackage[T1]{fontenc}
\usepackage[latin1]{inputenc}
\usepackage{amsmath}
\usepackage{graphicx}
\usepackage{amssymb}

\makeatletter

\usepackage{bm}

\usepackage{dcolumn}

\usepackage{endnotes}

\usepackage{amsfonts}

\providecommand{\U}[1]{\protect\rule{.1in}{.1in}}
\usepackage{soul}

\usepackage{cancel}

\@ifundefined{definecolor}
 {\usepackage{color}}{}


\makeatother

\usepackage{babel}

\begin{document}

\author{Alexey A. Kovalev}

\affiliation{Department of Physics and Astronomy, University of California, Los
Angeles, California 90095, USA}

\affiliation{Department of Physics, Texas A\&M University, College Station, TX
77843-4242, USA}

\author{Yaroslav Tserkovnyak}

\affiliation{Department of Physics and Astronomy, University of California, Los
Angeles, California 90095, USA}

\author{Karel Výborný}

\affiliation{Institute of Physics ASCR, Cukrovarnická 10, 162 53 Praha 6, Czech
Republic }

\author{Jairo Sinova}

\affiliation{Department of Physics, Texas A\&M University, College Station, TX
77843-4242, USA}

\affiliation{Institute of Physics ASCR, Cukrovarnická 10, 162 53 Praha 6, Czech
Republic }

\title{Transport theory for disordered multiple-band systems: Anomalous
Hall effect and anisotropic magnetoresistance}
\begin{abstract}
We present a study of transport in multiple-band non-interacting Fermi
metallic systems based on the Keldysh formalism, taking into account
the effects of Berry curvature due to spin-orbit coupling. We apply
this formalism to a Rashba 2DEG ferromagnet and calculate the anomalous
Hall effect (AHE) and anisotropic magnetoresistance (AMR). 
The numerical calculations reproduce analytical results in the metallic
regime revealing the crossover between the skew scattering mechanism
dominating in the clean systems and intrinsic mechanism dominating
in the moderately dirty systems. As we increase the disorder further,
the AHE starts to diminish due to the spectral broadening of the quasiparticles.
Although for certain parameters this reduction of the AHE can be approximated
as $\sigma_{xy}\thicksim\sigma_{xx}^{\varphi}$ with $\varphi$ varying
around $1.6$, this is found not to be true in general as $\sigma_{xy}$
can go through a change in sign as a function of disorder strength
in some cases. The reduction region in which the quasiparticle approximation
is meaningful is relatively narrow; therefore, a theory with a wider
range of applicability is called for. By considering the higher order
skew scattering processes, we resolve some discrepancies between the
AHE results obtained by using the Keldysh, Kubo and Boltzmann approaches.
We also show that similar higher order processes are important for
the AMR when the nonvertex and vertex parts cancel each other. We
calculate the AMR in anisotropic systems properly taking into account
the anisotropy of the non-equilibrium distribution function. These
calculations confirm recent findings on the unreliability of common
approximations to the Boltzmann equation. 
\end{abstract}

\date{\today{}}


\maketitle

\section{Introduction}

Recently, the interest in transport calculations in multiple-band
systems\citep{Onoda:apr2008,Shindou:jan2008} has been rekindled in
part due to the realization of diluted magnetic semiconductors (DMS)
that have strong spin-orbit interactions, variable carrier densities,
and ferromagnetic ordering. These properties imply the existence of
the anomalous Hall effect (AHE)\citep{Hall:jan1880} and the anisotropic
magnetoresistance (AMR).\citep{Thomson:jan1856} Even though the mechanisms
of the AHE and the AMR are different, they both have a similar description
based on the multiple-band transport theory. In this paper, we formulate
a relatively simple framework for doing such transport calculations.

The AHE is usually described in terms of the anomalous Hall resistivity
$\rho_{xy}$ that measures the transverse voltage with respect to
the transport direction and depends on the spontaneous magnetization
$M$ along the $z$ direction. Theoretical studies of the AHE have
a long history beginning with the work of Karplus and Luttinger.\citep{Karplus:sep1954}
A number of papers on the AHE also appeared not so long ago,\citep{Taguchi:mar2001,Jungwirth:may2002,Onoda:jan2002,Yao:jan2004,Lee:mar2004,Zeng:jan2006,Rashba:aug2008}
after the interpretation of the AHE based on the Berry phase\citep{Sundaram:jun1999}
was proposed. Nevertheless, theoretical description of the AHE is
far from being complete and it often involves cumbersome calculations
without transparent interpretations.\citep{Sinova:2004} The difficulties
appear due to the necessity to consider the off-diagonal elements
in Bloch band indices (the interband coherences induced by charge
currents). There is a general trend to focus on particular simple
models in order to overcome the common mistakes that are made in treating
the AHE. A number of recent publications concentrate on the simpler
but non-trivial Rashba 2D electron system,\citep{Culcer:jul2003,Dugaev:jun2005,Sinitsyn:jul2005,Liu:oct2006,Inoue:jul2006,Onoda:sep2006,Borunda:aug2007,Nunner:dec2007,Kato:sep2007,Onoda:apr2008}
yet arriving at contradictory predictions. Most of the disagreements
have been finally resolved\citep{Kato:sep2007,Nunner:dec2007,Kovalev:jul2008}
with some being addressed in this paper.

In calculating the AHE for a given material, the usual approximations
performed to leading order in $\hbar/\tau\varepsilon_{F}$ can fail,
where $\tau$ is the scattering time and $\varepsilon_{F}$ is the
Fermi energy. The semiclassical description of the Hall conductivity
within the usual Boltzmann equation leads to an AHE contribution due
to the scattering asymmetry in the collision term usually labeled
as skew scattering.\citep{Smit:1955} Other terms, arising from subtle
issues dealing with interband coherence during the collision and acceleration
by the electric field between collisions, are usually introduced by
hand through the so called anomalous velocity\citep{Nozieres:1973}
and side-jump.\citep{Sinitsyn:feb2006} This approach however, is
non-systematic and prone to errors from missing terms and wrong interpretations,
e.g. such as giving physical meaning to gauge dependent quantities.
A more systematic way to derive the correct semiclassical equations
is through the Keldysh formalism in which these interband coherences
effects are taken into account automatically.\citep{Onoda:apr2008,Kovalev:jul2008}

The system under consideration also allows us to study the diagonal
resistance as a function of the direction of the magnetization. The
change in the resistance as a function of the magnetization direction
relative to the current or crystallographic direction is called the
AMR effect. The microscopic origin of the AMR in transition metal
ferromagnets is still elusive\citep{Smit:jun1951,Berger:jun1964,McGuire:jul1975,Jaoul:mar1977}
and detailed calculations require consideration of complicated band
structures.\citep{Banhart:nov1995,Velev:mar2005} A relatively simple
host band structure in the DMS ferromagnets provides a possibility
for performing detailed microscopic calculations based on simple physical
models.\citep{Rushforth:oct2007} However, the relaxation time approximation
used in such calculations is not always reliable since it does not
fully take into account the anisotropies of the system.\citep{Vyborny:oct2008}
The Kubo formula approach has been applied to the AMR calculations
in Rashba systems and it has revealed the cancellation of the nonvertex
and vertex parts,\citep{Kato:jun2008} similar to the spin Hall effect
(SHE) and the AHE.

In this paper, we apply the Keldysh formalism for transport calculations
in multiple-band non-interacting Fermi systems. This treatment simultaneously
takes into account the Berry curvature effects (interband coherences)
and scattering, allowing us to immediately account for such physical
effects as side-jump scattering and skew scattering within the same
footing. We calculate the AHE analytically and numerically for the
Rashba model and find in agreement with Onoda \textit{et al.}\citep{Onoda:sep2006,Onoda:apr2008}
three distinct regimes: the skew scattering regime, the disorder independent
regime, and the dirty regime in which, although the basis of theory
is not as well established, a distinct rapid reduction of the AHE
is observed as the conductivity $\sigma_{xx}$ diminishes. Even though
almost all ferromagnetic systems are three dimensional, the findings
of this simple 2D model has been linked to higher dimensional systems
arguing that most likely the major contributions to the AHE come from
the band anti-crossing regions\citep{Onoda:apr2008} similar to one
observed in the Rashba model.

We further analyze the scaling found in the dirty regime\citep{Onoda:sep2006,Onoda:apr2008}
in which the AHE seems to diminish in a manner that can be approximated
as: $\sigma_{xy}\thicksim\sigma_{xx}^{\varphi}$ with $\varphi$ being
close to $1.6$. Some experimental results claim to confirm such scaling;\citep{Ueno:2007,Miyasato:aug2007,Fukumura:2007,Venkateshvaran:2008,Fernandez-Pacheco:2008}
however, treatment of some of these experimental results has to be
done with extra care as the region of interest is often restricted
to less than a single decade, the materials have strong mangetoresistances
and in-plane anisotropies associated with them, and most of the data
associated with the zero field calculation is in fact at very high
magnetic fields. Although our numerical results confirm this scaling,
it is found to be in a very narrow region as the quasiparticle approximation
fails when $\tau\varepsilon_{F}\thicksim1$. In addition, changing
the sign of the scatterer changes the sign of the skew effect and
no scaling is observed. Although this simple model seems to capture
qualitative aspects of the three regions, to make a quantitative link
to 3D materials with much more complex behavior seems premature at
this stage. In our calculations, we also identify the hybrid skew
scattering regime of the AHE resulting from the higher order scattering
processes. Such processes appear to be important for the AMR as the
nonvertex and vertex diagrammatic parts cancel each other for the
Rashba model.\citep{Kato:jun2008} Our results suggest that the relaxation
time approximation is not always reliable for the AMR calculations
as it has been shown recently within the Boltzmann equation treatment.\citep{Vyborny:oct2008}

The paper is organized as follows. In Sec. II, we develop a general
formulation of transport in multiple-band non-interacting Fermi systems
with further generalizations in Appendix A. In Sec. III, we calculate
the AHE in 2DEG ferromagnet with spin-orbit interaction. The analytical
and numerical results are followed by discussions and comparison to
other works. In Sec. IV, we calculate the AMR in 2DEG ferromagnet
with spin-orbit interaction. Finally in Sec. V, we present our conclusions.

\section{Transport in multiple-band systems }

The method presented in this section can be applied to a multiple-band
system described by a Hamiltonian $\hat{H}_{0}+\hat{V}(\mathbf{r})$
that is a matrix in the band (chiral) index. In this section, we first
derive general non-linear equations using nonequilibrium diagrammatic
technique, further restricting our consideration to a linear response
theory.

\subsection{Quantum kinetic equation}

We start by defining the following Green's functions:\citep{Rammer:apr1986}\begin{equation}
\begin{array}{c}
\hat{G}_{11}\equiv-i\left\langle T_{c}\boldsymbol{\Psi}(1_{+})\boldsymbol{\Psi}^{\dagger}(1_{+}^{'})\right\rangle =-i\left\langle \overrightarrow{T}\boldsymbol{\Psi}(1_{+})\boldsymbol{\Psi}^{\dagger}(1_{+}^{'})\right\rangle ,\\
\hat{G}_{21}\equiv-i\left\langle T_{c}\boldsymbol{\Psi}(1_{-})\boldsymbol{\Psi}^{\dagger}(1_{+}^{'})\right\rangle =-i\left\langle \boldsymbol{\Psi}(1_{-})\boldsymbol{\Psi}^{\dagger}(1_{+}^{'})\right\rangle ,\\
\hat{G}_{12}\equiv-i\left\langle T_{c}\boldsymbol{\Psi}(1_{+})\boldsymbol{\Psi}^{\dagger}(1_{-}^{'})\right\rangle =i\left\langle \boldsymbol{\Psi}^{\dagger}(1_{-}')\boldsymbol{\Psi}(1_{+})\right\rangle ,\\
\hat{G}_{22}\equiv-i\left\langle T_{c}\boldsymbol{\Psi}(1_{-})\boldsymbol{\Psi}^{\dagger}(1_{-}^{'})\right\rangle =-i\left\langle \overleftarrow{T}\boldsymbol{\Psi}(1_{-})\boldsymbol{\Psi}^{\dagger}(1_{-}^{'})\right\rangle ,\end{array}\label{Green'sFunctions}\end{equation}
 where $T_{c}$ is the generalized time ordering operator acting on
the Keldysh contour which can be split in two time axis $t_{+}$ (forward)
and $t_{-}$ (backward), $\boldsymbol{\Psi}$ is the vector in the
band (chiral) space corresponding to the Fermi field, and $1_{\pm}=(\mathbf{r},t_{\pm})$
is the variable that describes the spatial variable $\mathbf{r}$
and the time variable $t$. The generalized time ordering operator
performs an ordinary time ordering $\overrightarrow{T}$ for the time
$t_{+}$, an anti-time ordering $\overleftarrow{T}$ for the time
$t_{-}$ and in the mixed case $t_{-}$ occurs always after $t_{+}$
within the Keldysh time contour. We can now define the Green's function
in the Keldysh space:\begin{equation}
\tilde{G}=\left(\begin{array}{cc}
\hat{G}_{11} & \hat{G}_{12}\\
\hat{G}_{21} & \hat{G}_{22}\end{array}\right).\label{GKeldysh}\end{equation}
 The scattering potential due to impurities in the Keldysh space has
the form:\begin{equation}
\tilde{V}(1,1')=\left(\begin{array}{cc}
\hat{V}(\mathbf{r}) & 0\\
0 & -\hat{V}(\mathbf{r})\end{array}\right)\delta(1-1'),\label{ScatteringInKeldysh}\end{equation}
where $\hat{V}(\mathbf{r})$ describes the potential in the band (chiral)
space formed by many scatterers which for current consideration can
have any general matrix form. The negative sign arises here simply
because the lower branch integration is taken from $+\infty$ to $-\infty$
while in the Keldysh loop the time goes from $-\infty$ to $+\infty$.
The Green's function in Eq. (\ref{GKeldysh}) allows for a perturbation
expansion relying on the Feynman rules. However, the four matrix elements
of a so defined Green's function are linearly dependent, i.e. $\hat{G}_{12}+\hat{G}_{21}=\hat{G}_{11}+\hat{G}_{22}$.
Hence it is advantageous to perform a linear transformation in the
Keldysh space to eliminate one matrix element in Eq. (\ref{GKeldysh}):\[
\check{G}=\left(\begin{array}{cc}
1 & 0\\
1 & -1\end{array}\right)\left(\begin{array}{cc}
\hat{G}_{11} & \hat{G}_{12}\\
\hat{G}_{21} & \hat{G}_{22}\end{array}\right)\left(\begin{array}{cc}
1 & 0\\
-1 & 1\end{array}\right)=\left(\begin{array}{cc}
\hat{G}^{R} & \hat{G}^{<}\\
0 & \hat{G}^{A}\end{array}\right),\]
 which leads to the following scattering potential:\[
\check{V}=\left(\begin{array}{cc}
1 & 0\\
1 & 1\end{array}\right)\tilde{V}\left(\begin{array}{cc}
1 & 0\\
1 & -1\end{array}\right)=\left(\begin{array}{cc}
1 & 0\\
0 & 1\end{array}\right)\hat{V}(\mathbf{r})\delta(1-1'),\]
 where $\hat{G}^{R}=\hat{G}_{11}-\hat{G}_{12}$ is the retarded Green's
function, $\hat{G}^{A}=\hat{G}_{12}-\hat{G}_{22}$ is the advanced
Green's function and $\hat{G}^{<}=\hat{G}_{12}$. There are other
choices for the linear transformation, and our choice is dictated
by the fact that the Green's function $\hat{G}^{<}$ can be immediately
related to the distribution function in the Boltzmann equation.\citep{Mahan:1990}

As of now, it is assumed that $\hat{V}(\mathbf{r})$ describes some
disordered potential and all Green's functions are averaged over this
disorder. In the transformed Keldysh space, the Dyson equation\citep{Rammer:apr1986}
becomes: \begin{eqnarray}
\left(\begin{array}{cc}
\hat{G}_{0}^{-1}-\hat{\Sigma}^{R} & -\hat{\Sigma}^{<}\\
0 & \hat{G}_{0}^{-1}-\hat{\Sigma}^{A}\end{array}\right)\otimes\left(\begin{array}{cc}
\hat{G}^{R} & \hat{G}^{<}\\
0 & \hat{G}^{A}\end{array}\right)=\check{1},\label{Dyson}\end{eqnarray}
 where R, A, and < respectively stand for the retarded, advanced and
lesser components of the disorder averaged Green's functions and self-energies.
The symbol $\otimes$ denotes a convolution (in position, time and
band/spin). The diagonal components of Eq. (\ref{Dyson}), yield the
two equations for the retarded and advanced Green's functions: \begin{equation}
(\hat{G}_{0}^{-1}-\hat{\Sigma}^{R/A})\otimes\hat{G}^{R/A}=\hat{1}.\label{DiagonalEquations}\end{equation}
 The off-diagonal component of Eq. (\ref{Dyson}) yields the kinetic
equation (sometimes called quantum Boltzmann equation) which contains
the non-equilibrium information necessary to study transport: \begin{equation}
[\hat{G}^{R}]^{-1}\otimes\hat{G}^{<}-\hat{\Sigma}^{<}\otimes\hat{G}^{A}=0.\label{Kinetic0}\end{equation}

\begin{figure}[t]
\centerline{\includegraphics[scale=0.5]{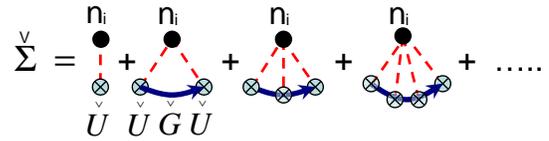}}

\caption{The non-equilibrium self-energy calculated using the self-consistent
$T$ matrix approximation in Keldysh space. }

\label{NEdiagrams} 
\end{figure}

In order to solve Eq. (\ref{Kinetic0}), one has to calculate the
self energy $\hat{\Sigma}^{<}$ of the particular problem. Here we
focus on scattering by randomly distributed identical impurities at
zero temperature with \begin{equation}
\hat{V}(\mathbf{r})=\sum_{i}\hat{\eta}U(\mathbf{r}-\mathbf{r}_{i}),\label{IdenticalImp}\end{equation}
where $\mathbf{r}_{i}$ describes the positions of random impurities
of density $n_{i}$ and $\hat{\eta}$ is some matrix in the band index
\textit{(e.g. }in section III, it is a unit matrix corresponding to
scalar impurities, and in section IV, it is a combination of unit
and unitary matrices corresponding to charged and magnetic impurities).
A common approximation to this problem is the self-consistent T-matrix
approximation (TMA) which takes into account all the non-crossing
scattering events from single impurities (see Fig. \ref{NEdiagrams}).
We assume here that the system is uniform and $\check{G}$ depends
on the difference of spatial variables $(\mathbf{r}-\mathbf{r}_{i})$
(however, this requirement can be lifted for the short-range disorder
as it is shown in Appendix A). In this case, we can sum up the infinite
series of diagrams in Fig. \ref{NEdiagrams} arriving at the following
expression for the self energy in the momentum representation (for
the sake of compact form we use the momentum representation here):
\begin{equation}
\left\langle \mathbf{k}|\check{\Sigma}|\mathbf{k}'\right\rangle =n_{i}\left\langle \mathbf{k}|\check{T}|\mathbf{k}\right\rangle \delta(\mathbf{k}-\mathbf{k}'),\label{NEslefEnergy}\end{equation}
with the following expression for the T matrix operator of impurity
placed in the origin:\begin{equation}
\check{T}\equiv\left(\check{V}+\check{V}\otimes\check{G}\otimes\check{V}+\dots\right),\label{TmatrixSeries}\end{equation}
where $\check{V}=\left(\begin{array}{cc}
\hat{\eta} & 0\\
0 & \hat{\eta}\end{array}\right)U(\mathbf{r})\delta(1-1')$. Combining the T-matrix structure $\check{T}=\check{V}\otimes[\check{1}+\check{G}\otimes\check{T}]$
and solving for the off-diagonal component we obtain the equation
for the lesser component of self energy: \begin{equation}
\left\langle \mathbf{k}|\hat{\Sigma}^{<}|\mathbf{k}'\right\rangle =n_{i}\left\langle \mathbf{k}|\hat{T}^{R}\otimes\hat{G}^{<}\otimes\hat{T}^{A}|\mathbf{k}\right\rangle \delta(\mathbf{k}-\mathbf{k}').\label{Lesser}\end{equation}

The retarded and advanced T-matrices are given by the usual form \begin{eqnarray}
\hat{T}^{R(A)}=\hat{V}\otimes(1+\hat{G}^{R(A)}\otimes\hat{T}^{R(A)})\nonumber \\
=(1+\hat{T}^{R(A)}\otimes\hat{G}^{R(A)})\otimes\hat{V}.\label{DiagonalEquations1}\end{eqnarray}
 Equations (\ref{Kinetic0}) and (\ref{Lesser}) form a general closed
set of equations for $\hat{G}^{<}$. In order to solve these equations,
we can further simplify them by looking for a solution of the form
\begin{equation}
\hat{G}^{<}=\hat{G}_{2}^{<}+\hat{G}_{1}^{<},\label{TryFunction}\end{equation}
 where \begin{equation}
\hat{G}_{2}^{<}=n_{F}\otimes\hat{G}^{A}-\hat{G}^{R}\otimes n_{F},\label{G2}\end{equation}
 and the operator $n_{F}$ is the Fermi distribution function. In
the case of zero temperature, $n_{F}$ is the step function in the
frequency representation $n_{F}(\omega)=\theta(-\omega)$ and $n_{F}(t,t')=i/\left[2\pi(t-t'+i0)\right]$
in the time representation. Equations (\ref{TryFunction}) and (\ref{G2})
will allow us to separate the Fermi sea and Fermi surface components
of the lesser Green's function. By substituting Eq. (\ref{TryFunction})
into Eq. (\ref{Kinetic0}), we obtain the kinetic equation for $\hat{G}_{1}^{<}$:
\begin{eqnarray}
[\hat{G}^{R}]^{-1}\otimes\hat{G}_{1}^{<}-\hat{\Sigma}_{1}^{<}\otimes\hat{G}^{A}=\left[\hat{H}_{0}\overset{\otimes}{,}n_{F}\right]\otimes\hat{G}^{A},\label{Kinetic1}\end{eqnarray}
 with $\left\langle \mathbf{k}|\hat{\Sigma}_{1}^{<}|\mathbf{k}'\right\rangle =n_{i}\left\langle \mathbf{k}|\hat{T}^{R}\otimes\hat{G}_{1}^{<}\otimes\hat{T}^{A}|\mathbf{k}\right\rangle \delta(\mathbf{k}-\mathbf{k}')$,
where $\left[...\overset{\otimes}{,}...\right]$ stands for a commutator.
In order to derive Eq. (\ref{Kinetic1}), Eqs. (\ref{DiagonalEquations})
and (\ref{DiagonalEquations1}) are used along with the fact that
$\hat{T}^{R}\otimes\hat{G}_{2}^{<}\otimes\hat{T}^{A}=n_{F}\otimes\hat{T}^{A}-\hat{T}^{R}\otimes n_{F}$,
 and \begin{equation}
\hat{\Sigma}_{2}^{<}=n_{F}\otimes\hat{\Sigma}^{A}-\hat{\Sigma}^{R}\otimes n_{F},\label{Sigma2}\end{equation}
which is a consequence of Eqs. (\ref{Lesser},\ref{DiagonalEquations1},\ref{G2}).

The lesser Green's function contains all the information about the
transport properties of our system and the charge current density
can be calculated as:\begin{equation}
\begin{array}{lll}
j_{x(y,z)} & = & \dfrac{e}{2}{\rm Tr}\left\langle \boldsymbol{\Psi}^{\dagger}(1^{'})\hat{\upsilon}_{x(y,z)}(1)\boldsymbol{\Psi}(1)\right\rangle _{1=1^{'}}+c.c.\\
 & = & -\dfrac{ie}{2}{\rm Tr}\left[\left(\hat{\upsilon}_{x(y,z)}(1)+\hat{\upsilon}_{x(y,z)}^{\dagger}(1^{'})\right)\hat{G}^{<}(1,1^{'})\right]_{1=1^{'}},\end{array}\label{Current}\end{equation}
where $\boldsymbol{\hat{\upsilon}}(1)=\left(-i\hbar\boldsymbol{\nabla}_{1}-e\boldsymbol{\mathcal{\hat{A}}}(1)/c\right)/m$,
$\boldsymbol{\hat{\upsilon}}^{\dagger}(1^{'})=\left(i\hbar\boldsymbol{\nabla}_{1^{'}}-e\boldsymbol{\mathcal{\hat{A}}}^{\dagger}(1^{'})/c\right)/m$
and $\boldsymbol{\mathcal{\hat{A}}}(1)$ is the generalized vector
potential matrix in the band index that also describes spin-orbit
interactions; $e=-|e|$ stands for an electron charge.

\subsection{Linearized Fermi surface contribution}

The kinetic Eq. (\ref{Kinetic1}) has not assumed linearity in electric
field strength nor any particular temporal dependence. Higher order
terms in the impurity density $n_{i}$ corresponding to non-crossed
diagrams have been taken into account as the retarded and advanced
Green's functions in Eq. (\ref{Kinetic1}) are calculated self-consistently.
In the following, we solve the problem for linear response theory
of a uniform and stationary system in the presence of a uniform electric
field. 

In the presence of slowly varying perturbations, it is useful to perform
the Wigner transformation, \textit{viz}. the center-of-mass coordinates
($X=(\mathbf{R},T)$) and the Fourier transform with respect to the
relative coordinates ($k=(\mathbf{k},\omega)$). However, the Wigner
coordinate $\mathbf{k}$ associated with the momentum operator $-i\boldsymbol{\nabla}$
is not gauge invariant and consequently it is not the correct choice
for describing our system. On the other hand, the kinetic momentum
$\mathbf{k}(T)=-i\boldsymbol{\nabla}-e\mathbf{A}_{E}(T)/(\hbar c)$
is gauge invariant, and as it will be shown below, for the stationary
case all time dependence can be conceived in $\mathbf{k}(T)$; here
the vector potential $\mathbf{A}_{E}(T)$ describes the external electric
field. The time derivative within the canonical coordinates (marked
by wave) becomes a combination of time and momentum derivatives within
the kinetic coordinates: $\partial_{\tilde{T}}=\partial_{T}+\partial_{T}\mathbf{k}(T)\partial_{\mathbf{k}}$,
$\partial_{\mathbf{\tilde{R}}}=\partial_{\mathbf{R}}$, $\partial_{\mathbf{\tilde{k}}}=\partial_{\mathbf{k}}$
and $\partial_{\tilde{\omega}}=\partial_{\omega}$.

In the Wigner representation with the kinetic momentum, the convolution
of two operators is approximated as: \[
\begin{array}{lcl}
\hat{A}\otimes\hat{B} & = & \exp^{i(\partial_{X}^{A}\partial_{k}^{B}-\partial_{k}^{A}\partial_{X}^{B})/2}\hat{A}(X,k)\hat{B}(X,k)\\
\\ & \thickapprox & \hat{A}\hat{B}+\frac{i}{2}\left(\partial_{X}\hat{A}\partial_{k}\hat{B}-\partial_{k}\hat{A}\partial_{X}\hat{B}\right),\end{array}\]
 where we use the four vector notations $\partial_{X}\partial_{k}=\partial_{\mathbf{R}}\partial_{\mathbf{k}}-\partial_{\tilde{T}}\partial_{\omega}$
and $\partial_{\tilde{T}}=\partial_{T}+\frac{e\mathbf{E}}{\hbar}\partial_{\mathbf{k}}$.
Here, we assume that a vector potential $\mathbf{A}_{E}(T)=-c\mathbf{E}T$
which corresponds to a uniform electric field $\mathbf{E}$. The first
order gradient expansion is sufficient for the linear response theory,
while the second order gradient expansion may be necessary for time
dependent problems and when the Hamiltonian $\hat{H}_{0}$ is spatially
dependent in order to account for the corresponding Berry curvature
effects.\citep{Shindou:jan2008} Since we are seeking homogeneous
solutions both in space and time with respect to the center-of-mass
coordinates, the only surviving terms in the expansion are \begin{eqnarray}
\hat{A}\otimes\hat{B}\thickapprox & \hat{A}\hat{B}-\frac{i}{2\hbar}e\mathbf{E}\left(\partial_{{\bf k}}\hat{A}\partial_{\omega}\hat{B}-\partial_{\omega}\hat{A}\partial_{{\bf k}}\hat{B}\right)\end{eqnarray}

Applying the above Wigner transformation to $\hat{G}_{2}^{<}$ in
Eq. (\ref{G2}), we obtain directly \begin{equation}
\hat{G}_{2}^{<}=n_{F}(\hat{G}^{A}-\hat{G}^{R})+\frac{i}{2\hbar}\partial_{\omega}n_{F}e\mathbf{E}(\partial_{\mathbf{k}}\hat{G}_{eq}^{A}+\partial_{\mathbf{k}}\hat{G}_{eq}^{R}),\end{equation}
 where $\hat{G}_{eq}^{R/A}$ are the Green's functions evaluated at
equilibrium, i.e. ${\bf E}=0$. $\hat{G}_{2}^{<}$ solves the Kinetic
Eq. (\ref{Kinetic0}) up to zeroth order in the electric field $E$,
and therefore the expansion in $E$ of $\hat{G}_{1}^{<}$ and $\hat{\Sigma}_{1}^{<}$
starts from the linear in ${\bf E}$ terms. With this knowledge, we
apply the Wigner transformation to Eq. (\ref{Kinetic1}), and find
the self-consistent simple form of the kinetic equation for $\hat{G}_{1}^{<}$:
\begin{eqnarray}
 & \hat{G}_{1}^{<}=\hat{G}_{eq}^{R}\hat{\Sigma}_{1}^{<}\hat{G}_{eq}^{A}-{i}e\mathbf{E}(\partial_{\omega}n_{F})\hat{G}_{eq}^{R}\boldsymbol{\hat{\upsilon}}\hat{G}_{eq}^{A}\label{Kinetic2}\\
 & \hat{\Sigma}_{1}^{<}=n_{i}\int{\displaystyle \frac{d^{2}k'}{(2\pi)^{2}}}\hat{T}_{eq}^{R}(\mathbf{k},\mathbf{k}')\hat{G}_{1}^{<}(\mathbf{k}')\hat{T}_{eq}^{A}(\mathbf{k}',\mathbf{k})\label{Kinetic3}\end{eqnarray}
 where $\boldsymbol{\hat{\upsilon}}=\partial\hat{H}_{0}/\partial{\hbar{\bf k}}$,
and $\hat{T}_{eq}^{R/A}$ are self-consistent T-matrices evaluated
at equilibrium. In the following section, we show how to solve the
kinetic Eqs. (\ref{Kinetic2}) and (\ref{Kinetic3}) for a simple
system described by the Rashba Hamiltonian. Whereas solving Eqs. (\ref{Kinetic2})
and (\ref{Kinetic3}) require only the equilibrium retarded and advance
Green's functions and T-matrices, note that for $\hat{G}_{2}^{<}$
we need to solve these Green's functions up to linear order in ${\bf E}$
(see below).

From the equations above, it is natural to decompose the contributions
to $\hat{G}^{<}$ into the Fermi sea and Fermi surface contributions\citep{Onoda:apr2008}
such that $\hat{G}^{<}=\hat{G}_{1}^{<}+\hat{G}_{2}^{<}=\hat{G}_{I}^{<}+\hat{G}_{II}^{<}$
where \begin{equation}
\begin{array}{ccl}
\hat{G}_{I}^{<} & = & \hat{G}_{1}^{<}+\frac{i}{2\hbar}(\partial_{\omega}n_{F})e\mathbf{E}(\partial_{\mathbf{k}}\hat{G}_{eq}^{A}+\partial_{\mathbf{k}}\hat{G}_{eq}^{R}),\end{array}\label{Separation1}\end{equation}
\begin{equation}
\begin{array}{lll}
\hat{G}_{II}^{<} & = & n_{F}(\hat{G}^{A}-\hat{G}^{R}).\end{array}\label{Separation}\end{equation}
Next, we linearize Eq. (\ref{Current}) in $\mathbf{E}$, carry out
the Wigner transformation and insert the two components of $\hat{G}^{<}$,
arriving at the two corresponding components of the current density:

\begin{equation}
j_{x(y,z)}^{I}=-ie{\displaystyle {\displaystyle \intop}}{\displaystyle \frac{d^{2}\mathbf{k}}{(2\pi)^{2}}}{\displaystyle \frac{d\omega}{2\pi}}{\rm Tr}\left(\hat{G}_{I}^{<}\hat{\upsilon}_{x(y,z)}\right),\label{CurrentI}\end{equation}
 \begin{equation}
j_{x(y,z)}^{II}=-ie{\displaystyle {\displaystyle \intop}}{\displaystyle \frac{d^{2}\mathbf{k}}{(2\pi)^{2}}}{\displaystyle \frac{d\omega}{2\pi}}{\rm Tr}\left(\hat{G}_{II}^{<}\hat{\upsilon}_{x(y,z)}\right),\label{CurrentII}\end{equation}
 where the Fermi surface ($j_{x(y,z)}^{I}$) and Fermi sea ($j_{x(y,z)}^{II}$)
contributions are identical to ones defined within Kubo-Streda formalism.\citep{Streda:aug1982}
Equations (\ref{Kinetic2}), (\ref{Kinetic3}) and (\ref{Separation1})
are the main results of this subsection.

\subsection{Linearized Fermi sea contribution}

In order to calculate the Fermi sea contribution using Eqs. (\ref{Separation})
and (\ref{CurrentII}), we expand the retarded (advanced) Green's
function and self-energy up to the first order in $\mathbf{E}$ following
the procedure of Onoda \textit{et al}.:\citep{Onoda:apr2008}\begin{equation}
\begin{array}{c}
\hat{G}^{R(A)}=\hat{G}_{eq}^{R(A)}+e\mathbf{E}\hat{G}_{\mathbf{E}}^{R(A)}+O(E^{2}),\\
\\\hat{\Sigma}^{R(A)}=\hat{\Sigma}_{eq}^{R(A)}+e\mathbf{E}\hat{\Sigma}_{\mathbf{E}}^{R(A)}+O(E^{2}),\end{array}\label{GIIexpand}\end{equation}
 where $\hat{G}_{\mathbf{E}}^{R}=\frac{1}{e}\partial_{\mathbf{E}}\hat{G}^{R}|_{E=0}$,
$\hat{\Sigma}_{\mathbf{E}}^{R}=\frac{1}{e}\partial_{\mathbf{E}}\hat{\Sigma}^{R}|_{E=0}$
and $\hat{G}_{eq}^{R/A}$ ($\hat{\Sigma}_{eq}^{R(A)}$) are the Green's
functions (self-energies) evaluated at equilibrium, i.e. ${\bf E}=0$.
The Fermi sea lesser Green's function $\hat{G}_{II}^{<}$ calculated
up to the first order in the electric field $\mathbf{E}$ becomes:\begin{equation}
\hat{G}_{II}^{<}=n_{F}(\hat{G}_{eq}^{A}-\hat{G}_{eq}^{R})+n_{F}e\mathbf{E}(\hat{G}_{\mathbf{E}}^{A}-\hat{G}_{\mathbf{E}}^{R}).\label{GIIexpand1}\end{equation}
 We now substitute Eqs. (\ref{GIIexpand}) into Eqs. (\ref{DiagonalEquations})
and (\ref{DiagonalEquations1}) only retaining linear terms in $\mathbf{E}$
in order to arrive at the following self-consistent equations: \begin{eqnarray}
 & \hat{G}_{\mathbf{E}}^{R(A)}(\omega)=\hat{G}_{eq}^{R}\hat{\Sigma}_{\mathbf{E}}\hat{G}_{eq}^{R}-\frac{i}{2}\left[\hat{G}_{eq}^{R}\left(\hat{{\bf \upsilon}}{+}\partial_{{\bf \hbar k}}\hat{\Sigma}_{eq}^{R}\right)\partial_{\omega}\hat{G}_{eq}^{R}\right.\nonumber \\
 & -\left.\partial_{\omega}\hat{G}_{eq}^{R}\left(\hat{{\bf \upsilon}}{+}\partial_{{\bf \hbar k}}\hat{\Sigma}_{eq}^{R}\right)\hat{G}_{eq}^{R}\right],\label{KineticBelow}\end{eqnarray}
 \begin{equation}
\hat{\Sigma}_{\mathbf{E}}^{R(A)}(\omega)=n_{i}\int{\displaystyle \frac{d^{2}k'}{(2\pi)^{2}}}\hat{T}_{eq}^{R(A)}(\mathbf{k},\mathbf{k}')\hat{G}_{\mathbf{E}}^{R(A)}(\mathbf{k}')\hat{T}_{eq}^{R(A)}(\mathbf{k}',\mathbf{k}),\label{BELOWselfenergy}\end{equation}
 where in Eq. (\ref{DiagonalEquations}) we also performed the gradient
expansion. Equations (\ref{GIIexpand1}), (\ref{KineticBelow}) and
(\ref{BELOWselfenergy}) are the main results of this subsection.

\section{AHE in Rashba systems }

In this section, we apply the above formalism to 2DEG with exchange
field and spin-orbit interaction. A general numerical procedure is
followed by analytical results valid in the metallic regime in the
limit of small impurity scattering broadening $\hbar/\tau$ with respect
to the Fermi energy $\varepsilon_{F}$. We end the section with a
discussion of the numerical and analytical results comparing them
to other approaches. For convenience, and in order to keep the expressions
more concise, we introduce here the dimensionless units that can easily
be transformed into dimensional units by following equations at the
beginning of this section. Note that our formalism cannot be used
close to the energies $\omega=\pm h$ in Fig. \ref{Spectrum}, as
$k_{F}l$ ($l$ is the mean-free path) can become very small and the
non-crossing approximation in Fig. \ref{NEdiagrams} may fail. Nevertheless,
we do not expect large corrections to our results around these singularities
as the non-diagonal conductivity seems not to be strongly affected
by including the crossed diagrams.\citep{Dugaev:sep2001}

\subsection{Calculational procedure}

We restrict ourselves here to 2DEG Rashba Hamiltonian with an exchange
field $\breve{h}$ (breve accent here means that $h$ is in dimensional
units) in order to obtain simple analytical results that connect directly
with other microscopic linear response calculations:\citep{Inoue:jul2006,Nunner:dec2007,Sinitsyn:jan2007}
\begin{equation}
\hat{H}_{R}=\hat{1}(\hbar\mathbf{\breve{k}})^{2}/2m+\breve{\alpha}\mathbf{\breve{k}}\cdot\boldsymbol{\hat{\sigma}}\times\mathbf{z}-\breve{h}\hat{\sigma}_{z}+\hat{1}V(\mathbf{\breve{r}}),\label{Hamiltonian}\end{equation}
where $\breve{\alpha}$ is the strength of spin-orbit interaction,
$\boldsymbol{\hat{\sigma}}$ are Pauli matrices, $\hbar\mathbf{\breve{k}}=-i\hbar\boldsymbol{\nabla}-e\mathbf{A}/c$,
$\mathbf{A}(t)=-c\mathbf{E}t$ describes the external electric field
and $V(\mathbf{r})$ describes the impurities. From symmetry considerations,
the most general form of the Hamiltonian in Eq. (\ref{Hamiltonian})
should treat the coordinate $\mathbf{r}$ as an operator $\mathbf{r}+\mathbf{\hat{r}}_{so}(k)$
with $\mathbf{\hat{r}}_{so}(\mathbf{k})=\lambda\boldsymbol{\hat{\sigma}}\times\mathbf{k}$
originating from the projection procedure onto the band under consideration.\citep{Pikus:1984}
The spin-orbit interaction can also include higher \textsl{e.g.} cubic
terms relevant for the bulk InSb and the HgTe quantum wells with an
inverted band structure.\citep{Rashba1991,Zhang:jun2001} Here, only
linear terms with Rashba symmetry are considered with $\mathbf{\hat{r}}_{so}(\mathbf{k})$
being disregarded as we expect effect of $H_{SO}=\mathbf{\hat{r}}_{so}(\mathbf{k})\boldsymbol{\nabla}V(\mathbf{r})$
on the AHE to be small for wide band semiconductors in which $\lambda$
is relatively small.\citep{Engel:oct2005} The disorder in the system
is modeled by impurity delta-scatterers: \begin{equation}
V(\mathbf{r})=\breve{V}_{0}\sum_{i}\delta(\mathbf{\breve{r}}-\mathbf{\breve{r}}_{i}),\label{Delta}\end{equation}
 where $\mathbf{\breve{r}}_{i}$ describes the positions of randomly
distributed impurities of density $\breve{n}_{i}$.

\begin{figure}[t]
\centerline{\includegraphics[scale=0.4]{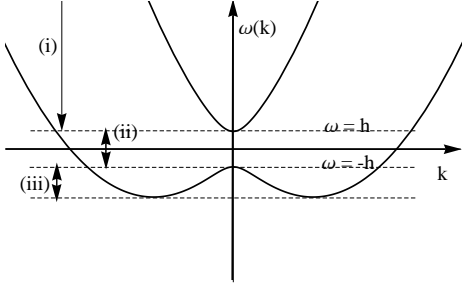}} 

\caption{Electronic band dispersions of the Rashba model; throughout the paper,
$\varepsilon_{F}$ is the Fermi energy measured from the bottom of
the lower band while $\omega_{F}$ is the Fermi energy measured from
the middle of the gap (region (ii)). }

\label{Spectrum} 
\end{figure}

We rewrite the Hamiltonian in dimensionless quantities: \begin{equation}
\frac{\hat{H}_{R}}{\varepsilon_{F}}=\hat{1}\frac{1}{2}\mathbf{k}^{2}+{\alpha}\mathbf{k}\cdot\boldsymbol{\hat{\sigma}}\times\mathbf{z}-h\hat{\sigma}_{z}+\hat{1}V_{0}\sum_{i}\delta(\mathbf{r}-\mathbf{r}_{i}),\label{HamiltonianDimLess}\end{equation}
 where $\varepsilon_{F}$ is the Fermi energy measured from the minimum
of energy, $k=\breve{k}l_{0}$ is the dimensionless momentum. The
dimensionality can be restored by substituting expressions for the
dimensionless units into the final formulas: \[
\begin{array}{c}
l_{0}={\displaystyle \sqrt{\frac{\hbar^{2}}{m\varepsilon_{F}}}},\;\alpha=\breve{\alpha}{\displaystyle \sqrt{\frac{m}{\hbar^{2}\varepsilon_{F}}}},\; V_{0}={\displaystyle \frac{m\breve{V}_{0}}{\hbar^{2}}},\\
h={\displaystyle \frac{\breve{h}}{\varepsilon_{F}}},\; n_{i}=\breve{n}_{i}l_{0}^{2},\; k=\breve{k}l_{0}.\end{array}\]
 Also note that whereas $\varepsilon_{F}$ is measured from the bottom
of the lower band, in the notation below, we introduce $\omega_{F}$
which is the Fermi energy measured from the middle of the gap (region
(ii) in Fig.\ref{Spectrum}).

In the following, we solve Eqs. (\ref{Kinetic2}) and (\ref{Kinetic3})
in order to find the non-equilibrium Green's function $\hat{G}_{1}^{<}$
describing processes at the Fermi surface, and Eqs. (\ref{KineticBelow})
and (\ref{BELOWselfenergy}) for the non-equilibrium Green's function
$\hat{G}_{2}^{<}$ - primarily Fermi sea contribution.

We calculate ${\textstyle \hat{\Sigma}_{eq}^{R(A)}}$ and the Green's
functions $\hat{G}_{eq}^{R(A)}$ using the self-consistent TMA, \textit{i.e.}
diagonal components of Eq. (\ref{NEslefEnergy}): \citealp{Onoda:sep2006,Onoda:apr2008}
\begin{equation}
\hat{T}_{eq}^{R(A)}=V_{0}(\hat{1}-V_{0}\hat{\gamma}^{R(A)})^{-1},\label{Tmatrix}\end{equation}
 \begin{equation}
\hat{\Sigma}_{eq}^{R(A)}=n_{i}\hat{T}_{eq}^{R(A)}(\omega)=\Sigma_{eq0}^{R(A)}\hat{\sigma}_{0}+\Sigma_{eq\mbox{z}}^{R(A)}\hat{\sigma}_{z},\label{SelfEnergy}\end{equation}
 \begin{equation}
\begin{array}{l}
{\textstyle \hat{G}_{eq}^{R(A)}}=(\omega\hat{1}-\hat{H}_{0}-\hat{\Sigma}_{eq}^{R(A)})^{-1}\\
={\displaystyle \frac{(\omega-\frac{k^{2}}{2}-\Sigma_{eq0}^{R(A)})\hat{\sigma}_{0}+\alpha k_{y}\hat{\sigma}_{x}-\alpha k_{x}\hat{\sigma}_{y}-(h-\Sigma_{eq\mbox{z}}^{R(A)})\hat{\sigma}_{z}}{(\omega-\frac{k^{2}}{2}-\Sigma_{eq0}^{R(A)})^{2}-(h-\Sigma_{eq\mbox{z}}^{R(A)})^{2}-\alpha^{2}k^{2}}},\end{array}\label{GreenRA}\end{equation}
 where $\hat{\gamma}^{R(A)}=\int{\displaystyle {d^{2}k}/{(2\pi)^{2}}}{\textstyle \hat{G}_{eq}^{R(A)}}(\mathbf{k},\omega)\equiv\gamma^{R(A)}\hat{\sigma}_{0}+\gamma_{z}^{R(A)}\hat{\sigma}_{z}$.
We calculate self-consistent value of the self energy $\hat{\Sigma}_{eq}^{R(A)}(\omega)$
for each $\omega$ by performing sufficient number of iterations in
Eq. (\ref{SelfEnergy}) in order to achieve the prescribed accuracy
(see Appendix B for details). 

With the knowledge of the equilibrium Green's function ${\textstyle \hat{G}_{eq}^{R}}(\mathbf{k},\omega)$,
we can calculate the local densities of states:\[
D(\omega)\equiv-\dfrac{1}{\pi}\int{\displaystyle \frac{d^{2}k}{(2\pi)^{2}}}\mbox{Im}\left\{ \mbox{Tr}\left[{\textstyle \hat{G}_{eq}^{R}}(\mathbf{k},\omega)\right]\right\} ,\]
and the total number of electrons:\begin{equation}
N={\displaystyle \int_{-\infty}^{\omega_{F}}}d\omega D(\omega).\label{NumberofElectrons}\end{equation}
The number of electrons changes as we increase the disorder, and following
Eq. (\ref{NumberofElectrons}), $\omega_{F}$ is always adjusted so
that the total number of electrons is constant.

\begin{figure}[t]
\centerline{\includegraphics[scale=0.4]{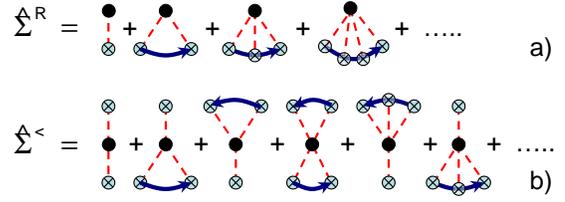}} 

\caption{(color online). An infinite set of diagrams representing the self-consistent
TMA in calculating; a) the retarded (advanced) self energy - $\hat{\Sigma}_{eq}^{R(A)}$
and b) the lesser component of self energy - $\hat{\Sigma}^{<}$ in
Eqs. (\ref{SelfEnergy}) and (\ref{NEselfenergy1Rshb}), respectively.}

\label{Diagrams1} 
\end{figure}

The same TMA is also used to calculate ${\textstyle \hat{\Sigma}^{<}}$
and $\hat{\Sigma}_{\mathbf{E}}^{R(A)}$ from Eqs. (\ref{Kinetic3})
and (\ref{BELOWselfenergy}), respectively:\begin{equation}
\hat{\Sigma}_{1}^{<}=n_{i}\int{\displaystyle \frac{d^{2}k}{(2\pi)^{2}}}\hat{T}_{eq}^{R}(\omega)\hat{G}_{1}^{<}(\mathbf{k},\omega)\hat{T}_{eq}^{A}(\omega)\label{NEselfenergy1Rshb}\end{equation}
 \begin{equation}
\hat{\Sigma}_{\mathbf{E}}^{R(A)}=n_{i}\int{\displaystyle \frac{d^{2}k}{(2\pi)^{2}}}\hat{T}_{eq}^{R(A)}(\omega)\hat{G}_{\mathbf{E}}^{R(A)}(\mathbf{k},\omega)\hat{T}_{eq}^{R(A)}(\omega)\label{NEselfenergy2Rshb}\end{equation}
 The TMA with self-consistent calculation of the equilibrium Green's
functions $\hat{G}_{eq}^{R(A)}$ described in Appendix B allows us
to take into account higher order non-crossed diagrams in the concentration
of impurities $n_{i}$, with weak localization diagrams being disregarded.
The procedure of calculating the retarded (advanced) and non-equilibrium
self energies in Eqs. (\ref{SelfEnergy}) and (\ref{NEselfenergy1Rshb})
is represented graphically in Fig. \ref{Diagrams1}. In this graphical
representation, the bold arrow corresponds to the self-consistently
calculated retarded (advanced) Green's function.

For the delta scatterers, $T$-matrix does not depend on momentum
$\mathbf{k}$ which allows us to perform momentum integrations in
Eqs. (\ref{NEselfenergy1Rshb}) and (\ref{NEselfenergy2Rshb}). It
is then useful to introduce the following $2\times2$ matrices:\begin{equation}
\hat{\rho}(\omega)\equiv\int{\displaystyle \frac{d^{2}k}{(2\pi)^{2}}}\hat{G}_{1}^{<}(\mathbf{k},\omega)\label{DensityM1}\end{equation}
 \begin{equation}
\hat{\rho}_{\mathbf{E}}^{R(A)}(\omega)\equiv\int{\displaystyle \frac{d^{2}k}{(2\pi)^{2}}}\hat{G}_{\mathbf{E}}^{R(A)}(\mathbf{k},\omega)\label{DensityM2}\end{equation}
 The elements of matrices $\hat{\rho}$ and $\hat{\rho}_{\mathbf{E}}$
satisfy a system of linear equations obtained by integrating in momentum
space the left and right hand sides of Eqs. (\ref{Kinetic2}) and
(\ref{KineticBelow}), respectively:\begin{equation}
\begin{array}{ccl}
\hat{\rho} & = & \int{\displaystyle \frac{d^{2}k}{(2\pi)^{2}}}\hat{G}_{eq}^{R}\hat{T}_{eq}^{R}(\omega)\hat{\rho}(\omega)\hat{T}_{eq}^{A}(\omega)\hat{G}_{eq}^{A}\\
 &  & -i\partial_{\omega}n_{F}e\mathbf{E}\int{\displaystyle \frac{d^{2}k}{(2\pi)^{2}}}\hat{G}_{eq}^{R}\boldsymbol{\hat{\upsilon}}\hat{G}_{eq}^{A}\end{array},\label{DensityEq1}\end{equation}
 \begin{equation}
\begin{array}{l}
\hat{\rho}_{\mathbf{E}}^{R(A)}=\int{\displaystyle \frac{d^{2}k}{(2\pi)^{2}}}\hat{G}_{eq}^{R(A)}\hat{T}_{eq}^{R(A)}(\omega)\hat{\rho}_{\mathbf{E}}^{R(A)}(\omega)\hat{T}_{eq}^{R(A)}(\omega)\hat{G}_{eq}^{R(A)}\\
-{\displaystyle \frac{i}{2}}\int{\displaystyle \frac{d^{2}k}{(2\pi)^{2}}}\left(\hat{G}_{eq}^{R(A)}\boldsymbol{\hat{\upsilon}}\partial_{\omega}\hat{G}_{eq}^{R(A)}-\partial_{\omega}\hat{G}_{eq}^{R(A)}\boldsymbol{\hat{\upsilon}}\hat{G}_{eq}^{R(A)}\right)\end{array}.\label{DensityEq2}\end{equation}
 The momentum integrations in the right hand side of Eqs. (\ref{DensityEq1})
and (\ref{DensityEq2}) are done analytically using the general form
of the Green's functions $\hat{G}_{eq}^{R(A)}(\mathbf{k},\omega)$
in Eq. (\ref{GreenRA}). Without loss of generality, we take the electric
field $\mathbf{E}$ along the $y$ axis $\mathbf{E}=(0,E_{y})$ and
solve the system of linear Eqs. (\ref{DensityEq1}) and (\ref{DensityEq2})
for the elements of matrices $\hat{\rho}$ and $\hat{\rho}_{\mathbf{E}}$
in Appendices C and D, respectively.

With this, we calculate the current from Eqs. (\ref{CurrentI}) and
(\ref{CurrentII}), respectively, with a use of Eqs. (\ref{Kinetic2}),
(\ref{Separation1}), (\ref{GIIexpand1}) and (\ref{KineticBelow}):\begin{equation}
\begin{array}{l}
j_{x(y)}^{I}=-ie\int{\displaystyle \frac{d^{2}k}{(2\pi)^{2}}}{\displaystyle \frac{d\omega}{2\pi}}{\rm Tr}\left\{ \hat{G}_{eq}^{R}\hat{T}_{eq}^{R}\hat{\rho}\hat{T}_{eq}^{A}\hat{G}_{eq}^{A}\hat{\upsilon}_{x(y)}\right.\\
\left.-ie\mathbf{E}\partial_{\omega}n_{F}\left(\hat{G}_{eq}^{R}\boldsymbol{\hat{\upsilon}}\hat{G}_{eq}^{A}-{\displaystyle \frac{1}{2}}(\hat{G}_{eq}^{A}\boldsymbol{\hat{\upsilon}}\hat{G}_{eq}^{A}-\hat{G}_{eq}^{R}\boldsymbol{\hat{\upsilon}}\hat{G}_{eq}^{R})\right)\hat{\upsilon}_{x(y)}\right\} \\
\\\end{array},\label{CurentI-Calc}\end{equation}
 \begin{equation}
\begin{array}{l}
j_{x(y)}^{II}=ie\int{\displaystyle \frac{d^{2}k}{(2\pi)^{2}}}{\displaystyle \frac{d\omega}{2\pi}}e\mathbf{E}n_{F}{\rm Tr}\left[\hat{G}_{eq}^{R}\hat{T}_{eq}^{R}\hat{\rho}_{\mathbf{E}}^{R}\hat{T}_{eq}^{R}\hat{G}_{eq}^{R}\hat{\upsilon}_{x(y)}\right.\\
-{\displaystyle \frac{i}{2}}\left(\hat{G}_{eq}^{R}\boldsymbol{\hat{\upsilon}}\partial_{\omega}\hat{G}_{eq}^{R}-\partial_{\omega}\hat{G}_{eq}^{R}\boldsymbol{\hat{\upsilon}}\hat{G}_{eq}^{R}\right)\hat{\upsilon}_{x(y)}\Bigr]+c.c.\end{array}.\label{CurrentII-Calc}\end{equation}
 where we use $\partial_{\mathbf{k}}\hat{G}_{eq}^{R(A)}=\hat{G}_{eq}^{R(A)}\boldsymbol{\hat{\upsilon}}\hat{G}_{eq}^{R(A)}$,
which holds for the model of delta impurities. In Eq. (\ref{CurentI-Calc}),
we perform analytical integrations over momentum $\mathbf{k}$ and
energy $\omega$ while in Eq. (\ref{CurrentII-Calc}), we only perform
analytical integration over momentum. The results of these integrations
are given in Appendices E and F for Eqs. (\ref{CurentI-Calc}) and
(\ref{CurrentII-Calc}), respectively.

\subsection{Analytical results in the metallic regime}

In the metallic regime, we are able to obtain analytical results as
it is sufficient to consider only finite number of terms in the expansion
with respect to the strength of impurity in Fig. \ref{Diagrams1}.
For the same reason, we are also able to generalize the disorder in
Eq. (\ref{Delta}) (generalization of the theory is given in Appendix
A) as follows: \begin{equation}
V(\mathbf{r})=\sum_{i}V_{0}^{i}\delta(\mathbf{r}-\mathbf{r}_{i}),\label{Delta1}\end{equation}
 where $\mathbf{r}_{i}$ is random, the strength of each impurity
has the same arbitrary distribution and all strength distributions
are independent leading to the first four cumulants: $\left\langle V_{0}^{i}\right\rangle _{dis}=0$,
$n_{i}\left\langle (V_{0}^{i})^{2}\right\rangle _{dis}=V_{2}$, $n_{i}\left\langle (V_{0}^{i})^{3}\right\rangle _{dis}=V_{3}$
and $n_{i}\left\langle (V_{0}^{i})^{4}\right\rangle _{dis}=V_{4}$
where $n_{i}$ is the concentration of impurities. For the disorder
described in Eq. (\ref{Delta}), we have $\sqrt{V_{2}/n_{i}}=\sqrt[3]{V_{3}/n_{i}}=\sqrt[4]{V_{4}/n_{i}}=V_{0}$
and for the telegraph white noise disorder we have $V_{3}=0$ as it
is mentioned in Appendix A.

In this section, we first expand the retarded (advanced) self energy
in Eq. (\ref{SelfEnergy}) up to the third order in $V_{0}$ (or up
to the terms $V_{3}$ in Eq. (\ref{SelfEnergyComp})). The lesser
component of the self-energy in Eq. (\ref{NEselfenergy1Rshb}) has
to be expanded up to the fourth order in $V_{0}$ (or up to the terms
$V_{4}$ in Eq. (\ref{NEselfenergyComp})) which corresponds to the
four legged diagrams in Fig. \ref{Diagrams1}b). This ensures that
the expansion of the conductivity $\sigma_{xy}^{I}$ following from
Eqs. (\ref{AppCondIxyb},\ref{AppCondIxySC}) captures all possible
terms proportional to $1/V_{0}$ and $1$. 

\begin{figure}[t]
\centerline{\includegraphics[scale=0.32,angle=90]{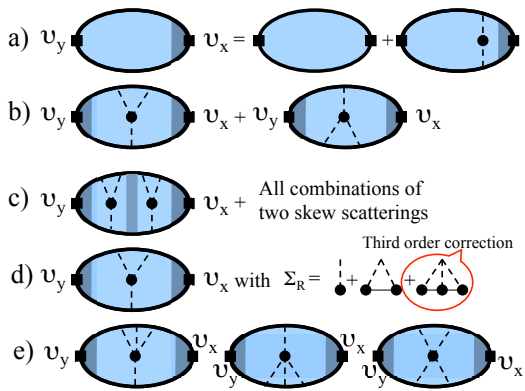}} 

\caption{Different diagrammatic contributions to $\sigma_{xy}^{I}$ within
the Kubo formula formalism; a) the ladder diagram (vertex) contribution
$\sim1$, b) the skew scattering contribution $\sim1/(n_{i}V_{0})\sim V_{3}/V_{2}^{2}$,
c) the double skew scattering contribution $\sim1/n_{i}\sim V_{3}^{2}/V_{2}^{3}$,
d) the skew scattering contribution in which the retarded (advanced)
self energy is calculated up to the third order $\sim1/n_{i}\sim V_{3}^{2}/V_{2}^{3}$
and e) the fourth order skew scattering contribution $\sim1/n_{i}\sim V_{4}/V_{2}^{2}$.}

\label{DiagramsNunner} 
\end{figure}

The expansion of $\sigma_{xy}^{II}$ following from Eq. (\ref{AppCondIIb})
is somewhat simpler as it only contains the terms proportional to
$1$ and its calculation requires consideration of only one bare bubble
diagram (\textit{e.g.} summation of vertices leads to higher order
corrections). In our discussion, we thus concentrate on the diagrams
for calculating $\sigma_{xy}^{I}$ and also present the result for
the bare bubble diagram of $\sigma_{xy}^{II}$. Note that in the expansion
of $\sigma^{I(II)}$, it is important to properly consider the branch
cut of the {}``$\ln$'' function taken as $(-\infty,0]$. The diagrams
in Fig. \ref{Diagrams1} have direct correspondence to the Kubo formalism
diagrams in Fig. \ref{DiagramsNunner} used in Ref. \onlinecite{Nunner:dec2007}.
This allows us to separate the conductivity into terms that directly
relate to each diagram in Fig. \ref{DiagramsNunner}. 

We distinguish three regimes for the position of the Fermi energy
with respect to the gap of the size $2h$; (i) $\omega_{F}>h$, (ii)
$-h<\omega_{F}<h$ and (iii) $\omega_{F}<-h$ (see Fig. \ref{Spectrum}).
To simplify formulas, we introduce the following notation: \[
k_{\pm}^{2}=2(\omega_{F}+\alpha^{2}\mp\sqrt{h^{2}+2\omega_{F}\alpha^{2}+\alpha^{4}}),\]
 \[
\lambda_{\pm}=\sqrt{(\alpha k_{\pm})^{2}+h^{2}},\:\lambda_{F}=\sqrt{2\omega_{F}\alpha^{2}+h^{2}},\]
 \[
\kappa_{\pm}=\sqrt{(\alpha k_{\pm})^{2}+4h^{2}},\]
 \[
\nu_{\pm}=k\left|\frac{d\omega(k)}{dk}\right|^{-1}=\left\{ \begin{array}{c}
{\displaystyle \frac{\lambda_{\pm}}{\lambda_{\pm}\pm\alpha^{2}}},\;\omega_{F}>h\\
{\displaystyle \frac{\lambda_{-}}{\lambda_{-}-\alpha^{2}}},\;-h<\omega_{F}<h\\
{\displaystyle \frac{\lambda_{\pm}}{\left|\lambda_{\pm}-\alpha^{2}\right|}},\;\omega_{F}<-h\end{array}\right.,\]
 where $\nu_{\pm}$ is the density of states at the Fermi level and
$k_{\pm}$ are the two Fermi wave numbers for the regimes (i) and
(iii). In the regime (ii), $k_{+}$ becomes pure imaginary and only
$k_{-}$ has the meaning of the Fermi wave number. Further, we introduce
the following parameter: \[
\Lambda=\frac{V_{3}}{V_{2}^{2}}\gamma_{z}^{i}+\frac{V_{4}}{V_{2}^{2}}(3\gamma^{r}\gamma_{z}^{i}+\gamma^{i}\gamma_{z}^{r}),\]
 where $\hat{\gamma}=\int{\displaystyle {d^{2}k}/{(2\pi)^{2}}}{\textstyle \hat{G}_{0}^{R}}\equiv\gamma\hat{\sigma}_{0}+\gamma_{z}\hat{\sigma}_{z}$,
with $\gamma=\gamma^{r}+i\gamma^{i}$, $\gamma_{z}=\gamma_{z}^{r}+i\gamma_{z}^{i}$.
Note that the two dimensional integral over momentum diverges and
$\hat{\gamma}$ is calculated by introducing the momentum cutoff,
see Appendix B. By expanding the result of Appendix B up to the zeroth
order in the strength of impurities, we obtain: \[
\begin{array}{c}
\gamma^{r}={\displaystyle \frac{(k_{-}^{2}-2\omega_{F})\ln\left|{\displaystyle \frac{k_{-}^{2}}{k_{0}^{2}-k_{-}^{2}}}\right|-(k_{+}^{2}-2\omega_{F})\ln\left|{\displaystyle \frac{k_{+}^{2}}{k_{0}^{2}-k_{+}^{2}}}\right|}{2\pi(k_{-}^{2}-k_{+}^{2})}},\\
\\\gamma_{z}^{r}={\displaystyle \frac{h}{\pi(k_{+}^{2}-k_{-}^{2})}}\ln\left|{\displaystyle \frac{k_{+}^{2}(k_{0}^{2}-k_{-}^{2})}{k_{-}^{2}(k_{0}^{2}-k_{+}^{2})}}\right|,\end{array}\]
 \[
\begin{array}{c}
\gamma^{i}=\left\{ \begin{array}{c}
-{\displaystyle \frac{\nu_{-}+\nu_{+}}{4}},\;\omega_{F}>h\\
-{\displaystyle \frac{\nu_{-}}{4}},\;-h<\omega_{F}<h\\
{\displaystyle -\frac{k_{-}^{2}+k_{+}^{2}-4\omega_{F}}{2(k_{-}^{2}-k_{+}^{2})}},\;\omega_{F}<-h\end{array}\right.,\\
\gamma_{z}^{i}=\left\{ \begin{array}{c}
{\displaystyle \frac{h}{4}(\frac{\nu_{+}}{\lambda_{+}}-\frac{\nu_{-}}{\lambda_{-}})},\;\omega_{F}>h\\
-{\displaystyle \frac{h}{4}\frac{\nu_{-}}{\lambda_{-}}},\;-h<\omega_{F}<h\\
-{\displaystyle \frac{2h}{k_{-}^{2}-k_{+}^{2}}},\;\omega_{F}<-h\end{array}\right.,\end{array}\]
where $k_{0}$ is the cutoff in the momentum integration. 

As it follows from the Appendices E and F, the non diagonal conductivities
$\sigma_{xy}^{I(II)}$ can be calculated by properly choosing the
{}``$\ln$'' branch that corresponds to the regimes (i), (ii) or
(iii), respectively. The result of expanding Eqs. (\ref{AppCondIxyb},\ref{AppCondIxySC})
and Eq. (\ref{AppCondIIb}) for conductivities $\sigma_{xy}^{I}$
and $\sigma_{xy}^{II}$, respectively, in the region (i) ($\omega_{F}>h$)
becomes: \begin{equation}
\begin{array}{c}
\sigma_{xy}^{I(i)}={\displaystyle \frac{2e^{2}\alpha^{2}}{\hbar\pi}}{\displaystyle \Lambda}=-{\displaystyle \frac{V_{4}}{V_{2}^{2}}}{\displaystyle \frac{e^{2}h\alpha^{2}\ln\left|{\displaystyle \frac{k_{+}^{2}(k_{0}^{2}-k_{-}^{2})}{k_{-}^{2}(k_{0}^{2}-k_{+}^{2})}}\right|}{\hbar\pi^{2}(k_{+}^{2}-k_{-}^{2})}},\\
\\\sigma_{xy}^{II(i)}=0,\end{array}\label{SigmaIresult(i)}\end{equation}
which reproduces result of Ref. \onlinecite{Kovalev:jul2008} in
the limit of large cutoff $k_{0}$. In reference to the Kubo formula
formalism, we can claim the following: the diagrams in Fig. \ref{DiagramsNunner}a)
vanish after summation (the intrinsic and side-jump contributions
defined in Ref. \onlinecite{Sinitsyn:jan2007} cancel each other),\citep{Kovalev:jul2008}
the diagrams in Figs. \ref{DiagramsNunner}b)-d) are all proportional
to ${\displaystyle \frac{\nu_{+}}{\lambda_{+}}}-{\displaystyle \frac{\nu_{-}}{\lambda_{-}}}\equiv0$
and also vanish, and the diagrams in Fig. \ref{DiagramsNunner}e)
lead to the result in Eq. (\ref{SigmaIresult(i)}). $\sigma_{xy}^{II(i)}$
is zero as the corresponding bare bubble contribution in Eq. (\ref{AppCondIIb})
vanishes. Repeating the same procedure for the region (ii) ($-h<\omega_{F}<h$),
we obtain: \begin{equation}
\begin{array}{c}
\sigma_{xy}^{I(ii)}={\displaystyle \frac{e^{2}}{4\pi\hbar}}\left({\displaystyle \frac{h\alpha^{2}\nu_{-}}{\lambda_{-}^{2}}}-{\displaystyle \frac{4hk_{-}^{2}\alpha^{2}}{\lambda_{-}\kappa_{-}^{2}}}+{\displaystyle \frac{3hk_{-}^{4}\alpha^{2}}{\kappa_{-}^{4}\nu_{-}}}+{\displaystyle \frac{8k_{-}^{4}\alpha^{2}\lambda_{-}^{2}}{\kappa_{-}^{4}\nu_{-}^{2}}}{\displaystyle \Lambda}\right.\\
\left.+\left[{\displaystyle \frac{8h(2h^{2}+2\omega_{F}\alpha^{2}+k_{-}^{2})}{\kappa_{-}^{2}}}\gamma_{z}^{i}+(k_{-}^{2}-k_{+}^{2})\gamma^{i}\right]{\displaystyle \frac{2hk_{-}^{4}\alpha^{2}}{\kappa_{-}^{4}}}{\displaystyle \frac{V_{3}^{2}}{V_{2}^{3}}}\right),\\
\\\sigma_{xy}^{II(ii)}={\displaystyle \frac{e^{2}}{4\pi\hbar}}(1-{\displaystyle \frac{h}{\sqrt{\alpha^{4}+\lambda_{F}^{2}}}}),\end{array}\label{SigmaIresult(ii)}\end{equation}
where the diagrams in Fig. \ref{DiagramsNunner}a) lead to the first
three disorder independent terms in Eq. (\ref{SigmaIresult(ii)})
(the intrinsic, the side-jump and the disorder independent skew scattering
terms, respectively),\citep{Kovalev:jul2008} the skew scattering
diagrams in Figs. \ref{DiagramsNunner}b) and e) lead to the term
in Eq. (\ref{SigmaIresult(ii)}) proportional to ${\displaystyle \Lambda}$,
and the diagrams in Figs. \ref{DiagramsNunner}c) and d) lead to the
terms in Eq. (\ref{SigmaIresult(ii)}) proportional to $V_{3}^{2}/V_{2}^{3}$.
$\sigma_{xy}^{II(ii)}$ is calculated from a bare bubble contribution
given by Eq. (\ref{AppCondIIb}) and also corresponds to the intrinsic
contribution. Finally for the region (iii) ($\omega_{F}<-h$), we
obtain:\begin{equation}
\begin{array}{c}
\sigma_{xy}^{I(iii)}={\displaystyle \frac{e^{2}}{4\pi\hbar}}\left({\displaystyle \frac{32h\omega_{F}^{2}\alpha^{4}}{(h^{2}+\alpha^{4})^{2}(k_{-}^{2}-k_{+}^{2})}}+{\displaystyle \frac{\alpha^{2}(k_{-}^{2}-k_{+}^{2})^{4}}{32(h^{2}+\alpha^{4})^{2}}}{\displaystyle \Lambda}\right.\\
+\left[{\displaystyle \frac{h(h^{2}\omega_{F}+2\alpha^{2}h^{2}-3\omega_{F}\alpha^{4})}{(h^{2}-\omega_{F}\alpha^{2})(h^{2}+\alpha^{4})}}\gamma_{z}^{i}+\gamma^{i}\right]{\displaystyle \frac{h\alpha^{2}(k_{-}^{2}-k_{+}^{2})^{3}}{4(h^{2}+\alpha^{4})^{2}}}{\displaystyle \frac{V_{3}^{2}}{V_{2}^{3}}},\\
\\\sigma_{xy}^{II(iii)}={\displaystyle \frac{e^{2}}{4\pi\hbar}}{\displaystyle \frac{h(\lambda_{-}-\lambda_{+})}{(\alpha^{2}-\lambda_{-})(\alpha^{2}-\lambda_{+})}},\end{array}\label{SigmaIresult(iii)}\end{equation}
 where the diagrams in Fig. \ref{DiagramsNunner}a) lead to the disorder
independent term in Eq. (\ref{SigmaIresult(iii)}) (it includes the
intrinsic, the side-jump and the disorder independent skew scattering
contributions), the skew scattering diagrams in Figs. \ref{DiagramsNunner}b)
and e) lead to the term in Eq. (\ref{SigmaIresult(iii)}) proportional
to ${\displaystyle \Lambda}$, and the diagrams in Figs. \ref{DiagramsNunner}c)
and d) lead to the terms in Eq. (\ref{SigmaIresult(iii)}) proportional
to $V_{3}^{2}/V_{2}^{3}$. $\sigma_{xy}^{II(ii)}$ is again calculated
from a bare bubble contribution given by Eq. (\ref{AppCondIIb}).

The diagonal conductivities can also be calculated by expanding Eqs.
(\ref{AppCondIyyb}) and (\ref{AppCondIyySC}):\[
\sigma_{yy}=\left\{ \begin{array}{c}
{\displaystyle \frac{e^{2}}{\hbar}}{\displaystyle \frac{\omega_{F}+\alpha^{2}}{\pi V_{2}}},\;\omega_{F}>h\\
{\displaystyle \frac{e^{2}}{\hbar}}{\displaystyle \frac{k_{-}^{2}\lambda_{-}^{2}}{\pi V_{2}\nu_{-}^{2}\kappa_{-}^{2}}},\;-h<\omega_{F}<h\\
{\displaystyle \frac{e^{2}}{\hbar}}{\displaystyle \frac{(\omega_{F}+\alpha^{2})(\alpha^{4}+\lambda_{F}^{2})}{\pi V_{2}(\alpha^{4}+h^{2})}},\;\omega_{F}<-h\end{array}\right.,\]
 where we only present the dominant non-vanishing terms $V_{2}^{-1}$
as the higher order terms are quite cumbersome.

\subsection{Numerical results and discussions}

\begin{figure}[t]
\centerline{\includegraphics[scale=0.7]{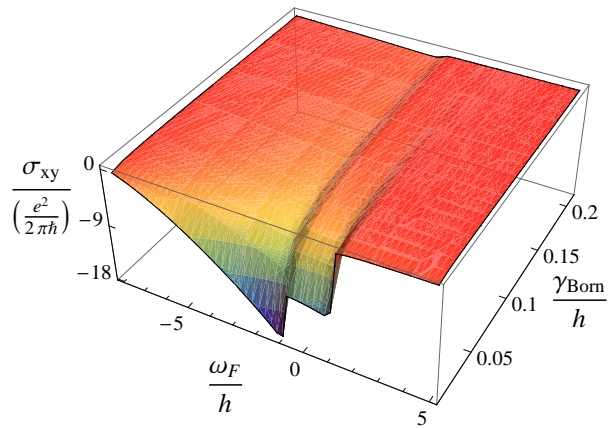}}

\caption{The anomalous Hall conductivity $\sigma_{xy}$ as a function of the
Fermi energy $\omega_{F}$ and the Born scattering amplitude $\gamma_{Born}$.
The parameters are chosen as $2\alpha^{2}/h=35.9$, $k_{0}=12$ and
$V_{0}=0.1$. The Fermi energy $\omega_{F}$ corresponds here to the
clean system and it is renormalized according to Eq. (\ref{NumberofElectrons})
in the presence of disorder.}

\label{Plot3D-1} 
\end{figure}

\begin{figure}[t]
\centerline{\includegraphics[scale=0.7]{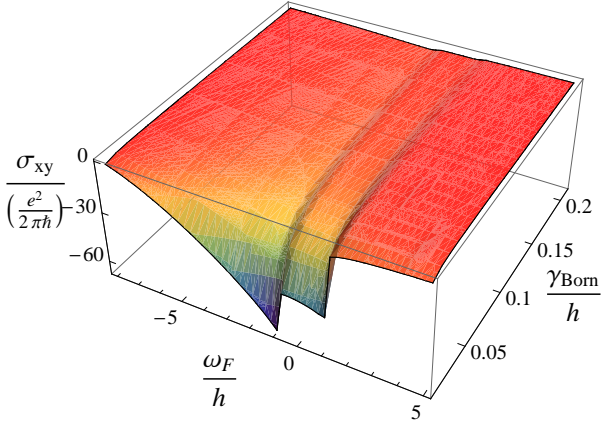}}

\caption{Identical to Fig. \ref{Plot3D-1} plot but for larger strength of
impurity $V_{0}=0.3$. }

\label{Plot3D-2} 
\end{figure}

\begin{figure}[t]
\centerline{\includegraphics[scale=0.7]{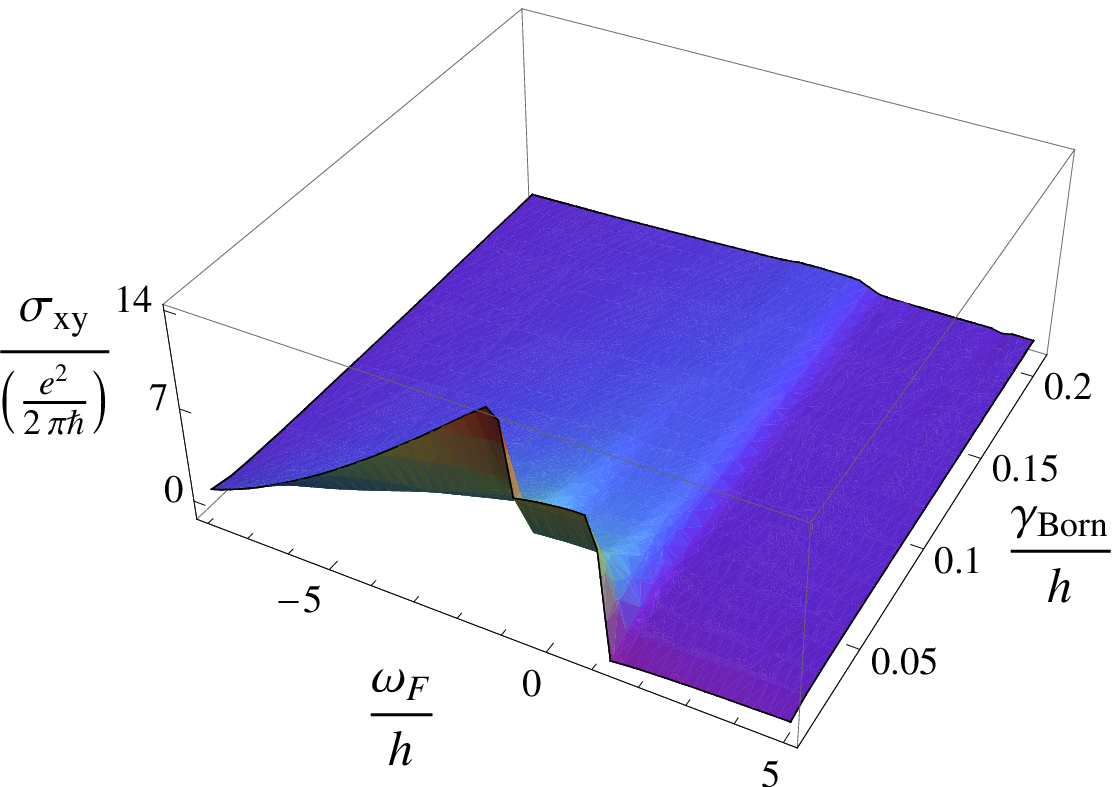}}

\caption{Identical to Fig. \ref{Plot3D-1} plot but for negative strength of
impurity $V_{0}=-0.1$.}

\label{Plot3D-3} 
\end{figure}

\begin{figure}[t]
\centerline{\includegraphics[scale=0.7]{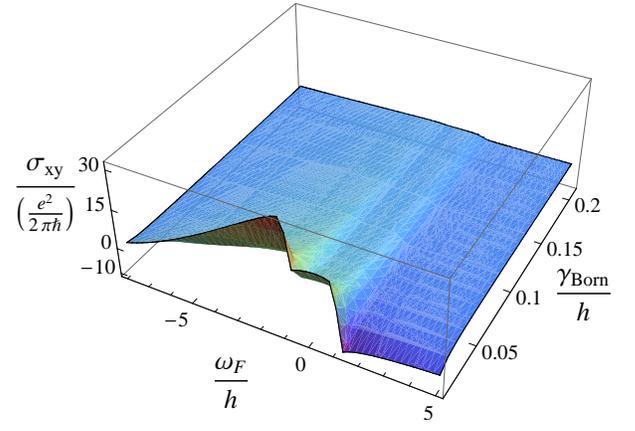}}

\caption{Identical to Fig. \ref{Plot3D-1} plot but for negative strength of
impurity $V_{0}=-0.3$.}

\label{Plot3D-4} 
\end{figure}

Here, we present results of our numerical calculations based on the
formalism developed in Section IIIA. Figures \ref{Plot3D-1},\ref{Plot3D-2},\ref{Plot3D-3}
and \ref{Plot3D-4} show the numerical results for the anomalous Hall
conductivity as a function of the Fermi energy $\omega_{F}$ and the
first Born scattering amplitude $\gamma_{Born}=n_{i}V_{0}^{2}m$.
The strength of the spin-orbit interaction is chosen to be the same
as in Ref. \onlinecite{Onoda:apr2008}, $2\alpha^{2}/h=35.9$ ($2\alpha^{2}/E_{\mbox{res}}=3.59$,
$E_{\mbox{res}}=10h$; and the strength of impurity is $V_{0}=0.1,\:0.3,\:-0.1$
and $-0.3$. For the retarded (advanced) self-energy, the cutoff in
the momentum integration is $k_{0}=12$ which corresponds to the energy
cutoff of Ref. \onlinecite{Onoda:apr2008}, $\varepsilon_{c}=3E_{\mbox{res}}$.
The Born scattering amplitude is varied by changing the impurity concentration
$n_{i}$.

In the clean limit, when $\gamma_{Born}\rightarrow0$, we observe
skew scattering behavior ($\sigma_{xy}\sim1/n_{i}V_{0}$) in which
$|\sigma_{xy}|$ rapidly increases. For repulsive scatterers ($V_{0}>0$,
see Figs. \ref{Plot3D-1} and \ref{Plot3D-2}), the negative conductivity
diminishes as we increase the Fermi energy, until the point $\omega_{F}=-h$
is reached. At this point, the conductivity suddenly increases without
a change of sign, in contrast to Ref. \onlinecite{Onoda:apr2008}
where the sign change has been observed but in agreement with Refs.
\onlinecite{Nunner:dec2007} and \onlinecite{Kovalev:jul2008}
(note that Fig. \ref{Plot3D-1} is calculated for exactly the same
parameters as Fig. 5(c) in Ref. \onlinecite{Onoda:apr2008}). As
we increase the Fermi energy further, the conductivity increases again
around $\omega_{F}=h$ acquiring a very small negative value. In this
regime, both subbands are partially occupied and only the higher order
skew scattering \citealp{Nunner:dec2007,Kovalev:jul2008} (hybrid
skew scattering) contributes to the anomalous Hall effect. Relatively
large hybrid skew scattering is present in Fig. \ref{Plot3D-2} compared
to Fig. \ref{Plot3D-1} as the hybrid skew scattering contribution
is proportional to $1/n_{i}\sim V_{0}^{2}/\gamma_{Born}$ and should
be larger for greater impurity strength.\citealp{Kovalev:jul2008}
The same is true for the conventional skew scattering proportional
to $1/V_{0}n_{i}\sim V_{0}/\gamma_{Born}$ , which can be immediately
seen from Figs. \ref{Plot3D-1}, \ref{Plot3D-2}, \ref{Plot3D-3}
and \ref{Plot3D-4}. For attractive scatterers ($V_{0}<0$, see Figs.
\ref{Plot3D-3} and \ref{Plot3D-4}) the sign of the ordinary skew
scattering dominating in the clean limit is opposite to the sign of
the ordinary skew scattering for the repulsive scatterers. The conductivity
now increases until we reach the point $\omega_{F}=-h$ in which we
observe a sudden drop. One more drop happens around the point $\omega_{F}=h$
where the anomalous Hall conductivity changes sign (see Figs. \ref{Plot3D-3}
and \ref{Plot3D-4}). This change of sign is consistent with the fact
that the higher order (hybrid) skew scattering (prevailing when both
subbands are partially occupied) does not change its sign as we change
the sign of disorder.\citealp{Kovalev:jul2008} Comparing Figs. \ref{Plot3D-3}
and \ref{Plot3D-4}, we again see that the hybrid skew scattering
is more pronounced for larger impurity strength.

As we increase the disorder by increasing $\gamma_{Born}$, the skew
scattering becomes less important while the other mechanisms, such
as intrinsic and side-jump, become more important. The intrinsic conductivity
only gradually decreases with the disorder because the only effect
of disorder on the intrinsic component comes from broadening of Green's
functions used in the calculation of the intrinsic component. For
repulsive scatterers ($V_{0}>0$), the skew scattering has sign opposite
to the sign of intrinsic and side jump contributions in the region
$-h<\omega_{F}<h$ (see \textit{e.g.} Refs. \onlinecite{Nunner:dec2007}
and \onlinecite{Kovalev:jul2008}). This explains the sign change
we observe in Figs. \ref{Plot3D-1}, \ref{Plot3D-2} and \ref{Scaling2}
in the region $-h<\omega_{F}<h$ as we increase $\gamma_{Born}$ (more
detailed plots are presented in Appendix G).

The positions of points in which the AHE vanishes can be estimated
by comparing the Fermi sea intrinsic term $\sigma_{xy}^{II}$ with
the skew scattering term in Eq. (\ref{SigmaIresult(ii)}) as those
two are the major contributions. Physically, the AHE vanishes because
the intrinsic deflection of electrons between the scattering events
can be balanced by the skew scattering events (in the cross-over region
between intrinsic and extrinsic mechanisms). As the former does not
rely on impurities and the latter does (and changes sign with impurities
changing sign), we can have full cancellation of the two by choosing
the proper sign and strength of impurities.

\subsubsection{Anticrossings and scalings}

As it can be seen from Figs. \ref{Plot3D-1}, \ref{Plot3D-2}, \ref{Plot3D-3}
and \ref{Plot3D-4}, the anomalous Hall effect is resonantly increased
around the band anticrossing which suggests that for more general
band structures, the major contribution to the AHE also comes from
the band anticrossings that happened to be in the vicinity of the
Fermi level.\citep{Onoda:apr2008} This view is well justified for
the intrinsic AHE in the metallic regime ($\tau\varepsilon_{F}>>1$)
as it follows from the Thouless-Kohmoto-Nightingale-Nijs formula\citep{Thouless:aug1982}
applied to the AHE.\citep{Onoda:apr2008} This leads to the intrinsic
AHE conductivity of the order of $e^{2}/(4\pi\hbar)$ within the region
(ii) in Fig. \ref{Spectrum}. The full conductivity that includes
the intrinsic, side-jump and skew-scattering contributions seems to
also have the resonant behavior around the anticrossing for the Rashba
model as it follows from our analysis. Whereas our analysis justifies
focusing the calculations on simplified phenomenological models near
the anti-crossing locations, we emphasize that it is unlikely that
these would be characterized universally by the Rashba geometry rather
than by a combination of Rashba and Dresselhauss symmetry.

\begin{figure}[t]
\centerline{\includegraphics[scale=0.8]{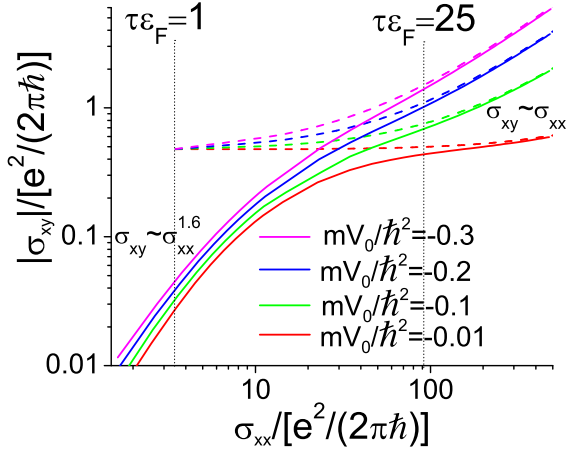}} 

\caption{The absolute value of the anomalous Hall conductivity $|\sigma_{xy}|$
versus the conductivity $\sigma_{xx}$ for the spin-orbit interaction
strength $2\alpha^{2}/h=35.9$. Dimensionality of quantities displayed
in this plot is restored.}

\label{Scaling1} 
\end{figure}

\begin{figure}[t]
\centerline{\includegraphics[scale=0.8]{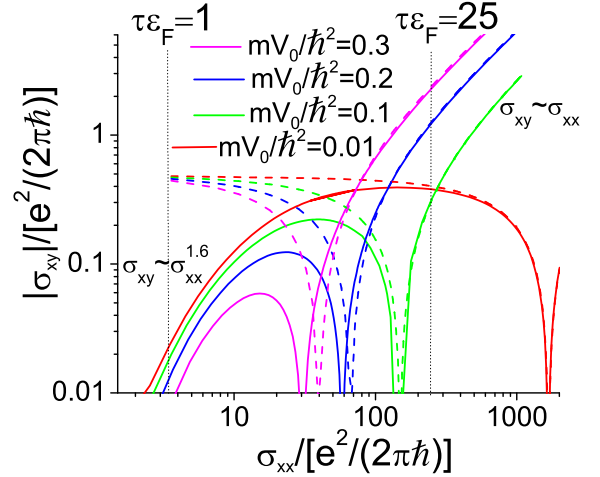}}

\caption{Identical to Fig. \ref{Scaling1} plot except for the disorder which
is repulsive here ($V_{0}>0$). Note that the conductivity $\sigma_{xy}$
changes sign around the cusps. }

\label{Scaling2} 
\end{figure}

In the regime of strong disorder, where this resonant behavior is
not pronounced, the disorder broadening of the Green's functions becomes
more dominant and the $\sigma_{xy}$ has dependence that can no longer
be expanded correctly in powers of $\tau$. This expectation can be
easily seen from the expressions for $\sigma_{xy}$ in our formulation
or the Kubo formulation, in which \begin{equation}
\sigma_{xy}\propto\sum_{\alpha,\beta}\frac{\langle\alpha|\hat{v}_{x}|\beta\rangle\langle\beta|\hat{v}_{y}|\alpha\rangle}{(E_{\alpha}-E_{\beta})^{2}}\label{Kubo}\end{equation}
 where $|\alpha\rangle$ are the exact eigenstates in the presence
of disorder and the major contribution for $\sigma_{xy}$ in the dirty
limit comes from interband matrix elements. When expanding things
in the momentum basis, the denominator is often approximated as $(E_{n}(\vec{k})-E_{n'}(\vec{k}))^{2}+(\hbar/\tau)^{2})$
while the matrix elements are evaluated within the disorder free eigenstates.
Hence, in the limit of large disorder broadening, the denominator
is simply replaced by $(\hbar/\tau)^{2})$ and $\sigma_{xy}\sim\tau^{2}$
(this is different for $\sigma_{xx}$ as the contribution from interband
matrix elements vanishes and $\sigma_{xx}\sim\tau$). This of course
gives an upper bound for the $\sigma_{xy}\sim\tau^{\eta}$ scaling
and in intermediate regimes one would expect $\eta$ to be lower than
2. In Figs. \ref{Scaling1} and \ref{Scaling2}, we study the AHE
calculated in the anticrossing region in order to examine in detail
the universal anomalous Hall effect regimes that could be valid for
more general band structures.

We now plot in the logarithmic scale $\sigma_{xy}$ as a function
of $\sigma_{xx}$ tuned via $n_{i}$ while all other parameters are
kept constant. In the clean limit, we recover the skew scattering
behavior ($\sigma_{xy}\sim1/n_{i}V_{0}\sim\sigma_{xx}/V_{0}$) and
our numerical results (bold line) agree well with the analytical results
(dashed line) obtained in Sec. IIIB. In the moderately dirty limit,
we observe the intrinsic-side-jump regime ($\sigma_{xy}=const$, this
regime is more pronounced for smaller $V_{0}$) in which the side-jump
and intrinsic mechanisms are dominant. All analytical curves (dashed
lines) asymptotically reach this regime when $\sigma_{xx}$ is very
small. In the stronger disorder regime, as reported in Ref. \onlinecite{Onoda:apr2008},
the numerical curves have downturn for smaller $\sigma_{xx}$ approaching
the third regime in which $\sigma_{xy}\sim\sigma_{xx}^{\varphi}$
with $\varphi\approx1.6$ in Fig. \ref{Scaling1}. However, a universal
scaling cannot be claimed since for large and positive strength of
impurities in Fig. \ref{Scaling2} we only observe the reduction of
the AHE. 

One should keep in mind that the gradient expansion is not fully justified
close to the line $\tau\varepsilon_{F}=1$ and our results are meaningful
only for $\tau\varepsilon_{F}>1$. Furthermore, since in this regime
the resonant behavior is strongly diminished, in realistic three-dimensional
systems, the result could be more accurately expressed via the averaged
matrix elements with some appropriate treatment of the disorder broadening.

Although some experimental works claim to confirm the scaling $\sigma_{xy}\sim\sigma_{xx}^{\varphi}$
with $\varphi$ around $1.6$.\citep{Ueno:2007,Miyasato:aug2007,Fukumura:2007,Venkateshvaran:2008,Fernandez-Pacheco:2008}
comparison of theory and experiments has to be done with care since
determining a scaling exponent over a single decade is often difficult
and has led to many errors in the past. For example, in DMS ferromagnets
(mentioned in Ref. \onlinecite{Onoda:apr2008} to support the scaling
hypothesis) the change of doping will cause change in the impurity
concentration, in the magnetization and even in the band structure.
The theoretical calculations only take into account the change in
the impurity concentration and further assume a Rashba symmetry at
the crossing points.\citep{Onoda:sep2006,Onoda:apr2008} Note also
that within the theoretical treatment, the Hall conductivity changes
its sign for repulsive impurities ($V_{0}>0$) in Fig. \ref{Scaling2}
which is expected as the skew scattering dominating in the clean limit
has the sign opposite to the sign of the intrinsic contribution dominating
in the dirty limit (see Eq. (\ref{SigmaIresult(ii)})). These types
of changes of signs have also been observed in experimental systems,
e.g. DMS,\citep{Mih'aly:mar2008} and of course at that stage scaling
is not justified. %

\section{AMR in Rashba systems}

In the previous section, we showed how the formalism developed in
Section II can be applied to calculations of the anomalous Hall effect
in multiple-band systems on the example of a Rashba system. In this
Section, we perform calculations of the anisotropic magnetoresistance
(AMR) in 2DEG with the in-plane exchange field, spin-orbit interaction
and magnetic impurities following the same formalism. A general numerical
procedure allows us to rigorously perform AMR calculations in multiple-band
anisotropic systems. Within the Boltzmann equation approach, such
calculations are usually performed by using the relaxation time approximation
in which the transport relaxation time $\tau$ is calculated from
the scattering amplitudes without fully taking into account the asymmetries.\citep{McGuire:jul1975,Rushforth:oct2007}
This approach was improved in Ref. \onlinecite{Schliemann:oct2003}
by introducing the perpendicular relaxation time $\tau_{\perp}$.
However, in some cases this improvement is yet not sufficient and
Vyborny \textit{et al.} formulated a procedure for finding an exact
solution to the Boltzmann equation in Ref. \onlinecite{Vyborny:oct2008}.
Here we propose an alternative approach for AMR calculations in multiple-band
anisotropic systems to the one proposed in Ref. \onlinecite{Vyborny:oct2008}.

We consider here a 2DEG Rashba Hamiltonian with additional in-plane
exchange field $h_{x}$ directed along the $x$- axis without any
loss of generality : \begin{equation}
\hat{H}_{R}=\mathbf{{\vec{k}}}^{2}/2+\alpha\mathbf{{\vec{k}}}\cdot\boldsymbol{\hat{\sigma}}\times\mathbf{z}-h_{x}\hat{\sigma}_{x}-h\hat{\sigma}_{z}+\hat{V}(\mathbf{r}),\label{Hamiltonian1}\end{equation}
 where now $\hat{V}(\mathbf{r})$ describes the disorder corresponding
to dilute charged magnetic impurities:\citep{Rushforth:oct2007,Nunner:jun2008,Rushforth2008}
\begin{equation}
\hat{V}(\mathbf{r})=V_{0}(a\hat{\sigma}_{0}+\hat{\sigma}_{x})\sum_{i}\delta(\mathbf{r}-\mathbf{r}_{i}),\label{Delta2}\end{equation}
 where $\mathbf{r}_{i}$ describes the positions of random impurities
and we assume that the magnetic impurities are magnetized along the
exchange field. The quantity $a$ describes the relative strength
of the electric part of impurity with respect to the magnetic part.
Note that the AMR is measured by changing the direction of electric
field $\mathbf{E}$ which is equivalent to changing the direction
of the exchange field.

For the AMR, we only need the diagonal conductivities, thus the Fermi
sea contribution given by Eq. (\ref{CurrentII-Calc}) vanishes. The
AMR can be calculated from Eq. (\ref{CurentI-Calc}) and we only need
to calculate Green's functions at the Fermi level. We calculate ${\textstyle \hat{\Sigma}_{eq}^{R(A)}}$
and Green's functions $\hat{G}_{eq}^{R(A)}$ using the self-consistent
TMA:\begin{equation}
\hat{T}_{eq}^{R(A)}=V_{0}(\hat{1}-V_{0}\hat{\gamma}^{R(A)})^{-1}\label{TmatrixAMR}\end{equation}
 \begin{equation}
\hat{\Sigma}_{eq}^{R(A)}=n_{i}\hat{T}_{eq}^{R(A)}(\omega)=\Sigma_{eq0}^{R(A)}\hat{\sigma}_{0}+\Sigma_{eq\mbox{x}}^{R(A)}\hat{\sigma}_{x}+\Sigma_{eq\mbox{z}}^{R(A)}\hat{\sigma}_{z},\label{SelfEnergyAMR}\end{equation}
 \begin{equation}
\begin{array}{l}
{\textstyle \hat{G}_{eq}^{R(A)}}=(\omega\hat{1}-\hat{H}_{0}-\hat{\Sigma}_{eq}^{R(A)})^{-1}\\
={\displaystyle \frac{(\omega-\frac{k^{2}}{2}-\Sigma_{eq0}^{R(A)})\hat{\sigma}_{0}+\alpha k_{y}\hat{\sigma}_{x}-\alpha k_{x}\hat{\sigma}_{y}}{(\omega-\frac{k^{2}}{2}-\Sigma_{eq0}^{R(A)})^{2}-(h-\Sigma_{eq\mbox{z}}^{R(A)})^{2}-\alpha^{2}k^{2}+2h_{x}\alpha k_{y}}}\\
+{\displaystyle \frac{-(h-\Sigma_{eq\mbox{z}}^{R(A)})\hat{\sigma}_{z}-(h_{x}-\Sigma_{eq\mbox{x}}^{R(A)})\hat{\sigma}_{x}}{(\omega-\frac{k^{2}}{2}-\Sigma_{eq0}^{R(A)})^{2}-(h-\Sigma_{eq\mbox{z}}^{R(A)})^{2}-\alpha^{2}k^{2}+2h_{x}\alpha k_{y}}}\end{array}\label{GreenRAAMR}\end{equation}
 where $\hat{\gamma}^{R(A)}=\int{\displaystyle {d^{2}k}/{(2\pi)^{2}}}{\textstyle \hat{G}_{eq}^{R(A)}}(\mathbf{k},\omega)\equiv\gamma^{R(A)}\hat{\sigma}_{0}+\gamma_{x}^{R(A)}\hat{\sigma}_{x}+\gamma_{z}^{R(A)}\hat{\sigma}_{z}$.
We calculate the self-consistent value of the self energy $\hat{\Sigma}_{eq}^{R(A)}(\omega_{F})$
by iterating Eq. (\ref{SelfEnergyAMR}) until the prescribed accuracy
is reached.

As soon as we know the $T$-matrix, we can substitute it into Eq.
(\ref{DensityEq1}) and find the matrix $\hat{\rho}$ by performing
the momentum integrations in the r.h.s.. Finally, by substituting
$\hat{\rho}$ into Eq. (\ref{CurentI-Calc}) we can calculate the
conductivity. Note that throughout this section, the angular part
of the momentum integrations is calculated analytically while the
radial part is calculated numerically.

\begin{figure}[t]
\centerline{\includegraphics[scale=0.8]{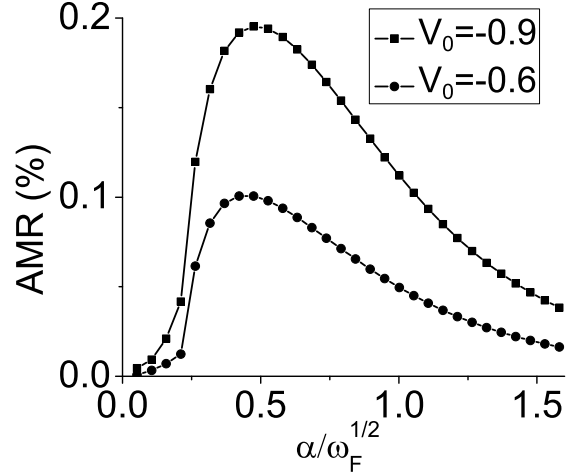}}

\caption{The AMR ($100\%$ corresponds to $\mbox{AMR}=1$) as a function of
the dimensionless spin-orbit interaction strength $\alpha/\sqrt{\omega_{F}}$
($\omega_{F}$ is counted from the middle of the gap in Fig. \ref{Spectrum}).
The parameters of the model are: $V_{0}=-0.6$ and $-0.9$, $n_{i}/\omega_{F}=0.01$
and $h_{x}/\omega_{F}=0.3$ .}

\label{AMR-Kato} 
\end{figure}

\begin{figure}[t]
\centerline{\includegraphics[scale=0.8]{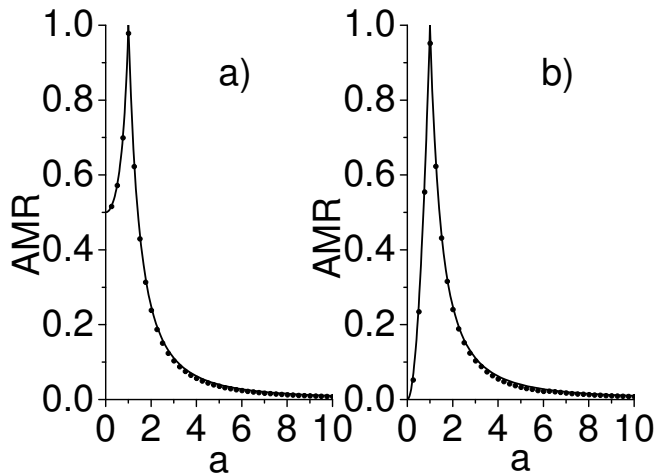}}

\caption{The AMR as a function of the relative strength $a$ of the electric
and magnetic parts of impurity potential, by solid line we plot analytical
results and dots represent numerical results; a) Fermi level crosses
only one band ($\omega_{F}=0$) with the following dimensionless parameters
$V_{0}=0.05$, $\alpha=1.4$, $n_{i}=0.0015$, $h_{x}=0.0015$ and
$h=0.015$; b) Fermi level crosses both bands $V_{0}=0.05$, $\alpha=0.03$,
$n_{i}=0.002$, $h_{x}=0.002$ and $h=0.001$.}

\label{AMR} 
\end{figure}

The anisotropic resistance in our system is defined as follows:\[
\mbox{AMR}=-\dfrac{\sigma_{xx}-\sigma_{yy}}{\sigma_{xx}+\sigma_{yy}}\]
 and it describes the relative difference in conductivity for current
flowing parallel or perpendicular to the magnetization (represented
by the exchange field and/or impurity magnetization).

First, we calculate the anisotropic magnetoresistance in Rashba system
with in-plane exchange field and non-magnetic delta scatterers (see
Eq. \ref{Delta}, the magnetic scatterers are absent in this model).
Kato \textit{et al.} found vanishing AMR in the regime (i) (see Fig.
\ref{Spectrum}) when both subbands are partially occupied due to
the cancellation of the nonvertex and vertex parts in the Kubo formulation.
In Fig. \ref{AMR-Kato}, we observe the non-vanishing AMR in the regime
(i) and this suggests the importance of the higher order diagrams
(such as plotted in Fig. \ref{DiagramsNunner}e)) not only for the
AHE but also for the AMR. The AMR effect resulting from the higher
order diagrams is more pronounced for the larger strength of impurities,
similar to the AHE. The AMR approaches its maximum around the point
at which the exchange energy is comparable to the spin orbit energy,
$2h_{x}\sim\alpha$. We note that the non-zero but comparatively weak
magnitude of the AMR here in the Rashba system is reminiscent of the
results in three-dimensional DMS ferromagnets.\citep{Rushforth:oct2007}
This agrees with physical intuition. Under comparison of two mechanisms
by which AMR can arise - carrier polarization/anisotropy in wavefunctions,
and impurity polarization/anisotropy in scattering operator (see Fig.
1 of Ref. \onlinecite{Rushforth:oct2007}) - the former implies
a competition between the exchange and spin-orbit terms (in the Hamiltonian)
resulting in reduced anisotropy strength.

Consequently, even though observation of the AMR effect is deemable
in the absence of magnetic scatterers, we expect much more pronounced
effect when the magnetic scatterers are present. Our numerical results
in Fig. \ref{AMR} (plotted together with the analytical results from
Ref. \onlinecite{Vyborny:oct2008}) confirm this. For the case when
the Fermi level crosses only one band (region (ii) in Fig. \ref{Spectrum}),
it was found in Ref. \onlinecite{Vyborny:oct2008} that $\mbox{AMR}=1/(2-a^{2})$
when $|a|<1$ and $\mbox{AMR}=1/a^{2}$ when $|a|>1$, provided the
exchange fields are small. For the case when the Fermi level crosses
two bands (region (i) in Fig. \ref{Spectrum}) it was found in Ref.
\onlinecite{Vyborny:oct2008} that $\mbox{AMR}=a^{2}$ when $|a|<1$
and $\mbox{AMR}=1/a^{2}$ when $|a|>1$, in the limit of large Fermi
energy (compared to the spin-orbit and exchange splitting). We observe
a perfect agreement between our numerical results and the analytical
results from Ref. \onlinecite{Vyborny:oct2008}. The result in Fig.
\ref{AMR}a) cannot be reproduced within the common approximate approaches\citep{McGuire:jul1975,Schliemann:oct2003,Rushforth:oct2007}
based on the relaxation time approximation as it was pointed out in
Ref. \onlinecite{Vyborny:oct2008}. The non-physical divergence
in $\sigma_{yy}$ at the point $a=1$ in Fig. \ref{AMR} is caused
by the special choice of the scattering potential.\citep{Vyborny:oct2008}
As soon as the spatial dependences of the electric and magnetic parts
cease to be identical (in Eq. (\ref{Delta2}), they correspond both
to delta-scatterers) the divergence of the $\sigma_{yy}$ is removed
(causing AMR$<1$).

\section{Conclusions}

We have developed a framework for transport calculations in multiple-band
non-interacting Fermi systems. By applying this framework to Rashba
2DEG, we have resolved some recent discrepancies related to the AHE
in such systems. The findings of this simple 2D model have been linked
to higher dimensional systems arguing that most likely the major contributions
to the AHE come from the band anti-crossing regions similar to one
observed in the Rashba model. Our analytical and numerical results
reveal the crossover between the skew scattering dominated regime
in clean systems ($\sigma_{xy}\sim V_{0}/\gamma_{Born}\sim\sigma_{xx}$)
and the intrinsic dominated regime in moderately dirty systems ($\sigma_{xy}\sim const$).
In dirty systems, we observe the third distinct regime also dominated
by the intrinsic contribution. In this regime, the AHE diminishes
in a manner similar to $\sigma_{xy}\sim\sigma_{xx}^{\varphi}$ with
$\varphi$ being close to $1.6$. This, however, cannot be called
by scaling as the theory is not meaningful in a sufficiently wide
range of $\sigma_{xy}$ and $\sigma_{xx}$ due to breakdown of the
quasiparticle approximation when $\tau\varepsilon_{F}\thicksim1$.
For the repulsive impurities, we observe that the intrinsic and skew
anomalous Hall effects have opposite signs. As a result, the crossover
between those two is also accompanied by the change of sign of the
AHE. We suggest to engineer samples with repulsive impurities in order
to see this change of sign in the AHE.

We have resolved some discrepancies between the AHE results obtained
by using the Keldysh, Kubo and Boltzmann approaches by considering
the higher order skew scattering processes. We have also shown that
similar higher order processes are also important for the AMR when
the nonvertex and vertex parts cancel each other. We have calculated
the AMR in anisotropic systems properly taking into account the anisotropy
of the non-equilibrium distribution function. These calculations confirm
recent findings on the unreliability of common approximate approaches
to the Boltzmann equation.
\begin{acknowledgments}
We gratefully acknowledge fruitful discussions with E. I. Rashba,
V. Dugaev, J. Inoue, T. Jungwirth, A. H. MacDonald, G.E.W. Bauer,
N. Nagaosa and S. Onoda. This work was supported by the Alfred P.
Sloan Foundation (YT), by ONR under grant ONR-N000140610122, by NSF
under grant DMR-0547875, by SWAN-NRI and grants KJB100100802, LC510
and AV0Z10100521. J.S. is a Cottrell Scholar of the Research Foundation. 
\end{acknowledgments}
\appendix

\section{Generalizations for short-range disorder}

In section IIA, we derive the kinetic equation with the self-energy
expression that is valid for uniform systems. Here, we generalize
this self-energy to non-uniform systems in the presence of a short
range disorder postulated by the following infinite set of correlators:

\begin{equation}
\begin{array}{c}
\left\langle VV\right\rangle =\left\langle V\right\rangle \left\langle V\right\rangle +V_{2}\delta_{\mathbf{r}_{1}\mathbf{r}_{2}},\\
\left\langle VVV\right\rangle =\sum\left\langle VV\right\rangle \left\langle V\right\rangle +V_{3}\delta_{\mathbf{r}_{1}\mathbf{r}_{2}\mathbf{r}_{3}},\\
\left\langle VVVV\right\rangle =\sum\left(\left\langle VVV\right\rangle \left\langle V\right\rangle +\left\langle VV\right\rangle \left\langle VV\right\rangle \right)+V_{4}\delta_{\mathbf{r}_{1}\mathbf{r}_{2}\mathbf{r}_{3}\mathbf{r}_{4}},\\
....\\
....\end{array}\label{Correlators}\end{equation}
where we sum all possible decouplings of the correlators into a product
of two lower order correlators and $\delta_{\mathbf{r}_{1}\mathbf{r}_{2}\mathbf{r}_{3}...r_{N}}=\prod_{i=1..N-1}\delta(\mathbf{r}_{i}-\mathbf{r}_{i+1})$.
Note that usually the averaged impurity potential is zero, $\left\langle V\right\rangle =V_{1}=0$. 

After performing the averaging procedure for the Green's function,
we again arrive at the kinetic Eq. (\ref{Kinetic0}) with the self-energy
given by the following formal expression:

\begin{equation}
\check{\Sigma}=\left(\check{V_{0}}+\check{V_{0}}\otimes\check{G}\otimes\check{V_{0}}+\dots\right)|_{V_{0}^{n}\rightarrow V_{n}},\label{TmatrixSeries1}\end{equation}
where $\check{V}_{0}=V_{0}\left(\begin{array}{cc}
\hat{\eta} & 0\\
0 & \hat{\eta}\end{array}\right)\delta(1-1')$ and in the term of $n$-th order proportional to $V_{0}^{n}$ we
replace $V_{0}^{n}$ by $V_{n}$ which ensures that the correlators
in Eq. (\ref{Correlators}) are properly considered. It is convenient
to introduce the notation:\begin{equation}
\check{E}=\left(\check{V_{0}}+\check{V_{0}}\otimes\check{G}\otimes\check{V_{0}}+\dots\right),\label{Eequation}\end{equation}
which, in analogy with the self-energy, also has retarded (advanced)
$E^{R(A)}$ and lesser $E^{<}$ components. Eq. (\ref{Eequation})
can be rewritten in the form of $T$-matrix equation, $\check{E}=\check{V}\otimes[\check{1}+\check{G}\otimes\check{E}]$,
which leads to the expressions for the self-energies:\begin{equation}
\begin{array}{c}
\Sigma^{R(A)}=E^{R(A)}|_{V_{0}^{n}\rightarrow V_{n}},\\
\\\Sigma^{<}=\left(E^{R}\otimes G^{<}\otimes E^{A}\right)|_{V_{0}^{n}\rightarrow V_{n}},\end{array}\label{SelfEnergyComp}\end{equation}
where the notation $|_{V_{0}^{n}\rightarrow V_{n}}$ is formal and
it means that $E^{R(A)}$ has to be fist expanded with respect to
$V_{0}$ and then the substitution has to be applied. Eqs. (\ref{Kinetic1})
and (\ref{Sigma2}) can now be rederived for non-uniform systems with
the disorder given by Eq. (\ref{Correlators}).

Nevertheless, for the purposes of this paper, it is sufficient to
consider the uniform and stationary case. This leads to substantial
simplifications outlined in sections IIB and IIC. Results of sections
IIB and IIC also hold for the disorder given by Eq. (\ref{Correlators})
with the exception of Eqs. (\ref{Kinetic3}) and (\ref{BELOWselfenergy})
that should be replaced by the following equations:\begin{equation}
\begin{array}{c}
\hat{\Sigma}_{1}^{<}=\left[\hat{E}_{eq}^{R}\left(\int{\displaystyle \frac{d^{2}k'}{(2\pi)^{2}}}\hat{G}_{1}^{<}(\mathbf{k}')\right)\hat{E}_{eq}^{A}\right]_{V_{0}^{n}\rightarrow V_{n}},\\
\\\hat{\Sigma}_{\mathbf{E}}^{R(A)}=\left[\hat{E}_{eq}^{R(A)}\left(\int{\displaystyle \frac{d^{2}k'}{(2\pi)^{2}}}\hat{G}_{\mathbf{E}}^{R(A)}(\mathbf{k}')\right)\hat{E}_{eq}^{R(A)}\right]_{V_{0}^{n}\rightarrow V_{n}}.\end{array}\label{NEselfenergyComp}\end{equation}

Finally, we would like to present several examples in which the disorder
given by Eq. (\ref{Correlators}) is realized. The simplest example
is given by Eq. (\ref{Delta}) and in this case $V_{n}=V_{0}^{n}$.
For the disorder given by Eq. (\ref{Delta1}), we have $V_{n}=n_{i}\left\langle (V_{0}^{i})^{n}\right\rangle _{dis}$.
For the Gaussian white-noise disorder, only $V_{2}$ is non-zero and
$V_{n}|_{n\neq2}=0$. For the telegraph white-noise disorder all odd
correlators vanish, $V_{2n+1}=0$.

\section{Calculation of self-consistent self energy $\hat{\Sigma}_{eq}^{R(A)}$}

The following relations can be calculated by a direct analytical integration
of Eq. (\ref{GreenRA}):\[
\hat{\gamma}^{R(A)}=\int{\displaystyle {d^{2}k}/{(2\pi)^{2}}}{\textstyle \hat{G}_{eq}^{R(A)}}(\mathbf{k},\omega)\equiv\gamma^{R(A)}\hat{\sigma}_{0}+\gamma_{z}^{R(A)}\hat{\sigma}_{z},\]
 \[
\begin{array}{ccl}
\gamma^{R} & = & \dfrac{(K_{+}-2W)\left[\ln(K_{0}-K_{+})-\ln(-K_{+})\right]}{2\pi(K_{-}-K_{+})}\\
 &  & -\dfrac{(K_{-}-2W)\left[\ln(K_{0}-K_{-})-\ln(-K_{-})\right]}{2\pi(K_{-}-K_{+})}\end{array},\]
 \[
\gamma_{z}^{R}=\dfrac{\ln(K_{0}-K_{+})-\ln(-K_{+})-\ln(K_{0}-K_{-})+\ln(-K_{-})}{\pi(K_{-}-K_{+})/H},\]
 \[
\gamma^{A}=(\gamma^{R})^{*};\;\gamma_{x(y,z)}^{A}=(\gamma_{x(y,z)}^{R})^{*}\;\gamma_{y(z)}^{R}=0,\]
 where $W=\omega-\Sigma_{eq0}^{R}$, $H=h-\Sigma_{eq\mbox{z}}^{R}$
, $K_{\pm}=2(W+\alpha^{2}\mp\sqrt{H^{2}+2W\alpha^{2}+\alpha^{4}})$
and $K_{0}=k_{0}^{2}$ describes the cut-off $k_{0}$ in momentum
integration.

For each energy, $\Sigma_{eq0}^{R}(\omega)$ and $\Sigma_{eq\mbox{z}}^{R}(\omega)$
are calculated by performing a number of iterations with the consequent
iteration according to\[
\Sigma_{eq0}^{R(A)}=\frac{1}{2}{\rm Tr}\left[n_{i}V_{0}(\hat{1}-V_{0}\hat{\gamma}^{R(A)})^{-1}\hat{\sigma}_{0}\right],\]
 \[
\Sigma_{eq\mbox{z}}^{R(A)}=\frac{1}{2}{\rm Tr}\left[n_{i}V_{0}(\hat{1}-V_{0}\hat{\gamma}^{R(A)})^{-1}\hat{\sigma}_{z}\right].\]
 The iterations are performed until the prescribed accuracy is reached.

\section{Calculation of the matrix $\hat{\rho}(\omega)$}

For the electric field $\mathbf{E}$ along the $y$ axis $\mathbf{E}=(0,E_{y})$,
we solve here the linear Eq. (\ref{DensityEq1}) for the elements
of the matrix $\hat{\rho}(\omega)$ by performing analytically the
momentum integrations of the Green's functions ${\textstyle \hat{G}_{0}^{R(A)}}(\mathbf{k},\omega)$
(given by Eq. (\ref{GreenRA})) in the right hand side. For each energy
$\omega$, we obtain the following expressions that also depend on
the self-consistent values of $\Sigma_{00}^{R}(\omega)$ and $\Sigma_{0z}^{R}(\omega)$:\[
\rho_{--}(\omega)=\rho_{++}(\omega)=0,\]

\begin{widetext}

\noindent \[
\begin{array}{c}
\rho_{+-}(\omega)=i\partial_{\omega}n_{F}eE_{y}\alpha\Biggl\{{\displaystyle \frac{\Bigl[K_{-}^{2}+4(H+W)(H^{*}-W^{*})\Bigr]\ln(-K_{-})}{(K_{-}-K_{-}^{*})(K_{-}-K_{+})(K_{-}-K_{+}^{*})\pi}}-{\displaystyle \frac{\Bigl[K_{-}^{*2}+4(H+W)(H^{*}-W^{*})\Bigr]\ln(-K_{-}^{*})}{(K_{-}-K_{-}^{*})(K_{-}^{*}-K_{+})(K_{-}^{*}-K_{+}^{*})\pi}}\\
\left.\left.+{\displaystyle \frac{\left[K_{+}^{2}+4(H+W)(H^{*}-W^{*})\right]\ln(-K_{+})}{(K_{-}-K_{+})(K_{-}^{*}-K_{+})(K_{+}-K_{+}^{*})\pi}}-{\displaystyle \frac{\left[K_{+}^{*2}+4(H+W)(H^{*}-W^{*})\right]\ln(-K_{+}^{*})}{(K_{-}-K_{+}^{*})(K_{-}^{*}-K_{+}^{*})(K_{+}-K_{+}^{*})\pi}}\right\} \right/\\
\left\{ -1+\left[{\displaystyle \frac{(2H-K_{-}+2W)(2H^{*}+K_{-}-2W^{*})\ln(-K_{-})}{(K_{-}-K_{-}^{*})(K_{-}-K_{+})(K_{-}-K_{+}^{*})\pi}}-{\displaystyle \frac{(2H-K_{-}^{*}+2W)(2H^{*}+K_{-}^{*}-2W^{*})\ln(-K_{-}^{*})}{(K_{-}-K_{-}^{*})(K_{-}^{*}-K_{+})(K_{-}^{*}-K_{+}^{*})\pi}}\right.\right.\\
\left.\left.+{\displaystyle \frac{(2H-K_{+}+2W)(2H^{*}+K_{+}-2W^{*})\ln(-K_{+})}{(K_{-}-K_{+})(K_{-}^{*}-K_{+})(K_{+}-K_{+}^{*})\pi}}-{\displaystyle \frac{(2H-K_{+}^{*}+2W)(2H^{*}+K_{+}^{*}-2W^{*})\ln(-K_{+}^{*})}{(K_{-}-K_{+}^{*})(K_{-}^{*}-K_{+}^{*})(K_{+}-K_{+}^{*})\pi}}\right]n_{i}T_{++}T_{--}^{*}\right\} \end{array}\]

\noindent \[
\begin{array}{c}
\rho_{-+}(\omega)=i\partial_{\omega}n_{F}eE_{y}\alpha\Biggl\{{\displaystyle \frac{\Bigl[K_{-}^{2}+4(H-W)(H^{*}+W^{*})\Bigr]\ln(-K_{-})}{(K_{-}-K_{-}^{*})(K_{-}-K_{+})(K_{-}-K_{+}^{*})\pi}}-{\displaystyle \frac{\Bigl[K_{-}^{*2}+4(H-W)(H^{*}+W^{*})\Bigr]\ln(-K_{-}^{*})}{(K_{-}-K_{-}^{*})(K_{-}^{*}-K_{+})(K_{-}^{*}-K_{+}^{*})\pi}}\\
\left.\left.+{\displaystyle \frac{\left(K_{+}^{2}+4(H-W)(H^{*}+W^{*})\right)\ln(-K_{+})}{(K_{-}-K_{+})(K_{-}^{*}-K_{+})(K_{+}-K_{+}^{*})\pi}}-{\displaystyle \frac{\left(K_{+}^{*2}+4(H-W)(H^{*}+W^{*})\right)\ln(-K_{+}^{*})}{(K_{-}-K_{+}^{*})(K_{-}^{*}-K_{+}^{*})(K_{+}-K_{+}^{*})\pi}}\right\} \right/\\
\left\{ -1+\left[{\displaystyle \frac{(2H+K_{-}-2W)(2H^{*}-K_{-}+2W^{*})\ln(-K_{-})}{(K_{-}-K_{-}^{*})(K_{-}-K_{+})(K_{-}-K_{+}^{*})\pi}}-{\displaystyle \frac{(2H+K_{-}^{*}-2W)(2H^{*}-K_{-}^{*}+2W^{*})\ln(-K_{-}^{*})}{(K_{-}-K_{-}^{*})(K_{-}^{*}-K_{+})(K_{-}^{*}-K_{+}^{*})\pi}}\right.\right.\\
\left.\left.+{\displaystyle \frac{(2H+K_{+}-2W)(2H^{*}-K_{+}+2W^{*})\ln(-K_{+})}{(K_{-}-K_{+})(K_{-}^{*}-K_{+})(K_{+}-K_{+}^{*})\pi}}-{\displaystyle \frac{(2H+K_{+}^{*}-2W)(2H^{*}-K_{+}^{*}+2W^{*})\ln(-K_{+}^{*})}{(K_{-}-K_{+}^{*})(K_{-}^{*}-K_{+}^{*})(K_{+}-K_{+}^{*})\pi}}\right]n_{i}T_{--}T_{++}^{*}\right\} \end{array}\]

\end{widetext}

\section{Calculation of the matrix $\hat{\rho}_{\mathbf{E}}^{R(A)}(\omega)$}

For the electric field $\mathbf{E}$ along the $y$ axis $\mathbf{E}=(0,E_{y})$,
we solve here the linear Eq. (\ref{DensityEq2}) for the elements
of the matrix $\hat{\rho}_{\mathbf{E}}^{R(A)}(\omega)$ by performing
analytically the momentum integrations of the Green's functions ${\textstyle \hat{G}_{0}^{R(A)}}(\mathbf{k},\omega)$
(given by Eq. (\ref{GreenRA})) in the right hand side. For each energy
$\omega$, we obtain the following expressions that also depend on
the self-consistent values of $\Sigma_{00}^{R}(\omega)$ and $\Sigma_{0z}^{R}(\omega)$:\[
\rho_{E_{y}--}^{R}(\omega)=\rho_{E_{y}++}^{R}(\omega)=0,\]

\begin{widetext}

\noindent \[
\begin{array}{c}
\rho_{E_{y}+-}^{R}(\omega)=\left.4i\alpha\Bigl[H(-1+\partial_{\omega}\Sigma_{00}^{R})-W\partial_{\omega}\Sigma_{0z}^{R}\Bigr]\biggl\{ K_{-}^{2}-K_{+}^{2}+2K_{-}K_{+}\Bigl[-\ln(-K_{-})+\ln(-K_{+})\Bigr]\biggr\}\right/\\
\biggl\{(K_{-}-K_{+})\Bigl[K_{-}^{3}K_{+}\pi+K_{+}^{3}K_{-}\pi+K_{-}^{2}K_{+}n_{i}T_{--}T_{++}-4K_{+}n_{i}T_{--}T_{++}\left(H^{2}-W^{2}\right)\\
+K_{+}^{2}K_{-}n_{i}T_{--}T_{++}-4K_{-}n_{i}T_{--}T_{++}\left(H^{2}-W^{2}\right)-2\pi K_{+}^{2}K_{-}^{2}-8K_{+}K_{-}n_{i}T_{--}T_{++}W\Bigr]\\
+2K_{-}K_{+}n_{i}T_{--}T_{++}\Bigl[4H^{2}+(K_{-}-2W)(-K_{+}+2W)\Bigr]\Bigl[\ln(-K_{-})-\ln(-K_{+})\Bigr]\biggr\}\end{array}\]
 \[
\rho_{E_{y}-+}^{R}(\omega)=-\rho_{E_{y}+-}^{R}(\omega)\]

\end{widetext}

\section{Calculation of the Fermi surface conductivity}

For the electric field $\mathbf{E}$ along the $y$ axis $\mathbf{E}=(0,E_{y})$,
we perform momentum $k$ and frequency $\omega$ integrations in Eq.
(\ref{CurentI-Calc}). It is convenient to divide the resultant conductivity
into two parts; the bare bubble part $\sigma_{xy(yy)}^{Ib}$ that
corresponds to calculating only the second line in Eq. (\ref{CurentI-Calc})
effectively assuming that $\hat{\rho}(\omega_{F})=0$, and self consistent
part $\sigma_{xy(yy)}^{Isc}$ that corresponds to calculating the
first line in Eq. (\ref{CurentI-Calc}) that takes into account correction
due to self-consistent calculation of $\hat{\rho}(\omega_{F})$:\begin{equation}
\sigma_{xy}^{I}=\sigma_{xy}^{Ib}+\sigma_{xy}^{Isc},\label{BubbleSC}\end{equation}

\begin{equation}
\sigma_{yy}^{I}=\sigma_{yy}^{Ib}+\sigma_{yy}^{Isc},\label{BubbleSC1}\end{equation}

We arrive at analytical expressions for the bare bubble contributions
to the conductivities $\sigma_{xy}^{Ib}$ and $\sigma_{yy}^{Ib}$
that depend on the self-consistent values of $\Sigma_{00}^{R}(\omega_{F})$
and $\Sigma_{0z}^{R}(\omega_{F})$ at the Fermi surface:

\begin{widetext}

\noindent \begin{equation}
\begin{array}{c}
\dfrac{\sigma_{xy}^{Ib}}{e^{2}/\hbar}={\displaystyle \frac{2i\alpha^{2}\left[-H^{*}(K_{-}+2W)+H(K_{-}+2W^{*})\right]\ln(-K_{-})}{(K_{-}-K_{-}^{*})(K_{-}-K_{+})(K_{-}-K_{+}^{*})\pi^{2}}}+{\displaystyle \frac{2i\alpha^{2}\left[H^{*}(K_{-}^{*}+2W)-H(K_{-}^{*}+2W^{*})\right]\ln(-K_{-}^{*})}{(K_{-}-K_{-}^{*})(K_{-}^{*}-K_{+})(K_{-}^{*}-K_{+}^{*})\pi^{2}}}\\
+{\displaystyle \frac{2i\alpha^{2}\left[-H^{*}(K_{+}+2W)+H(K_{+}+2W^{*})\right]\ln(-K_{+})}{(K_{-}-K_{+})(K_{-}^{*}-K_{+})(K_{+}-K_{+}^{*})\pi^{2}}}+{\displaystyle \frac{2i\alpha^{2}\left[H^{*}(K_{+}^{*}+2W)-H(K_{+}^{*}+2W^{*})\right]\ln(-K_{+}^{*})}{(K_{-}-K_{+}^{*})(K_{-}^{*}-K_{+}^{*})(K_{+}-K_{+}^{*})\pi^{2}}}\end{array},\label{AppCondIxyb}\end{equation}

\noindent \begin{equation}
\begin{array}{l}
\dfrac{\sigma_{yy}^{Ib}}{e^{2}/\hbar}=-{\displaystyle \frac{K_{-}^{2}(K_{-}-3K_{+})K_{+}^{*}\ln(-K_{-})}{4(K_{-}-K_{+})^{3}(K_{-}-K_{+}^{*})\pi^{2}}}-{\displaystyle \frac{K_{-}^{3}\left(K_{-}(K_{-}+K_{-}^{*})-(K_{-}+3K_{-}^{*})K_{+}+2K_{+}^{2}\right)\ln(-K_{-})}{4(K_{-}-K_{-}^{*})(K_{-}-K_{+})^{3}(K_{-}-K_{+}^{*})\pi^{2}}}\\
+{\displaystyle \frac{K_{-}K_{-}^{*2}(K_{-}^{*}-3K_{+}^{*})\ln(-K_{-}^{*})}{4(K_{-}-K_{-}^{*})(K_{-}^{*}-K_{+}^{*})^{3}\pi^{2}}}+{\displaystyle \frac{K_{-}^{*3}\left(K_{-}^{*}(K_{-}^{*}+K_{+})-(K_{-}^{*}+3K_{+})K_{+}^{*}+2K_{+}^{*2}\right)\ln(-K_{-}^{*})}{4(K_{-}-K_{-}^{*})(K_{-}^{*}-K_{+})(K_{-}^{*}-K_{+}^{*})^{3}\pi^{2}}}\\
-{\displaystyle \frac{K_{+}^{2}(3K_{-}-K_{+})K_{+}^{*}\ln(-K_{+})}{4(K_{-}-K_{+})^{3}(K_{+}-K_{+}^{*})\pi^{2}}}-{\displaystyle \frac{K_{+}^{3}\left(2K_{-}^{2}+K_{+}(K_{-}^{*}+K_{+})-K_{-}(3K_{-}^{*}+K_{+})\right)\ln(-K_{+})}{4(K_{-}^{*}-K_{+})(K_{-}-K_{+})^{3}(K_{+}-K_{+}^{*})\pi^{2}}}\\
+{\displaystyle \frac{K_{-}K_{+}^{*2}(3K_{-}^{*}-K_{+}^{*})\ln(-K_{+}^{*})}{4(K_{-}-K_{+}^{*})(K_{-}^{*}-K_{+}^{*})^{3}\pi^{2}}}+{\displaystyle \frac{K_{+}^{*3}\left(2K_{-}^{*2}+K_{+}^{*}(K_{+}+K_{+}^{*})-K_{-}^{*}(3K_{+}+K_{+}^{*})\right)\ln(-K_{+}^{*})}{4(K_{-}-K_{+}^{*})(K_{-}^{*}-K_{+}^{*})^{3}(K_{+}-K_{+}^{*})\pi^{2}}}\\
-{\displaystyle \frac{K_{-}K_{+}\left(K_{-}^{*2}+K_{+}^{*2}\right)-K_{-}^{2}\left(K_{-}^{*2}-K_{-}^{*}K_{+}^{*}+K_{+}^{*2}\right)-K_{+}^{2}\left(K_{-}^{*2}-K_{-}^{*}K_{+}^{*}+K_{+}^{*2}\right)}{2(K_{-}-K_{+})^{2}(K_{-}^{*}-K_{+}^{*})^{2}\pi^{2}}}\\
-{\displaystyle \frac{2\left(4WW^{*}\alpha^{2}+2HH^{*}\left(K_{-}-2\alpha^{2}\right)-K_{-}^{2}\left(W+W^{*}+\alpha^{2}\right)+2K_{-}\left(WW^{*}+(W+W^{*})\alpha^{2}\right)\right)\ln(-K_{-})}{2(K_{-}-K_{-}^{*})(K_{-}-K_{+})(K_{-}-K_{+}^{*})\pi^{2}}}\\
+{\displaystyle \frac{2\left(4WW^{*}\alpha^{2}+2HH^{*}\left(K_{-}^{*}-2\alpha^{2}\right)-K_{-}^{*2}\left(W+W^{*}+\alpha^{2}\right)+2K_{-}^{*}\left(WW^{*}+(W+W^{*})\alpha^{2}\right)\right)\ln(-K_{-}^{*})}{2(K_{-}-K_{-}^{*})(K_{-}^{*}-K_{+})(K_{-}^{*}-K_{+}^{*})\pi^{2}}}\\
-{\displaystyle \frac{2\left(4WW^{*}\alpha^{2}+2HH^{*}\left(K_{+}-2\alpha^{2}\right)-K_{+}^{2}\left(W+W^{*}+\alpha^{2}\right)+2K_{+}\left(WW^{*}+(W+W^{*})\alpha^{2}\right)\right)\ln(-K_{+})}{2(K_{-}-K_{+})(K_{-}^{*}-K_{+})(K_{+}-K_{+}^{*})\pi^{2}}}\\
+{\displaystyle \frac{2\left(4WW^{*}\alpha^{2}+2HH^{*}\left(K_{+}^{*}-2\alpha^{2}\right)-K_{+}^{*2}\left(W+W^{*}+\alpha^{2}\right)+2K_{+}^{*}\left(WW^{*}+(W+W^{*})\alpha^{2}\right)\right)\ln(-K_{+}^{*})}{2(K_{-}-K_{+}^{*})(-K_{-}^{*}+K_{+}^{*})(-K_{+}+K_{+}^{*})\pi^{2}}}\\
-{\displaystyle \frac{2\left(-2H^{2}+(K_{-}+K_{+}-2W)W\right)}{2(K_{-}-K_{+})^{2}\pi^{2}}}-{\displaystyle \frac{\left((K_{-}+K_{+})\left(4H^{2}+K_{-}K_{+}\right)-8K_{-}K_{+}W-4(K_{-}+K_{+})W^{2}\right)\alpha^{2}}{2K_{-}(K_{-}-K_{+})^{2}K_{+}\pi^{2}}}\\
-\left[\ln(-K_{-})-\ln(-K_{+})\right]\\
\times{\displaystyle \frac{2\left(W(-2K_{-}K_{+}+(K_{-}+K_{+})W)+\left(-K_{-}K_{+}+2(K_{-}+K_{+})W+4W^{2}\right)\alpha^{2}+H^{2}\left(K_{-}+K_{+}-4\alpha^{2}\right)\right)}{2(K_{-}-K_{+})^{3}\pi^{2}}}\\
-{\displaystyle \frac{2\left(-2H^{*2}+(K_{-}^{*}+K_{+}^{*}-2W^{*})W^{*}\right)}{2(K_{-}^{*}-K_{+}^{*})^{2}\pi^{2}}}-{\displaystyle \frac{\left((K_{-}^{*}+K_{+}^{*})\left(4H^{*2}+K_{-}^{*}K_{+}^{*}\right)-8K_{-}^{*}K_{+}^{*}W^{*}-4(K_{-}^{*}+K_{+}^{*})W^{*2}\right)\alpha^{2}}{2K_{-}^{*}(K_{-}^{*}-K_{+}^{*})^{2}K_{+}^{*}\pi^{2}}}\\
-\left[\ln(-K_{-}^{*})-\ln(-K_{+}^{*})\right]\\
\times{\displaystyle \frac{2\left(W^{*}(-2K_{-}^{*}K_{+}^{*}+(K_{-}^{*}+K_{+}^{*})W^{*})+\left(-K_{-}^{*}K_{+}^{*}+2(K_{-}^{*}+K_{+}^{*})W^{*}+4W^{*2}\right)\alpha^{2}+H^{*2}\left(K_{-}^{*}+K_{+}^{*}-4\alpha^{2}\right)\right)}{2(K_{-}^{*}-K_{+}^{*})^{3}\pi^{2}}}\end{array}\label{AppCondIyyb}\end{equation}

\end{widetext} where in this Appendix all parameter are taken at
the Fermi surface: $W=\omega_{F}-\Sigma_{00}^{R}(\omega_{F})$, $H=h-\Sigma_{0z}^{R}(\omega_{F})$
, $K_{\pm}=2(W+\alpha^{2}\mp\sqrt{H^{2}+2W\alpha^{2}+\alpha^{4}})$.

The analytical expressions for the self-consistent contributions to
the conductivities $\sigma_{xy}^{Isc}$ and $\sigma_{yy}^{Isc}$ become:

\begin{widetext}

\noindent \begin{equation}
\begin{array}{c}
\dfrac{\sigma_{xy}^{Isc}}{e^{2}/\hbar}={\displaystyle \frac{\alpha n_{i}\left(\rho_{+-}T_{++}T_{--}^{*}\left(K_{-}^{2}+4(H+W)(H^{*}-W^{*})\right)-\rho_{-+}T_{--}T_{++}^{*}\left(K_{-}^{2}+4(H-W)(H^{*}+W^{*})\right)\right)\ln(-K_{-})}{2(K_{-}-K_{-}^{*})(K_{-}-K_{+})(K_{-}-K_{+}^{*})\pi^{2}eE_{y}}}\\
-{\displaystyle \frac{\alpha n_{i}\left(\rho_{+-}T_{++}T_{--}^{*}\left(K_{-}^{*2}+4(H+W)(H^{*}-W^{*})\right)-\rho_{-+}T_{--}T_{++}^{*}\left(K_{-}^{*2}+4(H-W)(H^{*}+W^{*})\right)\right)\ln(-K_{-}^{*})}{2(K_{-}-K_{-}^{*})(K_{-}^{*}-K_{+})(K_{-}^{*}-K_{+}^{*})\pi^{2}eE_{y}}}\\
+{\displaystyle \frac{\alpha n_{i}\left(\rho_{+-}T_{++}T_{--}^{*}\left(K_{+}^{2}+4(H+W)(H^{*}-W^{*})\right)-\rho_{-+}T_{--}T_{++}^{*}\left(K_{+}^{2}+4(H-W)(H^{*}+W^{*})\right)\right)\ln(-K_{+})}{2(K_{-}-K_{+})(K_{-}^{*}-K_{+})(K_{+}-K_{+}^{*})\pi^{2}eE_{y}}}\\
-{\displaystyle \frac{\alpha n_{i}\left(\rho_{+-}T_{++}T_{--}^{*}\left(K_{+}^{*2}+4(H+W)(H^{*}-W^{*})\right)-\rho_{-+}T_{--}T_{++}^{*}\left(K_{+}^{*2}+4(H-W)(H^{*}+W^{*})\right)\right)\ln(-K_{+}^{*})}{2(K_{-}-K_{+}^{*})(-K_{-}^{*}+K_{+}^{*})(-K_{+}+K_{+}^{*})\pi^{2}eE_{y}}}\end{array},\label{AppCondIxySC}\end{equation}

\noindent \begin{equation}
\begin{array}{c}
\dfrac{\sigma_{yy}^{Isc}}{e^{2}/\hbar}={\displaystyle \frac{i\alpha n_{i}\left(\rho_{+-}T_{++}T_{--}^{*}\left(K_{-}^{2}+4(H+W)(H^{*}-W^{*})\right)+\rho_{-+}T_{--}T_{++}^{*}\left(K_{-}^{2}+4(H-W)(H^{*}+W^{*})\right)\right)\ln(-K_{-})}{2(K_{-}-K_{-}^{*})(K_{-}-K_{+})(K_{-}-K_{+}^{*})\pi^{2}eE_{y}}}\\
-{\displaystyle \frac{i\alpha n_{i}\left(\rho_{+-}T_{++}T_{--}^{*}\left(K_{-}^{*2}+4(H+W)(H^{*}-W^{*})\right)+\rho_{-+}T_{--}T_{++}^{*}\left(K_{-}^{*2}+4(H-W)(H^{*}+W^{*})\right)\right)\ln(-K_{-}^{*})}{2(K_{-}-K_{-}^{*})(K_{-}^{*}-K_{+})(K_{-}^{*}-K_{+}^{*})\pi^{2}eE_{y}}}\\
+{\displaystyle \frac{i\alpha n_{i}\left(\rho_{+-}T_{++}T_{--}^{*}\left(K_{+}^{2}+4(H+W)(H^{*}-W^{*})\right)+\rho_{-+}T_{--}T_{++}^{*}\left(K_{+}^{2}+4(H-W)(H^{*}+W^{*})\right)\right)\ln(-K_{+})}{2(K_{-}-K_{+})(K_{-}^{*}-K_{+})(K_{+}-K_{+}^{*})\pi^{2}eE_{y}}}\\
-{\displaystyle \frac{i\alpha n_{i}\left(\rho_{+-}T_{++}T_{--}^{*}\left(K_{+}^{*2}+4(H+W)(H^{*}-W^{*})\right)+\rho_{-+}T_{--}T_{++}^{*}\left(K_{+}^{*2}+4(H-W)(H^{*}+W^{*})\right)\right)\ln(-K_{+}^{*})}{2(K_{-}-K_{+}^{*})(-K_{-}^{*}+K_{+}^{*})(-K_{+}+K_{+}^{*})\pi^{2}eE_{y}}}\end{array}\label{AppCondIyySC}\end{equation}

\end{widetext} where again all parameters are calculated at the Fermi
surface.

\section{Calculation of the Fermi sea conductivity}

For the electric field $\mathbf{E}$ along the $y$ axis $\mathbf{E}=(0,E_{y})$,
we perform momentum integrations in Eq. (\ref{CurrentII-Calc}) arriving
at the following expressions for conductivities $\sigma_{xy}^{II}$
and $\sigma_{yy}^{II}$:\begin{equation}
\sigma_{xy}^{II}=\sigma_{xy}^{IIb}+\sigma_{xy}^{IIsc},\label{BubbleSCII}\end{equation}

\begin{equation}
\sigma_{yy}^{II}=0,\label{BubbleSCII1}\end{equation}

\begin{widetext}

\begin{equation}
\begin{array}{c}
\dfrac{\sigma_{xy}^{IIb}}{e^{2}/\hbar}={\displaystyle {\displaystyle \intop}}d\omega n_{F}\biggl\{{\displaystyle \frac{4i\alpha^{2}\left[H(1-\partial_{\omega}\Sigma_{00}^{R})(K_{-}+K_{+})+\partial_{\omega}\Sigma_{0z}^{R}(K_{+}W+K_{-}(K_{+}+W))\right]}{K_{-}K_{+}(K_{-}-K_{+})^{2}\pi^{2}}}\\
-{\displaystyle \frac{2i\alpha^{2}\left[4(1-\partial_{\omega}\Sigma_{00}^{R})H+\partial_{\omega}\Sigma_{0z}^{R}(K_{-}+K_{+}+4W)\right](\ln(-K_{-})-\ln(-K_{+}))}{(K_{-}-K_{+})^{3}\pi^{2}}}\biggr\}+c.c.\end{array}\label{AppCondIIb}\end{equation}
 \begin{equation}
\begin{array}{c}
\dfrac{\sigma_{xy}^{IIsc}}{e^{2}/\hbar}={\displaystyle {\displaystyle \intop}}d\omega n_{F}{\displaystyle \frac{\alpha(\rho_{E_{y}-+}^{R}-\rho_{E_{y}+-}^{R})T_{--}T_{++}\left(4H^{2}+K_{-}K_{+}-4W^{2}\right)\left(K_{-}^{2}-K_{+}^{2}+2K_{-}K_{+}(-\ln(-K_{-})+\ln(-K_{+}))\right)}{2K_{-}K_{+}(K_{-}-K_{+})^{3}\pi^{2}}}\\
+c.c.=0\end{array}\label{AppCondIIsc}\end{equation}

\end{widetext}

The fact that $\sigma_{xy}^{IIsc}=0$ follows from the identity $4H^{2}+K_{-}K_{+}-4W^{2}\equiv0$.

As one can see, $\sigma_{yy}^{II}$ and $\sigma_{xy}^{IIsc}$ contributions
to the Fermi sea Hall conductivity vanish and the non-vanishing contribution
$\sigma_{xy}^{IIb}$ depends on the self-consistent values of $\Sigma_{00}^{R}(\omega)$
and $\Sigma_{0z}^{R}(\omega)$ and its calculation from Eq. (\ref{AppCondIIb})
requires numerical integration over $\omega$.

\begin{widetext}

%
\begin{figure}
\centerline{\includegraphics[scale=0.8]{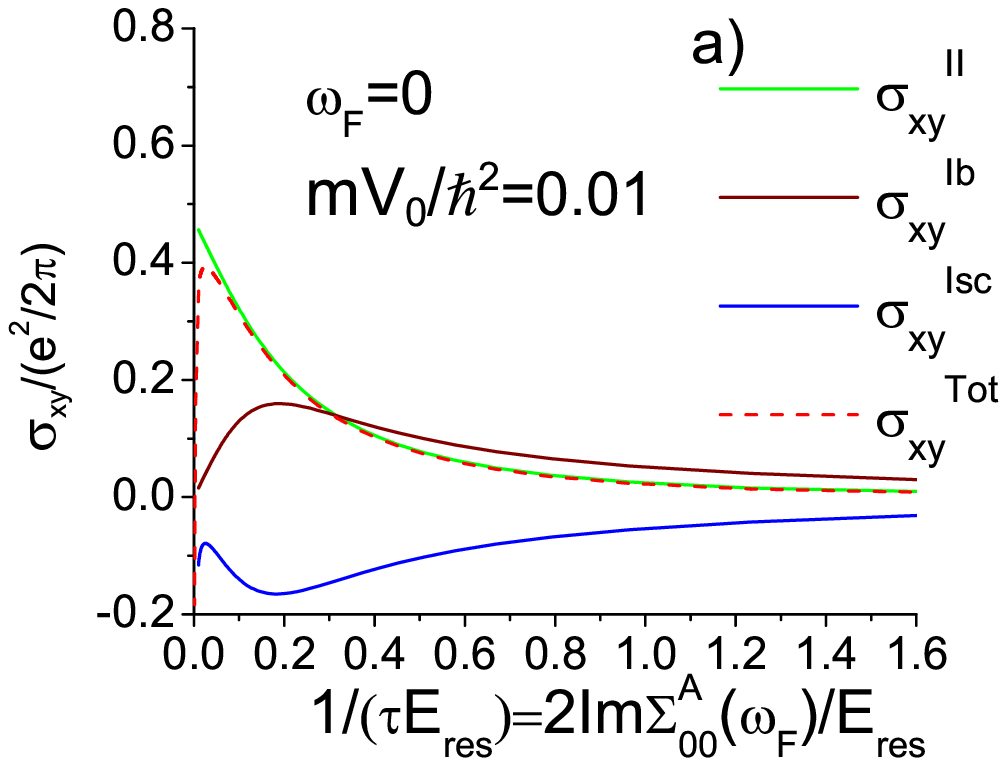}\includegraphics[scale=0.8]{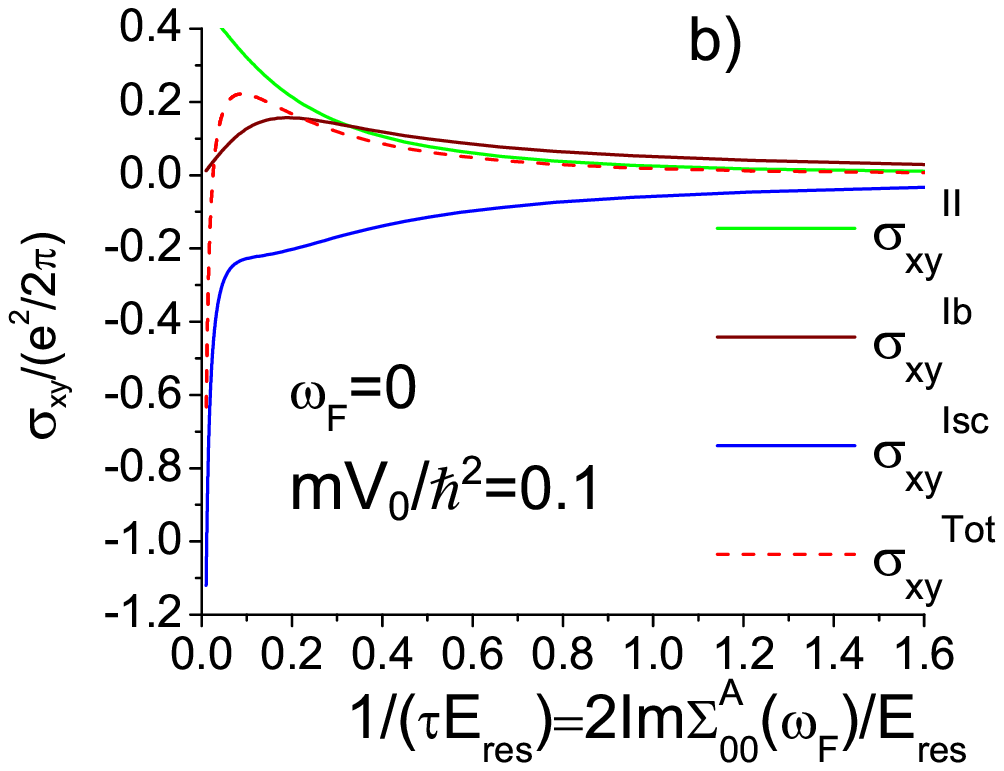}}

\centerline{\includegraphics[scale=0.8]{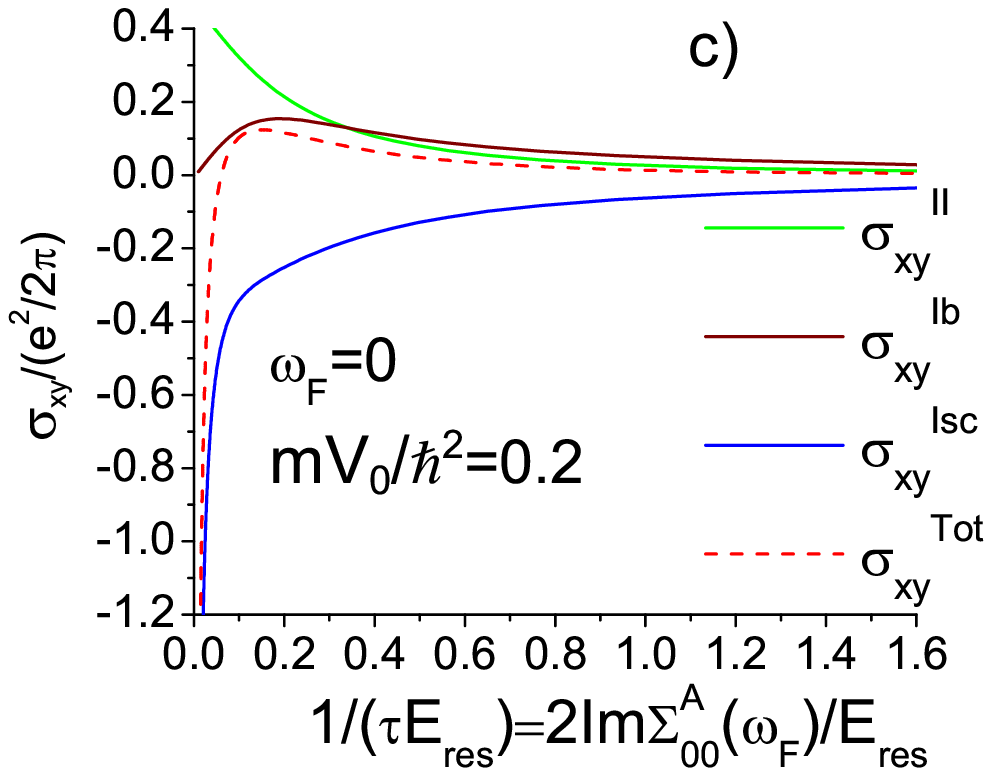}\includegraphics[scale=0.8]{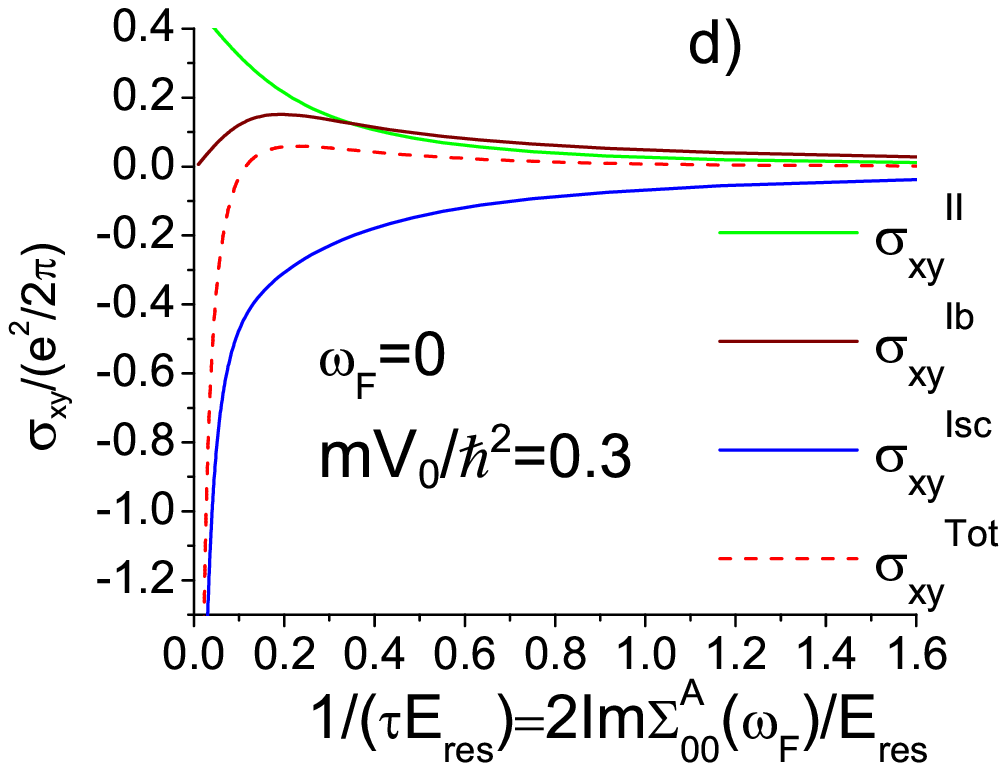}}

\centerline{\includegraphics[scale=0.8]{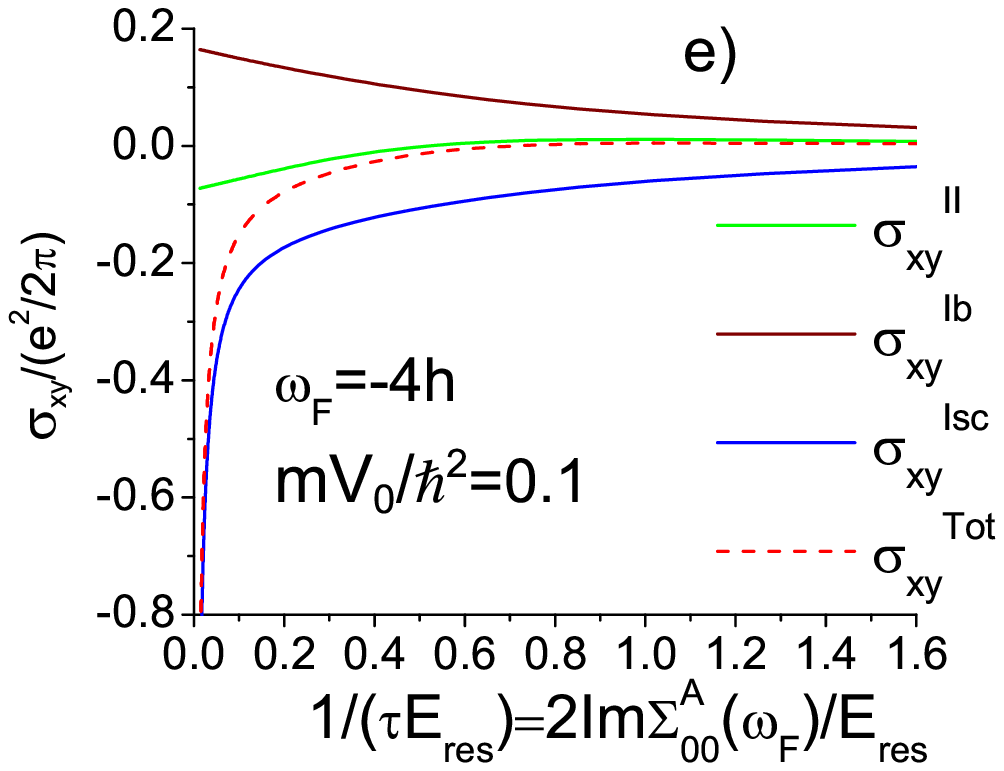}\includegraphics[scale=0.8]{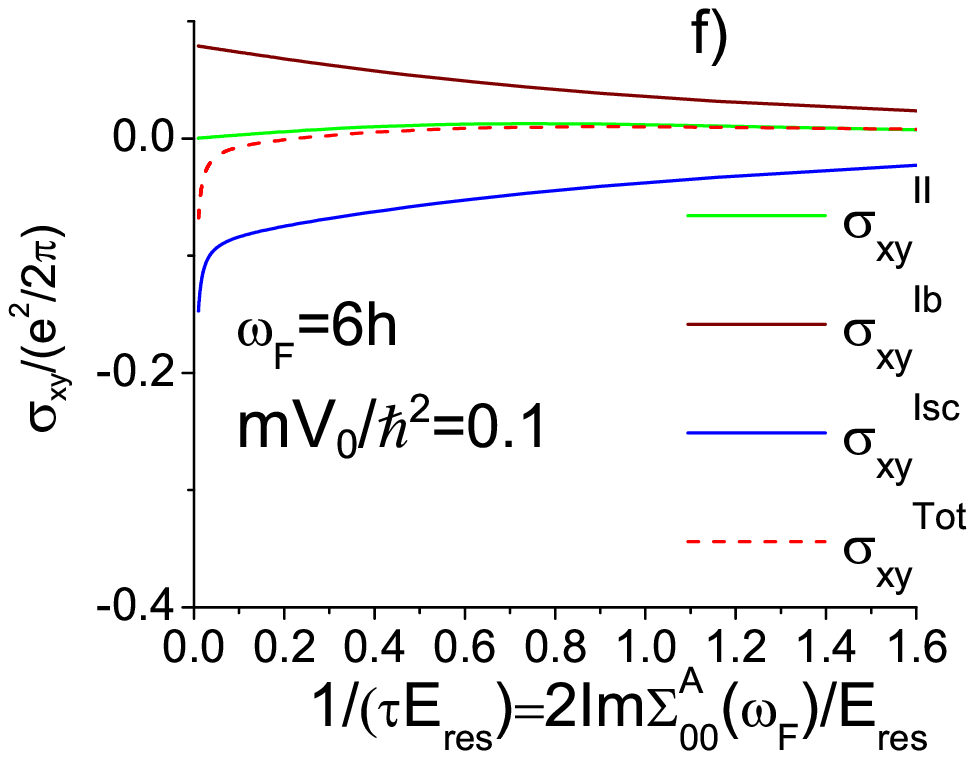}}

\caption{The anomalous Hall conductivity $\sigma_{xy}^{Tot}=\sigma_{xy}^{Ib}+\sigma_{xy}^{Isc}+\sigma_{xy}^{II}$
and its components ($\sigma_{xy}^{Ib}$, $\sigma_{xy}^{Isc}$, $\sigma_{xy}^{II}$)
versus the averaged relaxation rate $1/\tau=2\mbox{Im}\Sigma_{00}^{A}$
(defined in Appendix B). The spin-orbit interaction strength is $2m\alpha^{2}/E_{\mbox{res}}=3.59$
($E_{\mbox{res}}=10h$); the strength of impurities: $V_{0}=0.01$,
$0.1$, $0.2$, $0.3$; the Fermi energy $\varepsilon_{F}/E_{\mbox{res}}=0.9$
for $\omega_{F}=0$, $\varepsilon_{F}/E_{\mbox{res}}=0.5$ for $\omega_{F}=-4h$
and $\varepsilon_{F}/E_{\mbox{res}}=1.5$ for $\omega_{F}=6h$. Dimensionality
of quantities displayed in this plot is restored.}

\label{ComparisonP} 
\end{figure}

%
\begin{figure}
\centerline{\includegraphics[scale=0.8]{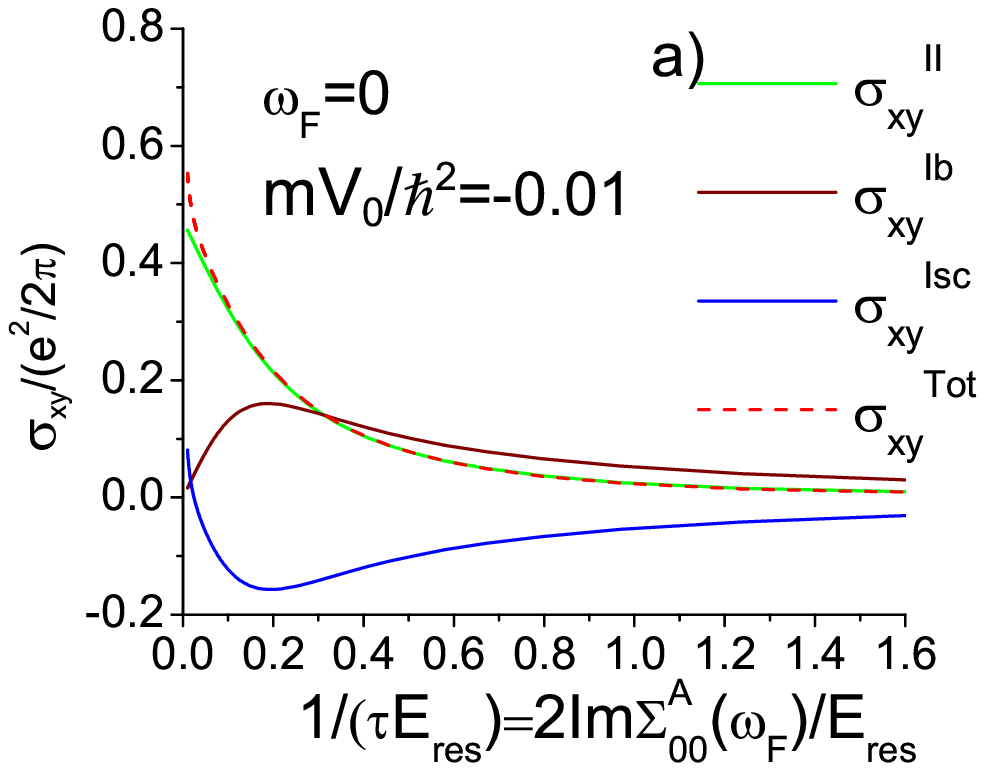}\includegraphics[scale=0.8]{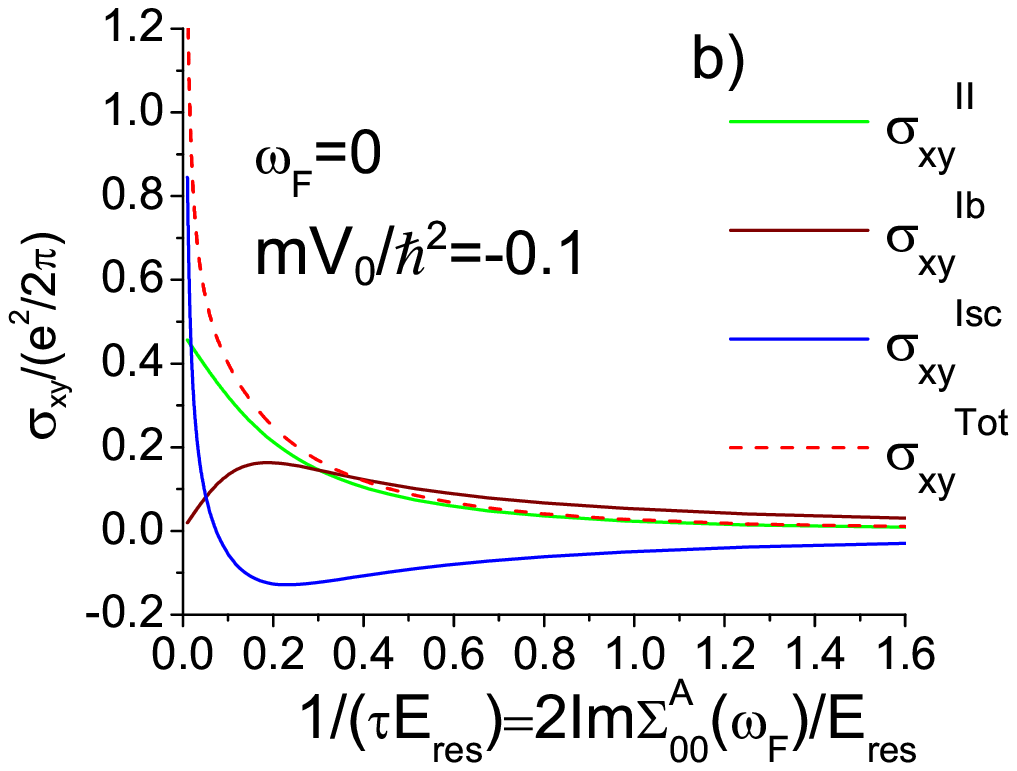}}

\centerline{\includegraphics[scale=0.8]{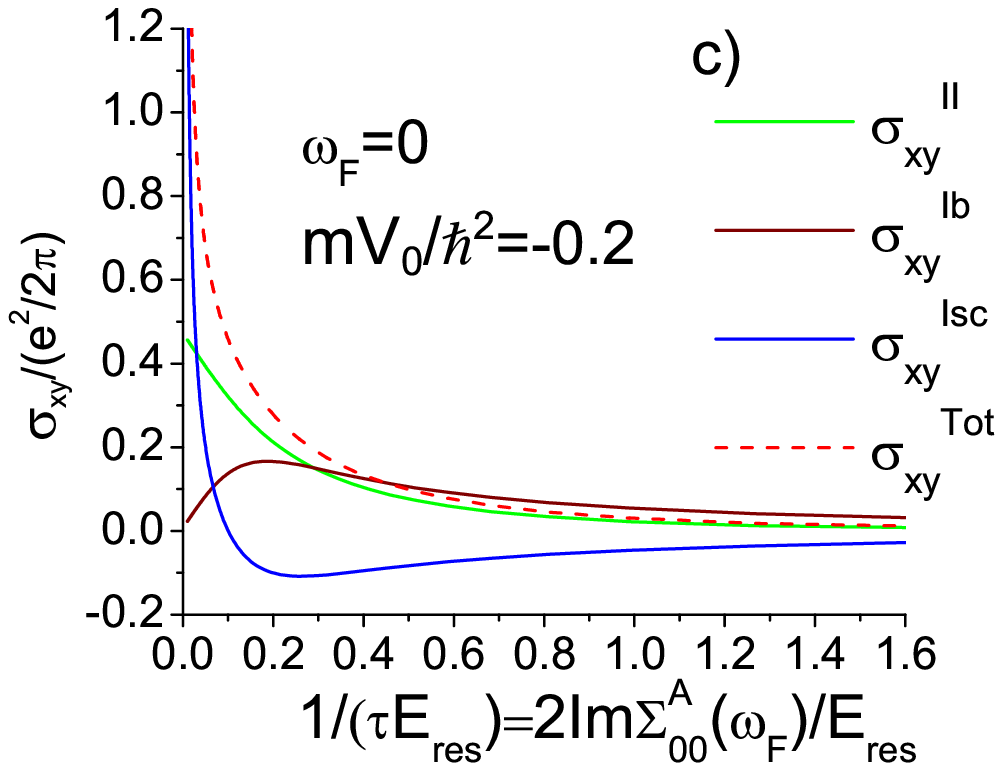}\includegraphics[scale=0.8]{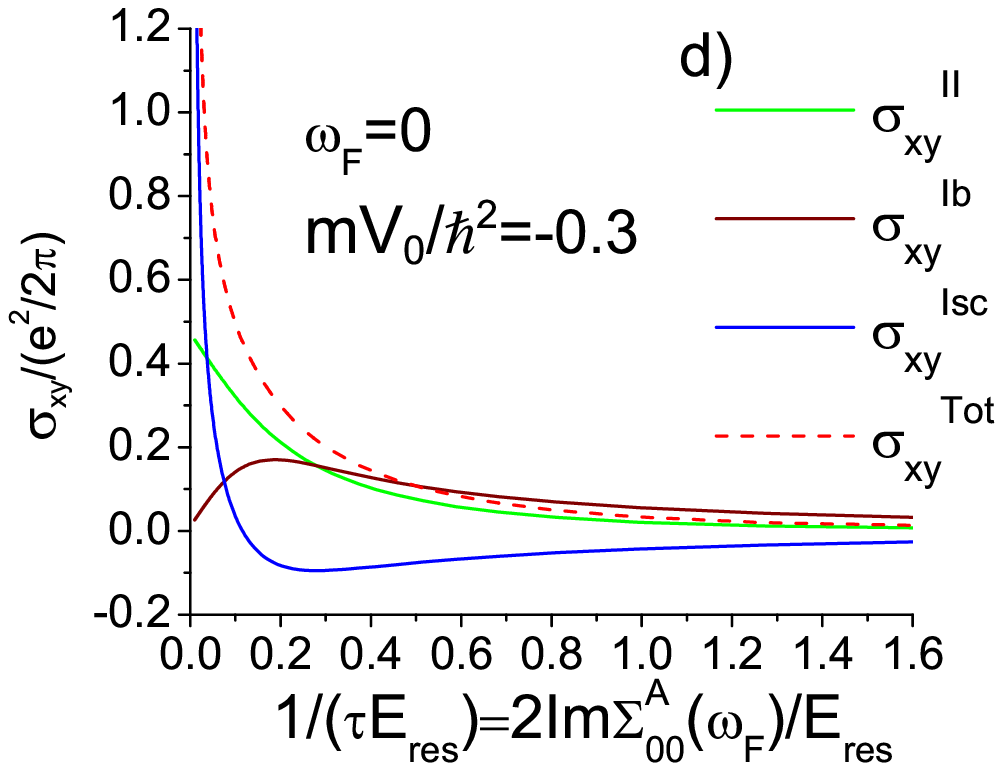}}

\centerline{\includegraphics[scale=0.8]{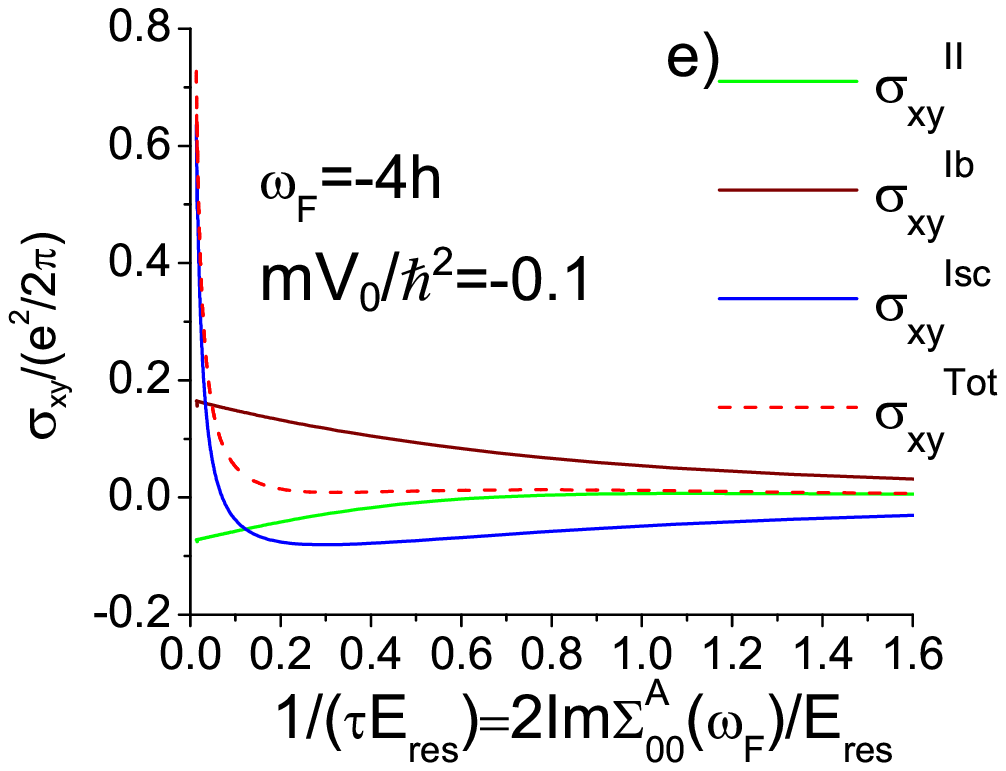}\includegraphics[scale=0.8]{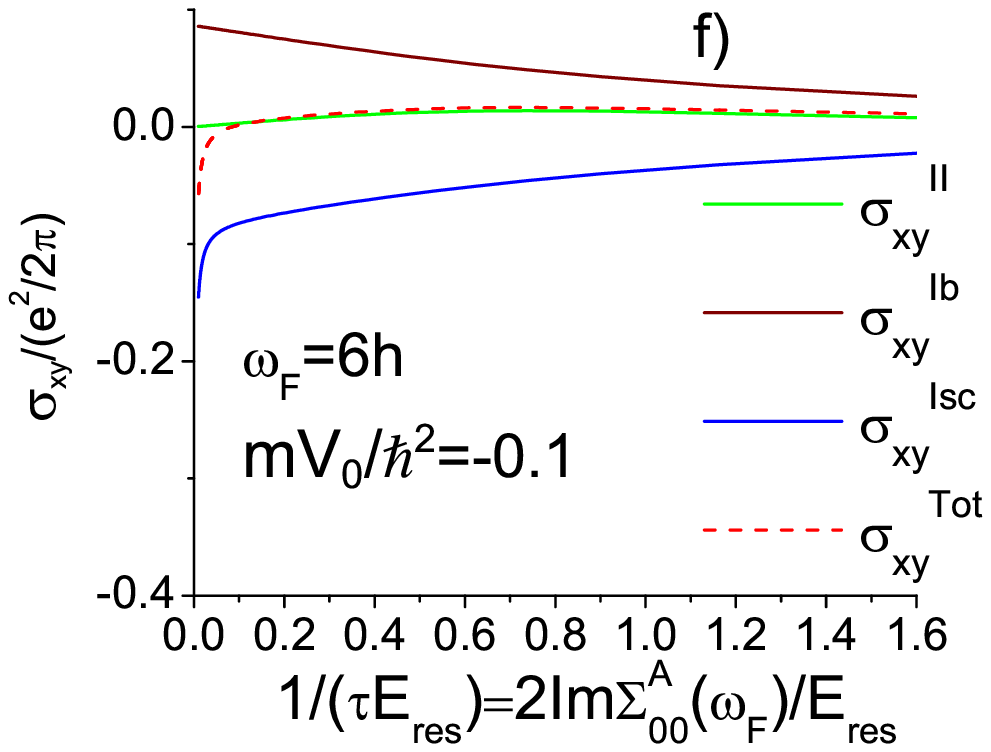}}

\caption{Identical to Fig. \ref{ComparisonP} plot with attractive disorder
($V_{0}=-0.01$, $-0.1$, $-0.2$, $-0.3$).}

\label{ComparisonM} 
\end{figure}

\end{widetext}

\section{Detailed results for the Hall conductivity}

In order to gain more insight into the behavior of the anomalous Hall
effect, in Figs. \ref{ComparisonP} and \ref{ComparisonM} we plot
different components of the AHE conductivity, particularly the Fermi
sea contribution $\sigma_{xy}^{II}$, the bare bubble contribution
$\sigma_{xy}^{Ib}$ (this corresponds to $\sigma_{xy}^{\mbox{I int}}$
in Ref. \onlinecite{Onoda:apr2008}) and the self consistent contribution
$\sigma_{xy}^{Isc}$ (this corresponds to $\sigma_{xy}^{\mbox{ext}}$
in Ref. \onlinecite{Onoda:apr2008}). In Fig. \ref{ComparisonP},
we take the same parameters as in Figs. 7 and 8 of Ref. \onlinecite{Onoda:apr2008}
and we find disagreement with Ref. \onlinecite{Onoda:apr2008} in
the results for the contribution $\sigma_{xy}^{\mbox{ext}}$ ($\sigma_{xy}^{Isc}$).
The contributions $\sigma_{xy}^{Ib}$ and $\sigma_{xy}^{II}$ perfectly
agree with Ref. \onlinecite{Onoda:apr2008}.

In the clean limit $\tau\rightarrow\infty$, we see that $\sigma_{xy}^{Isc}$
and thus the total Hall conductivity $\sigma_{xy}^{Tot}$ diverge.
This divergence ($\sigma_{xy}^{Isc}\sim1/n_{i}V_{0}$ in the regions
(ii) and (iii) and $\sigma_{xy}^{Isc}\sim1/n_{i}$ in the region (i),
see Fig. \ref{Spectrum}) is due to the skew scattering. The conductivity
$\sigma_{xy}^{Isc}$ also contains the side-jump contribution which
can be best seen in Fig. \ref{ComparisonP}a) in the sharp peak in
the conductivity for small $1/\tau$. The skew scattering contribution
decays much faster compared to the side-jump and intrinsic mechanisms
as we go to larger $1/\tau$. As a result, we can expect a cross-over
between the region dominated by the skew scattering and the region
dominated by the side-jump-intrinsic mechanisms. When both subbands
are partially occupied (see Figs. \ref{ComparisonP}f) and \ref{ComparisonM}f)),
the higher order skew scattering is still present. However, we do
not expect a well pronounced cross-over as the intrinsic contribution
cancels the side-jump contribution in the metallic regime (see Eq.
(\ref{SigmaIresult(i)})). By comparing Figs. \ref{ComparisonP}f)
and \ref{ComparisonM}f), one can see that the higher order skew scattering
(hybrid skew scattering)\citep{Kovalev:jul2008} does not change sign
when we change the sign of impurities.

When the side-jump-intrinsic and the skew scattering components have
opposite signs, as in Fig. (\ref{ComparisonP}), we observe the AHE
sign change instead of the cross-over. In Figs. \ref{ComparisonP}a)-d),
the skew scattering is negative in the clean limit while the side-jump-intrinsic
part is positive. This inevitably leads to the sign change of the
conductivity $\sigma_{xy}$ as we increase the disorder. 

\bibliographystyle{apsrev} \bibliographystyle{apsrev}
\bibliography{AHE}

\begin{thebibliography}{55}
\expandafter\ifx\csname natexlab\endcsname\relax\def\natexlab#1{#1}\fi
\expandafter\ifx\csname bibnamefont\endcsname\relax
  \def\bibnamefont#1{#1}\fi
\expandafter\ifx\csname bibfnamefont\endcsname\relax
  \def\bibfnamefont#1{#1}\fi
\expandafter\ifx\csname citenamefont\endcsname\relax
  \def\citenamefont#1{#1}\fi
\expandafter\ifx\csname url\endcsname\relax
  \def\url#1{\texttt{#1}}\fi
\expandafter\ifx\csname urlprefix\endcsname\relax\def\urlprefix{URL }\fi
\providecommand{\bibinfo}[2]{#2}
\providecommand{\eprint}[2][]{\url{#2}}

\bibitem[{\citenamefont{{Onoda} et~al.}(2008)\citenamefont{{Onoda}, {Sugimoto},
  and {Nagaosa}}}]{Onoda:apr2008}
\bibinfo{author}{\bibfnamefont{S.}~\bibnamefont{{Onoda}}},
  \bibinfo{author}{\bibfnamefont{N.}~\bibnamefont{{Sugimoto}}},
  \bibnamefont{and}
  \bibinfo{author}{\bibfnamefont{N.}~\bibnamefont{{Nagaosa}}},
  \bibinfo{journal}{Phys. Rev. B} \textbf{\bibinfo{volume}{77}},
  \bibinfo{pages}{165103} (\bibinfo{year}{2008}).

\bibitem[{\citenamefont{{Shindou} and {Balents}}(2008)}]{Shindou:jan2008}
\bibinfo{author}{\bibfnamefont{R.}~\bibnamefont{{Shindou}}} \bibnamefont{and}
  \bibinfo{author}{\bibfnamefont{L.}~\bibnamefont{{Balents}}},
  \bibinfo{journal}{Phys. Rev. B} \textbf{\bibinfo{volume}{77}},
  \bibinfo{pages}{035110} (\bibinfo{year}{2008}).

\bibitem[{\citenamefont{Hall}(1880)}]{Hall:jan1880}
\bibinfo{author}{\bibfnamefont{E.~H.} \bibnamefont{Hall}},
  \bibinfo{journal}{Philos. Mag.} \textbf{\bibinfo{volume}{19}},
  \bibinfo{pages}{301} (\bibinfo{year}{1880}).

\bibitem[{\citenamefont{Thomson}(1856)}]{Thomson:jan1856}
\bibinfo{author}{\bibfnamefont{W.}~\bibnamefont{Thomson}},
  \bibinfo{journal}{Proc. R. Soc. London} \textbf{\bibinfo{volume}{8}},
  \bibinfo{pages}{546} (\bibinfo{year}{1856}).

\bibitem[{\citenamefont{{Karplus} and {Luttinger}}(1954)}]{Karplus:sep1954}
\bibinfo{author}{\bibfnamefont{R.}~\bibnamefont{{Karplus}}} \bibnamefont{and}
  \bibinfo{author}{\bibfnamefont{J.~M.} \bibnamefont{{Luttinger}}},
  \bibinfo{journal}{Phys. Rev.} \textbf{\bibinfo{volume}{95}},
  \bibinfo{pages}{1154} (\bibinfo{year}{1954}).

\bibitem[{\citenamefont{{Taguchi} et~al.}(2001)\citenamefont{{Taguchi},
  {Oohara}, {Yoshizawa}, {Nagaosa}, and {Tokura}}}]{Taguchi:mar2001}
\bibinfo{author}{\bibfnamefont{Y.}~\bibnamefont{{Taguchi}}},
  \bibinfo{author}{\bibfnamefont{Y.}~\bibnamefont{{Oohara}}},
  \bibinfo{author}{\bibfnamefont{H.}~\bibnamefont{{Yoshizawa}}},
  \bibinfo{author}{\bibfnamefont{N.}~\bibnamefont{{Nagaosa}}},
  \bibnamefont{and} \bibinfo{author}{\bibfnamefont{Y.}~\bibnamefont{{Tokura}}},
  \bibinfo{journal}{Science} \textbf{\bibinfo{volume}{291}},
  \bibinfo{pages}{2573} (\bibinfo{year}{2001}).

\bibitem[{\citenamefont{{Jungwirth} et~al.}(2002)\citenamefont{{Jungwirth},
  {Niu}, and {MacDonald}}}]{Jungwirth:may2002}
\bibinfo{author}{\bibfnamefont{T.}~\bibnamefont{{Jungwirth}}},
  \bibinfo{author}{\bibfnamefont{Q.}~\bibnamefont{{Niu}}}, \bibnamefont{and}
  \bibinfo{author}{\bibfnamefont{A.~H.} \bibnamefont{{MacDonald}}},
  \bibinfo{journal}{Phys. Rev. Lett.} \textbf{\bibinfo{volume}{88}},
  \bibinfo{pages}{207208} (\bibinfo{year}{2002}).

\bibitem[{\citenamefont{{Onoda} and {Nagaosa}}(2002)}]{Onoda:jan2002}
\bibinfo{author}{\bibfnamefont{M.}~\bibnamefont{{Onoda}}} \bibnamefont{and}
  \bibinfo{author}{\bibfnamefont{N.}~\bibnamefont{{Nagaosa}}},
  \bibinfo{journal}{J. Phys. Soc. Jpn.} \textbf{\bibinfo{volume}{71}},
  \bibinfo{pages}{19} (\bibinfo{year}{2002}).

\bibitem[{\citenamefont{{Yao} et~al.}(2004)\citenamefont{{Yao}, {Kleinman},
  {MacDonald}, {Sinova}, {Jungwirth}, {Wang}, {Wang}, and {Niu}}}]{Yao:jan2004}
\bibinfo{author}{\bibfnamefont{Y.}~\bibnamefont{{Yao}}},
  \bibinfo{author}{\bibfnamefont{L.}~\bibnamefont{{Kleinman}}},
  \bibinfo{author}{\bibfnamefont{A.~H.} \bibnamefont{{MacDonald}}},
  \bibinfo{author}{\bibfnamefont{J.}~\bibnamefont{{Sinova}}},
  \bibinfo{author}{\bibfnamefont{T.}~\bibnamefont{{Jungwirth}}},
  \bibinfo{author}{\bibfnamefont{D.-S.} \bibnamefont{{Wang}}},
  \bibinfo{author}{\bibfnamefont{E.}~\bibnamefont{{Wang}}}, \bibnamefont{and}
  \bibinfo{author}{\bibfnamefont{Q.}~\bibnamefont{{Niu}}},
  \bibinfo{journal}{Phys. Rev. Lett.} \textbf{\bibinfo{volume}{92}},
  \bibinfo{pages}{037204} (\bibinfo{year}{2004}).

\bibitem[{\citenamefont{{Lee} et~al.}(2004)\citenamefont{{Lee}, {Watauchi},
  {Miller}, {Cava}, and {Ong}}}]{Lee:mar2004}
\bibinfo{author}{\bibfnamefont{W.-L.} \bibnamefont{{Lee}}},
  \bibinfo{author}{\bibfnamefont{S.}~\bibnamefont{{Watauchi}}},
  \bibinfo{author}{\bibfnamefont{V.~L.} \bibnamefont{{Miller}}},
  \bibinfo{author}{\bibfnamefont{R.~J.} \bibnamefont{{Cava}}},
  \bibnamefont{and} \bibinfo{author}{\bibfnamefont{N.~P.} \bibnamefont{{Ong}}},
  \bibinfo{journal}{Science} \textbf{\bibinfo{volume}{303}},
  \bibinfo{pages}{1647} (\bibinfo{year}{2004}).

\bibitem[{\citenamefont{{Zeng} et~al.}(2006)\citenamefont{{Zeng}, {Yao}, {Niu},
  and {Weitering}}}]{Zeng:jan2006}
\bibinfo{author}{\bibfnamefont{C.}~\bibnamefont{{Zeng}}},
  \bibinfo{author}{\bibfnamefont{Y.}~\bibnamefont{{Yao}}},
  \bibinfo{author}{\bibfnamefont{Q.}~\bibnamefont{{Niu}}}, \bibnamefont{and}
  \bibinfo{author}{\bibfnamefont{H.~H.} \bibnamefont{{Weitering}}},
  \bibinfo{journal}{Phys. Rev. Lett.} \textbf{\bibinfo{volume}{96}},
  \bibinfo{pages}{037204} (\bibinfo{year}{2006}).

\bibitem[{\citenamefont{{Rashba}}(2008)}]{Rashba:aug2008}
\bibinfo{author}{\bibfnamefont{E.~I.} \bibnamefont{{Rashba}}},
  \bibinfo{journal}{Semiconductors} \textbf{\bibinfo{volume}{42}},
  \bibinfo{pages}{905} (\bibinfo{year}{2008}), \eprint{0804.4181}.

\bibitem[{\citenamefont{{Sundaram} and {Niu}}(1999)}]{Sundaram:jun1999}
\bibinfo{author}{\bibfnamefont{G.}~\bibnamefont{{Sundaram}}} \bibnamefont{and}
  \bibinfo{author}{\bibfnamefont{Q.}~\bibnamefont{{Niu}}},
  \bibinfo{journal}{Phys. Rev. B} \textbf{\bibinfo{volume}{59}},
  \bibinfo{pages}{14915} (\bibinfo{year}{1999}).

\bibitem[{\citenamefont{{Sinova} et~al.}(2004)\citenamefont{{Sinova},
  {Jungwirth}, and {{\v C}erne}}}]{Sinova:2004}
\bibinfo{author}{\bibfnamefont{J.}~\bibnamefont{{Sinova}}},
  \bibinfo{author}{\bibfnamefont{T.}~\bibnamefont{{Jungwirth}}},
  \bibnamefont{and} \bibinfo{author}{\bibfnamefont{J.}~\bibnamefont{{{\v
  C}erne}}}, \bibinfo{journal}{Int. J. Mod. Phys. B}
  \textbf{\bibinfo{volume}{18}}, \bibinfo{pages}{1083} (\bibinfo{year}{2004}).

\bibitem[{\citenamefont{{Culcer} et~al.}(2003)\citenamefont{{Culcer},
  {MacDonald}, and {Niu}}}]{Culcer:jul2003}
\bibinfo{author}{\bibfnamefont{D.}~\bibnamefont{{Culcer}}},
  \bibinfo{author}{\bibfnamefont{A.}~\bibnamefont{{MacDonald}}},
  \bibnamefont{and} \bibinfo{author}{\bibfnamefont{Q.}~\bibnamefont{{Niu}}},
  \bibinfo{journal}{Phys. Rev. B} \textbf{\bibinfo{volume}{68}},
  \bibinfo{pages}{045327} (\bibinfo{year}{2003}).

\bibitem[{\citenamefont{{Dugaev} et~al.}(2005)\citenamefont{{Dugaev}, {Bruno},
  {Taillefumier}, {Canals}, and {Lacroix}}}]{Dugaev:jun2005}
\bibinfo{author}{\bibfnamefont{V.~K.} \bibnamefont{{Dugaev}}},
  \bibinfo{author}{\bibfnamefont{P.}~\bibnamefont{{Bruno}}},
  \bibinfo{author}{\bibfnamefont{M.}~\bibnamefont{{Taillefumier}}},
  \bibinfo{author}{\bibfnamefont{B.}~\bibnamefont{{Canals}}}, \bibnamefont{and}
  \bibinfo{author}{\bibfnamefont{C.}~\bibnamefont{{Lacroix}}},
  \bibinfo{journal}{Phys. Rev. B} \textbf{\bibinfo{volume}{71}},
  \bibinfo{pages}{224423} (\bibinfo{year}{2005}).

\bibitem[{\citenamefont{{Sinitsyn} et~al.}(2005)\citenamefont{{Sinitsyn},
  {Niu}, {Sinova}, and {Nomura}}}]{Sinitsyn:jul2005}
\bibinfo{author}{\bibfnamefont{N.~A.} \bibnamefont{{Sinitsyn}}},
  \bibinfo{author}{\bibfnamefont{Q.}~\bibnamefont{{Niu}}},
  \bibinfo{author}{\bibfnamefont{J.}~\bibnamefont{{Sinova}}}, \bibnamefont{and}
  \bibinfo{author}{\bibfnamefont{K.}~\bibnamefont{{Nomura}}},
  \bibinfo{journal}{Phys. Rev. B} \textbf{\bibinfo{volume}{72}},
  \bibinfo{pages}{045346} (\bibinfo{year}{2005}).

\bibitem[{\citenamefont{{Liu} et~al.}(2006)\citenamefont{{Liu}, {Horing}, and
  {Lei}}}]{Liu:oct2006}
\bibinfo{author}{\bibfnamefont{S.~Y.} \bibnamefont{{Liu}}},
  \bibinfo{author}{\bibfnamefont{N.~J.~M.} \bibnamefont{{Horing}}},
  \bibnamefont{and} \bibinfo{author}{\bibfnamefont{X.~L.} \bibnamefont{{Lei}}},
  \bibinfo{journal}{Phys. Rev. B} \textbf{\bibinfo{volume}{74}},
  \bibinfo{pages}{165316} (\bibinfo{year}{2006}).

\bibitem[{\citenamefont{{Inoue} et~al.}(2006)\citenamefont{{Inoue}, {Kato},
  {Ishikawa}, {Itoh}, {Bauer}, and {Molenkamp}}}]{Inoue:jul2006}
\bibinfo{author}{\bibfnamefont{J.-I.} \bibnamefont{{Inoue}}},
  \bibinfo{author}{\bibfnamefont{T.}~\bibnamefont{{Kato}}},
  \bibinfo{author}{\bibfnamefont{Y.}~\bibnamefont{{Ishikawa}}},
  \bibinfo{author}{\bibfnamefont{H.}~\bibnamefont{{Itoh}}},
  \bibinfo{author}{\bibfnamefont{G.~E.~W.} \bibnamefont{{Bauer}}},
  \bibnamefont{and} \bibinfo{author}{\bibfnamefont{L.~W.}
  \bibnamefont{{Molenkamp}}}, \bibinfo{journal}{Phys. Rev. Lett.}
  \textbf{\bibinfo{volume}{97}}, \bibinfo{pages}{046604}
  (\bibinfo{year}{2006}).

\bibitem[{\citenamefont{Onoda et~al.}(2006)\citenamefont{Onoda, Sugimoto, and
  Nagaosa}}]{Onoda:sep2006}
\bibinfo{author}{\bibfnamefont{S.}~\bibnamefont{Onoda}},
  \bibinfo{author}{\bibfnamefont{N.}~\bibnamefont{Sugimoto}}, \bibnamefont{and}
  \bibinfo{author}{\bibfnamefont{N.}~\bibnamefont{Nagaosa}},
  \bibinfo{journal}{Phys. Rev. Lett.} \textbf{\bibinfo{volume}{97}},
  \bibinfo{pages}{126602} (\bibinfo{year}{2006}).

\bibitem[{\citenamefont{{Borunda} et~al.}(2007)\citenamefont{{Borunda},
  {Nunner}, {L{\"u}ck}, {Sinitsyn}, {Timm}, {Wunderlich}, {Jungwirth},
  {MacDonald}, and {Sinova}}}]{Borunda:aug2007}
\bibinfo{author}{\bibfnamefont{M.}~\bibnamefont{{Borunda}}},
  \bibinfo{author}{\bibfnamefont{T.~S.} \bibnamefont{{Nunner}}},
  \bibinfo{author}{\bibfnamefont{T.}~\bibnamefont{{L{\"u}ck}}},
  \bibinfo{author}{\bibfnamefont{N.~A.} \bibnamefont{{Sinitsyn}}},
  \bibinfo{author}{\bibfnamefont{C.}~\bibnamefont{{Timm}}},
  \bibinfo{author}{\bibfnamefont{J.}~\bibnamefont{{Wunderlich}}},
  \bibinfo{author}{\bibfnamefont{T.}~\bibnamefont{{Jungwirth}}},
  \bibinfo{author}{\bibfnamefont{A.~H.} \bibnamefont{{MacDonald}}},
  \bibnamefont{and} \bibinfo{author}{\bibfnamefont{J.}~\bibnamefont{{Sinova}}},
  \bibinfo{journal}{Phys. Rev. Lett.} \textbf{\bibinfo{volume}{99}},
  \bibinfo{pages}{066604} (\bibinfo{year}{2007}).

\bibitem[{\citenamefont{{Nunner} et~al.}(2007)\citenamefont{{Nunner},
  {Sinitsyn}, {Borunda}, {Dugaev}, {Kovalev}, {Abanov}, {Timm}, {Jungwirth},
  {Inoue}, {MacDonald} et~al.}}]{Nunner:dec2007}
\bibinfo{author}{\bibfnamefont{T.~S.} \bibnamefont{{Nunner}}},
  \bibinfo{author}{\bibfnamefont{N.~A.} \bibnamefont{{Sinitsyn}}},
  \bibinfo{author}{\bibfnamefont{M.~F.} \bibnamefont{{Borunda}}},
  \bibinfo{author}{\bibfnamefont{V.~K.} \bibnamefont{{Dugaev}}},
  \bibinfo{author}{\bibfnamefont{A.~A.} \bibnamefont{{Kovalev}}},
  \bibinfo{author}{\bibfnamefont{A.}~\bibnamefont{{Abanov}}},
  \bibinfo{author}{\bibfnamefont{C.}~\bibnamefont{{Timm}}},
  \bibinfo{author}{\bibfnamefont{T.}~\bibnamefont{{Jungwirth}}},
  \bibinfo{author}{\bibfnamefont{J.-I.} \bibnamefont{{Inoue}}},
  \bibinfo{author}{\bibfnamefont{A.~H.} \bibnamefont{{MacDonald}}},
  \bibnamefont{et~al.}, \bibinfo{journal}{Phys. Rev. B}
  \textbf{\bibinfo{volume}{76}}, \bibinfo{pages}{235312}
  (\bibinfo{year}{2007}).

\bibitem[{\citenamefont{{Kato} et~al.}(2007)\citenamefont{{Kato}, {Ishikawa},
  {Itoh}, and {Inoue}}}]{Kato:sep2007}
\bibinfo{author}{\bibfnamefont{T.}~\bibnamefont{{Kato}}},
  \bibinfo{author}{\bibfnamefont{Y.}~\bibnamefont{{Ishikawa}}},
  \bibinfo{author}{\bibfnamefont{H.}~\bibnamefont{{Itoh}}}, \bibnamefont{and}
  \bibinfo{author}{\bibfnamefont{J.-i.} \bibnamefont{{Inoue}}},
  \bibinfo{journal}{New J. Phys.} \textbf{\bibinfo{volume}{9}},
  \bibinfo{pages}{350} (\bibinfo{year}{2007}).

\bibitem[{\citenamefont{{Kovalev} et~al.}(2008)\citenamefont{{Kovalev},
  {V{\'y}born{\'y}}, and {Sinova}}}]{Kovalev:jul2008}
\bibinfo{author}{\bibfnamefont{A.~A.} \bibnamefont{{Kovalev}}},
  \bibinfo{author}{\bibfnamefont{K.}~\bibnamefont{{V{\'y}born{\'y}}}},
  \bibnamefont{and} \bibinfo{author}{\bibfnamefont{J.}~\bibnamefont{{Sinova}}},
  \bibinfo{journal}{Phys. Rev. B} \textbf{\bibinfo{volume}{78}},
  \bibinfo{pages}{041305} (\bibinfo{year}{2008}).

\bibitem[{\citenamefont{{Smit}}(1955)}]{Smit:1955}
\bibinfo{author}{\bibfnamefont{J.}~\bibnamefont{{Smit}}},
  \bibinfo{journal}{Physica} \textbf{\bibinfo{volume}{21}},
  \bibinfo{pages}{877} (\bibinfo{year}{1955}).

\bibitem[{\citenamefont{Nozieres and Lewiner}(1973)}]{Nozieres:1973}
\bibinfo{author}{\bibfnamefont{P.}~\bibnamefont{Nozieres}} \bibnamefont{and}
  \bibinfo{author}{\bibfnamefont{C.}~\bibnamefont{Lewiner}},
  \bibinfo{journal}{Journal De Physique} \textbf{\bibinfo{volume}{34}},
  \bibinfo{pages}{901} (\bibinfo{year}{1973}).

\bibitem[{\citenamefont{{Sinitsyn} et~al.}(2006)\citenamefont{{Sinitsyn},
  {Niu}, and {MacDonald}}}]{Sinitsyn:feb2006}
\bibinfo{author}{\bibfnamefont{N.~A.} \bibnamefont{{Sinitsyn}}},
  \bibinfo{author}{\bibfnamefont{Q.}~\bibnamefont{{Niu}}}, \bibnamefont{and}
  \bibinfo{author}{\bibfnamefont{A.~H.} \bibnamefont{{MacDonald}}},
  \bibinfo{journal}{Phys. Rev. B} \textbf{\bibinfo{volume}{73}},
  \bibinfo{pages}{075318} (\bibinfo{year}{2006}).

\bibitem[{\citenamefont{{Smit}}(1951)}]{Smit:jun1951}
\bibinfo{author}{\bibfnamefont{J.}~\bibnamefont{{Smit}}},
  \bibinfo{journal}{Physica} \textbf{\bibinfo{volume}{17}},
  \bibinfo{pages}{612} (\bibinfo{year}{1951}).

\bibitem[{\citenamefont{{Berger}}(1964)}]{Berger:jun1964}
\bibinfo{author}{\bibfnamefont{L.}~\bibnamefont{{Berger}}},
  \bibinfo{journal}{Physica} \textbf{\bibinfo{volume}{30}},
  \bibinfo{pages}{1141} (\bibinfo{year}{1964}).

\bibitem[{\citenamefont{{McGuire} and {Potter}}(1975)}]{McGuire:jul1975}
\bibinfo{author}{\bibfnamefont{T.}~\bibnamefont{{McGuire}}} \bibnamefont{and}
  \bibinfo{author}{\bibfnamefont{R.}~\bibnamefont{{Potter}}},
  \bibinfo{journal}{IEEE Trans. Magn.} \textbf{\bibinfo{volume}{11}},
  \bibinfo{pages}{1018} (\bibinfo{year}{1975}).

\bibitem[{\citenamefont{{Jaoul} et~al.}(1977)\citenamefont{{Jaoul}, {Campbell},
  and {Fert}}}]{Jaoul:mar1977}
\bibinfo{author}{\bibfnamefont{O.}~\bibnamefont{{Jaoul}}},
  \bibinfo{author}{\bibfnamefont{I.~A.} \bibnamefont{{Campbell}}},
  \bibnamefont{and} \bibinfo{author}{\bibfnamefont{A.}~\bibnamefont{{Fert}}},
  \bibinfo{journal}{J. Magn. Magn. Mater.} \textbf{\bibinfo{volume}{5}},
  \bibinfo{pages}{23} (\bibinfo{year}{1977}).

\bibitem[{\citenamefont{{Banhart} and {Ebert}}(1995)}]{Banhart:nov1995}
\bibinfo{author}{\bibfnamefont{J.}~\bibnamefont{{Banhart}}} \bibnamefont{and}
  \bibinfo{author}{\bibfnamefont{H.}~\bibnamefont{{Ebert}}},
  \bibinfo{journal}{Europhys. Lett.} \textbf{\bibinfo{volume}{32}},
  \bibinfo{pages}{517} (\bibinfo{year}{1995}).

\bibitem[{\citenamefont{{Velev} et~al.}(2005)\citenamefont{{Velev},
  {Sabirianov}, {Jaswal}, and {Tsymbal}}}]{Velev:mar2005}
\bibinfo{author}{\bibfnamefont{J.}~\bibnamefont{{Velev}}},
  \bibinfo{author}{\bibfnamefont{R.~F.} \bibnamefont{{Sabirianov}}},
  \bibinfo{author}{\bibfnamefont{S.~S.} \bibnamefont{{Jaswal}}},
  \bibnamefont{and} \bibinfo{author}{\bibfnamefont{E.~Y.}
  \bibnamefont{{Tsymbal}}}, \bibinfo{journal}{Phys. Rev. Lett.}
  \textbf{\bibinfo{volume}{94}}, \bibinfo{pages}{127203}
  (\bibinfo{year}{2005}).

\bibitem[{\citenamefont{{Rushforth} et~al.}(2007)\citenamefont{{Rushforth},
  {V{\'y}born{\'y}}, {King}, {Edmonds}, {Campion}, {Foxon}, {Wunderlich},
  {Irvine}, {Va{\v s}ek}, {Nov{\'a}k} et~al.}}]{Rushforth:oct2007}
\bibinfo{author}{\bibfnamefont{A.~W.} \bibnamefont{{Rushforth}}},
  \bibinfo{author}{\bibfnamefont{K.}~\bibnamefont{{V{\'y}born{\'y}}}},
  \bibinfo{author}{\bibfnamefont{C.~S.} \bibnamefont{{King}}},
  \bibinfo{author}{\bibfnamefont{K.~W.} \bibnamefont{{Edmonds}}},
  \bibinfo{author}{\bibfnamefont{R.~P.} \bibnamefont{{Campion}}},
  \bibinfo{author}{\bibfnamefont{C.~T.} \bibnamefont{{Foxon}}},
  \bibinfo{author}{\bibfnamefont{J.}~\bibnamefont{{Wunderlich}}},
  \bibinfo{author}{\bibfnamefont{A.~C.} \bibnamefont{{Irvine}}},
  \bibinfo{author}{\bibfnamefont{P.}~\bibnamefont{{Va{\v s}ek}}},
  \bibinfo{author}{\bibfnamefont{V.}~\bibnamefont{{Nov{\'a}k}}},
  \bibnamefont{et~al.}, \bibinfo{journal}{Phys. Rev. Lett.}
  \textbf{\bibinfo{volume}{99}}, \bibinfo{pages}{147207}
  (\bibinfo{year}{2007}).

\bibitem[{\citenamefont{{Vyborny} et~al.}(2009)\citenamefont{{Vyborny},
  {Kovalev}, {Sinova}, and {Jungwirth}}}]{Vyborny:oct2008}
\bibinfo{author}{\bibfnamefont{K.}~\bibnamefont{{Vyborny}}},
  \bibinfo{author}{\bibfnamefont{A.~A.} \bibnamefont{{Kovalev}}},
  \bibinfo{author}{\bibfnamefont{J.}~\bibnamefont{{Sinova}}}, \bibnamefont{and}
  \bibinfo{author}{\bibfnamefont{T.}~\bibnamefont{{Jungwirth}}},
  \bibinfo{journal}{Phys. Rev. B} \textbf{\bibinfo{volume}{79}},
  \bibinfo{pages}{045427} (\bibinfo{year}{2009}).

\bibitem[{\citenamefont{{Kato} et~al.}(2008)\citenamefont{{Kato}, {Ishikawa},
  {Itoh}, and {Inoue}}}]{Kato:jun2008}
\bibinfo{author}{\bibfnamefont{T.}~\bibnamefont{{Kato}}},
  \bibinfo{author}{\bibfnamefont{Y.}~\bibnamefont{{Ishikawa}}},
  \bibinfo{author}{\bibfnamefont{H.}~\bibnamefont{{Itoh}}}, \bibnamefont{and}
  \bibinfo{author}{\bibfnamefont{J.-I.} \bibnamefont{{Inoue}}},
  \bibinfo{journal}{Phys. Rev. B} \textbf{\bibinfo{volume}{77}},
  \bibinfo{pages}{233404} (\bibinfo{year}{2008}).

\bibitem[{\citenamefont{Ueno et~al.}(2007)\citenamefont{Ueno, Fukumura,
  Toyosaki, Nakano, and Kawasaki}}]{Ueno:2007}
\bibinfo{author}{\bibfnamefont{K.}~\bibnamefont{Ueno}},
  \bibinfo{author}{\bibfnamefont{T.}~\bibnamefont{Fukumura}},
  \bibinfo{author}{\bibfnamefont{H.}~\bibnamefont{Toyosaki}},
  \bibinfo{author}{\bibfnamefont{M.}~\bibnamefont{Nakano}}, \bibnamefont{and}
  \bibinfo{author}{\bibfnamefont{M.}~\bibnamefont{Kawasaki}},
  \bibinfo{journal}{Appl. Phys. Lett.} \textbf{\bibinfo{volume}{90}},
  \bibinfo{eid}{072103} (pages~\bibinfo{numpages}{3}) (\bibinfo{year}{2007}).

\bibitem[{\citenamefont{{Miyasato} et~al.}(2007)\citenamefont{{Miyasato},
  {Abe}, {Fujii}, {Asamitsu}, {Onoda}, {Onose}, {Nagaosa}, and
  {Tokura}}}]{Miyasato:aug2007}
\bibinfo{author}{\bibfnamefont{T.}~\bibnamefont{{Miyasato}}},
  \bibinfo{author}{\bibfnamefont{N.}~\bibnamefont{{Abe}}},
  \bibinfo{author}{\bibfnamefont{T.}~\bibnamefont{{Fujii}}},
  \bibinfo{author}{\bibfnamefont{A.}~\bibnamefont{{Asamitsu}}},
  \bibinfo{author}{\bibfnamefont{S.}~\bibnamefont{{Onoda}}},
  \bibinfo{author}{\bibfnamefont{Y.}~\bibnamefont{{Onose}}},
  \bibinfo{author}{\bibfnamefont{N.}~\bibnamefont{{Nagaosa}}},
  \bibnamefont{and} \bibinfo{author}{\bibfnamefont{Y.}~\bibnamefont{{Tokura}}},
  \bibinfo{journal}{Phys. Rev. Lett.} \textbf{\bibinfo{volume}{99}},
  \bibinfo{pages}{086602} (\bibinfo{year}{2007}).

\bibitem[{\citenamefont{Fukumura et~al.}(2007)\citenamefont{Fukumura, Toyosaki,
  Ueno, Nakano, Yamasaki, and Kawasaki}}]{Fukumura:2007}
\bibinfo{author}{\bibfnamefont{T.}~\bibnamefont{Fukumura}},
  \bibinfo{author}{\bibfnamefont{H.}~\bibnamefont{Toyosaki}},
  \bibinfo{author}{\bibfnamefont{K.}~\bibnamefont{Ueno}},
  \bibinfo{author}{\bibfnamefont{M.}~\bibnamefont{Nakano}},
  \bibinfo{author}{\bibfnamefont{T.}~\bibnamefont{Yamasaki}}, \bibnamefont{and}
  \bibinfo{author}{\bibfnamefont{M.}~\bibnamefont{Kawasaki}},
  \bibinfo{journal}{Jpn. J. Appl. Phys.} \textbf{\bibinfo{volume}{46}},
  \bibinfo{pages}{L642} (\bibinfo{year}{2007}).

\bibitem[{\citenamefont{Venkateshvaran
  et~al.}(2008)\citenamefont{Venkateshvaran, Kaiser, Boger, Althammer, Rao,
  Goennenwein, Opel, and Gross}}]{Venkateshvaran:2008}
\bibinfo{author}{\bibfnamefont{D.}~\bibnamefont{Venkateshvaran}},
  \bibinfo{author}{\bibfnamefont{W.}~\bibnamefont{Kaiser}},
  \bibinfo{author}{\bibfnamefont{A.}~\bibnamefont{Boger}},
  \bibinfo{author}{\bibfnamefont{M.}~\bibnamefont{Althammer}},
  \bibinfo{author}{\bibfnamefont{M.~S.~R.} \bibnamefont{Rao}},
  \bibinfo{author}{\bibfnamefont{S.~T.~B.} \bibnamefont{Goennenwein}},
  \bibinfo{author}{\bibfnamefont{M.}~\bibnamefont{Opel}}, \bibnamefont{and}
  \bibinfo{author}{\bibfnamefont{R.}~\bibnamefont{Gross}},
  \bibinfo{journal}{Phys. Rev. B} \textbf{\bibinfo{volume}{78}},
  \bibinfo{eid}{092405} (\bibinfo{year}{2008}).

\bibitem[{\citenamefont{Fernandez-Pacheco
  et~al.}(2008)\citenamefont{Fernandez-Pacheco, Teresa, Orna, Morellon,
  Algarabel, Pardo, and Ibarra}}]{Fernandez-Pacheco:2008}
\bibinfo{author}{\bibfnamefont{A.}~\bibnamefont{Fernandez-Pacheco}},
  \bibinfo{author}{\bibfnamefont{J.~M.~D.} \bibnamefont{Teresa}},
  \bibinfo{author}{\bibfnamefont{J.}~\bibnamefont{Orna}},
  \bibinfo{author}{\bibfnamefont{L.}~\bibnamefont{Morellon}},
  \bibinfo{author}{\bibfnamefont{P.~A.} \bibnamefont{Algarabel}},
  \bibinfo{author}{\bibfnamefont{J.~A.} \bibnamefont{Pardo}}, \bibnamefont{and}
  \bibinfo{author}{\bibfnamefont{M.~R.} \bibnamefont{Ibarra}},
  \bibinfo{journal}{Phys. Rev. B} \textbf{\bibinfo{volume}{77}},
  \bibinfo{eid}{100403} (\bibinfo{year}{2008}).

\bibitem[{\citenamefont{{Rammer} and {Smith}}(1986)}]{Rammer:apr1986}
\bibinfo{author}{\bibfnamefont{J.}~\bibnamefont{{Rammer}}} \bibnamefont{and}
  \bibinfo{author}{\bibfnamefont{H.}~\bibnamefont{{Smith}}},
  \bibinfo{journal}{Rev. Mod. Phys.} \textbf{\bibinfo{volume}{58}},
  \bibinfo{pages}{323} (\bibinfo{year}{1986}).

\bibitem[{\citenamefont{Mahan}(1990)}]{Mahan:1990}
\bibinfo{author}{\bibfnamefont{G.~D.} \bibnamefont{Mahan}},
  \emph{\bibinfo{title}{Many-Particle Physics}} (\bibinfo{publisher}{Plenum
  Press, New York}, \bibinfo{year}{1990}).

\bibitem[{\citenamefont{{Streda}}(1982)}]{Streda:aug1982}
\bibinfo{author}{\bibfnamefont{P.}~\bibnamefont{{Streda}}},
  \bibinfo{journal}{J. Phys. C} \textbf{\bibinfo{volume}{15}},
  \bibinfo{pages}{L717} (\bibinfo{year}{1982}).

\bibitem[{\citenamefont{{Dugaev} et~al.}(2001)\citenamefont{{Dugaev},
  {Cr{\'e}pieux}, and {Bruno}}}]{Dugaev:sep2001}
\bibinfo{author}{\bibfnamefont{V.~K.} \bibnamefont{{Dugaev}}},
  \bibinfo{author}{\bibfnamefont{A.}~\bibnamefont{{Cr{\'e}pieux}}},
  \bibnamefont{and} \bibinfo{author}{\bibfnamefont{P.}~\bibnamefont{{Bruno}}},
  \bibinfo{journal}{Phys. Rev. B} \textbf{\bibinfo{volume}{64}},
  \bibinfo{pages}{104411} (\bibinfo{year}{2001}).

\bibitem[{\citenamefont{Sinitsyn et~al.}(2007)\citenamefont{Sinitsyn,
  MacDonald, Jungwirth, Dugaev, and Sinova}}]{Sinitsyn:jan2007}
\bibinfo{author}{\bibfnamefont{N.~A.} \bibnamefont{Sinitsyn}},
  \bibinfo{author}{\bibfnamefont{A.~H.} \bibnamefont{MacDonald}},
  \bibinfo{author}{\bibfnamefont{T.}~\bibnamefont{Jungwirth}},
  \bibinfo{author}{\bibfnamefont{V.~K.} \bibnamefont{Dugaev}},
  \bibnamefont{and} \bibinfo{author}{\bibfnamefont{J.}~\bibnamefont{Sinova}},
  \bibinfo{journal}{Phys. Rev. B} \textbf{\bibinfo{volume}{75}},
  \bibinfo{pages}{045315} (\bibinfo{year}{2007}).

\bibitem[{\citenamefont{Pikus and Titkov}(1984)}]{Pikus:1984}
\bibinfo{author}{\bibfnamefont{G.~E.} \bibnamefont{Pikus}} \bibnamefont{and}
  \bibinfo{author}{\bibfnamefont{A.~N.} \bibnamefont{Titkov}},
  \emph{\bibinfo{title}{in Optical Orientation}}
  (\bibinfo{publisher}{North-Holland, Amsterdam}, \bibinfo{year}{1984}).

\bibitem[{\citenamefont{Rashba and Sheka}(1991)}]{Rashba1991}
\bibinfo{author}{\bibfnamefont{E.}~\bibnamefont{Rashba}} \bibnamefont{and}
  \bibinfo{author}{\bibfnamefont{V.}~\bibnamefont{Sheka}},
  \emph{\bibinfo{title}{in Landau Level Spectroscopy}}
  (\bibinfo{publisher}{North-Holland, Amsterdam}, \bibinfo{year}{1991}),
  \bibinfo{note}{p.167}.

\bibitem[{\citenamefont{Zhang et~al.}(2001)\citenamefont{Zhang,
  Pfeuffer-Jeschke, Ortner, Hock, Buhmann, Becker, and
  Landwehr}}]{Zhang:jun2001}
\bibinfo{author}{\bibfnamefont{X.~C.} \bibnamefont{Zhang}},
  \bibinfo{author}{\bibfnamefont{A.}~\bibnamefont{Pfeuffer-Jeschke}},
  \bibinfo{author}{\bibfnamefont{K.}~\bibnamefont{Ortner}},
  \bibinfo{author}{\bibfnamefont{V.}~\bibnamefont{Hock}},
  \bibinfo{author}{\bibfnamefont{H.}~\bibnamefont{Buhmann}},
  \bibinfo{author}{\bibfnamefont{C.~R.} \bibnamefont{Becker}},
  \bibnamefont{and} \bibinfo{author}{\bibfnamefont{G.}~\bibnamefont{Landwehr}},
  \bibinfo{journal}{Phys. Rev. B} \textbf{\bibinfo{volume}{6324}},
  \bibinfo{pages}{245305} (\bibinfo{year}{2001}).

\bibitem[{\citenamefont{{Engel} et~al.}(2005)\citenamefont{{Engel}, {Halperin},
  and {Rashba}}}]{Engel:oct2005}
\bibinfo{author}{\bibfnamefont{H.-A.} \bibnamefont{{Engel}}},
  \bibinfo{author}{\bibfnamefont{B.~I.} \bibnamefont{{Halperin}}},
  \bibnamefont{and} \bibinfo{author}{\bibfnamefont{E.~I.}
  \bibnamefont{{Rashba}}}, \bibinfo{journal}{Phys. Rev. Lett.}
  \textbf{\bibinfo{volume}{95}}, \bibinfo{pages}{166605}
  (\bibinfo{year}{2005}).

\bibitem[{\citenamefont{{Thouless} et~al.}(1982)\citenamefont{{Thouless},
  {Kohmoto}, {Nightingale}, and {den Nijs}}}]{Thouless:aug1982}
\bibinfo{author}{\bibfnamefont{D.~J.} \bibnamefont{{Thouless}}},
  \bibinfo{author}{\bibfnamefont{M.}~\bibnamefont{{Kohmoto}}},
  \bibinfo{author}{\bibfnamefont{M.~P.} \bibnamefont{{Nightingale}}},
  \bibnamefont{and} \bibinfo{author}{\bibfnamefont{M.}~\bibnamefont{{den
  Nijs}}}, \bibinfo{journal}{Phys. Rev. Lett.} \textbf{\bibinfo{volume}{49}},
  \bibinfo{pages}{405} (\bibinfo{year}{1982}).

\bibitem[{\citenamefont{{Mih{\'a}ly} et~al.}(2008)\citenamefont{{Mih{\'a}ly},
  {Csontos}, {Bord{\'a}cs}, {K{\'e}zsm{\'a}rki}, {Wojtowicz}, {Liu},
  {Jank{\'o}}, and {Furdyna}}}]{Mih'aly:mar2008}
\bibinfo{author}{\bibfnamefont{G.}~\bibnamefont{{Mih{\'a}ly}}},
  \bibinfo{author}{\bibfnamefont{M.}~\bibnamefont{{Csontos}}},
  \bibinfo{author}{\bibfnamefont{S.}~\bibnamefont{{Bord{\'a}cs}}},
  \bibinfo{author}{\bibfnamefont{I.}~\bibnamefont{{K{\'e}zsm{\'a}rki}}},
  \bibinfo{author}{\bibfnamefont{T.}~\bibnamefont{{Wojtowicz}}},
  \bibinfo{author}{\bibfnamefont{X.}~\bibnamefont{{Liu}}},
  \bibinfo{author}{\bibfnamefont{B.}~\bibnamefont{{Jank{\'o}}}},
  \bibnamefont{and} \bibinfo{author}{\bibfnamefont{J.~K.}
  \bibnamefont{{Furdyna}}}, \bibinfo{journal}{Phys. Rev. Lett.}
  \textbf{\bibinfo{volume}{100}}, \bibinfo{pages}{107201}
  (\bibinfo{year}{2008}).

\bibitem[{\citenamefont{{Schliemann} and {Loss}}(2003)}]{Schliemann:oct2003}
\bibinfo{author}{\bibfnamefont{J.}~\bibnamefont{{Schliemann}}}
  \bibnamefont{and} \bibinfo{author}{\bibfnamefont{D.}~\bibnamefont{{Loss}}},
  \bibinfo{journal}{Phys. Rev. B} \textbf{\bibinfo{volume}{68}},
  \bibinfo{pages}{165311} (\bibinfo{year}{2003}).

\bibitem[{\citenamefont{{Nunner} et~al.}(2008)\citenamefont{{Nunner},
  {Zar{\'a}nd}, and {von Oppen}}}]{Nunner:jun2008}
\bibinfo{author}{\bibfnamefont{T.~S.} \bibnamefont{{Nunner}}},
  \bibinfo{author}{\bibfnamefont{G.}~\bibnamefont{{Zar{\'a}nd}}},
  \bibnamefont{and} \bibinfo{author}{\bibfnamefont{F.}~\bibnamefont{{von
  Oppen}}}, \bibinfo{journal}{Phys. Rev. Lett.} \textbf{\bibinfo{volume}{100}},
  \bibinfo{pages}{236602} (\bibinfo{year}{2008}).

\bibitem[{\citenamefont{{Rushforth} et~al.}(2008)\citenamefont{{Rushforth},
  {V{\'y}born{\'y}}, {King}, {Edmonds}, {Campion}, {Foxon}, {Wunderlich},
  {Irvine}, {Nov{\'a}k}, {Olejn{\'{\i}}k} et~al.}}]{Rushforth2008}
\bibinfo{author}{\bibfnamefont{A.~W.} \bibnamefont{{Rushforth}}},
  \bibinfo{author}{\bibfnamefont{K.}~\bibnamefont{{V{\'y}born{\'y}}}},
  \bibinfo{author}{\bibfnamefont{C.~S.} \bibnamefont{{King}}},
  \bibinfo{author}{\bibfnamefont{K.~W.} \bibnamefont{{Edmonds}}},
  \bibinfo{author}{\bibfnamefont{R.~P.} \bibnamefont{{Campion}}},
  \bibinfo{author}{\bibfnamefont{C.~T.} \bibnamefont{{Foxon}}},
  \bibinfo{author}{\bibfnamefont{J.}~\bibnamefont{{Wunderlich}}},
  \bibinfo{author}{\bibfnamefont{A.~C.} \bibnamefont{{Irvine}}},
  \bibinfo{author}{\bibfnamefont{V.}~\bibnamefont{{Nov{\'a}k}}},
  \bibinfo{author}{\bibfnamefont{K.}~\bibnamefont{{Olejn{\'{\i}}k}}},
  \bibnamefont{et~al.}, \bibinfo{journal}{J. Magn. Magn. Mater.}
  (\bibinfo{year}{2008}), \bibinfo{note}{doi:10.1016/j.jmmm.2008.04.070}.

\end{thebibliography}

\end{document}